\documentclass{article}
\usepackage{cite}
\usepackage{amsmath,amssymb,amsfonts}
\usepackage{graphicx}
\usepackage{calc}
\usepackage{textcomp}
\usepackage[justification=centering]{caption}
\usepackage[justification=centering]{subcaption}
\usepackage[font=footnotesize]{caption}
\usepackage{float}
\usepackage{pifont}
\usepackage{algorithm}
\usepackage{algpseudocode}
\usepackage{multicol}
\usepackage{mathtools}
\usepackage{tabularray}
\usepackage{forloop}
\usepackage{pgffor}
\usepackage{stackengine}
\usepackage{xcolor}
\definecolor{Links}{HTML}{7A0022}
\usepackage{hyperref}
\hypersetup{
    colorlinks=true,
    citecolor=Links,
    linkcolor=Links,
    filecolor=Links,      
    urlcolor=Links,
    pdftitle={Triangular Automata: The 256 Elementary Cellular Automata of the 2D Plane},
    pdfpagemode=FullScreen,
}

\def\BibTeX{{\rm B\kern-.05em{\sc i\kern-.025em b}\kern-.08em
    T\kern-.1667em\lower.7ex\hbox{E}\kern-.125emX}}
    
\providecommand{\keywords}[1]
{
  \small	
  \textbf{\textit{Keywords ---}} #1
}
    
\begin{document}

\title{\Huge Triangular Automata \\
\Large The 256 Elementary Cellular Automata of the 2D Plane}

\author{\href{https://orcid.org/0000-0002-3866-7615}{Paul Cousin} \\
        \small Université Paris Cité, France
}

\date{}

\maketitle

\begin{abstract}
Cellular automata on the triangular grid are here referred to as\linebreak Triangular Automata (TA). This paper focuses on the simplest class of TA, called Elementary Triangular Automata (ETA). They are argued to be the two-dimensional counterpart of Wolfram's Elementary Cellular Automata. Conceptual and computational tools for their study are presented, along with an initial analysis. This paper is accompanied by a website\footnote{Website: \href{https://triangular-automata.net}{https://triangular-automata.net}} where the results can be explored interactively. The source code is available in the form of a Mathematica package\footnote{Package demonstration: \href{https://triangular-automata.net/code}{https://triangular-automata.net/code}}. \smallskip

\end{abstract}

\keywords{cellular automata, triangular grid, dynamical systems, complexity.}

\section{Introduction} \label{introduction}
\begin{figure}[H]
    \centering
    \includegraphics[width=.75\textwidth]{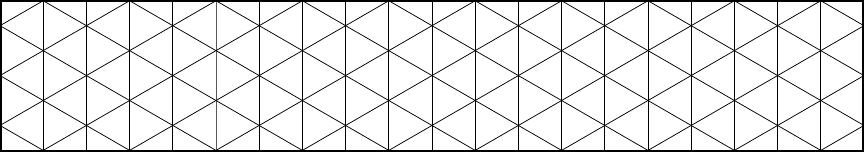}
    \caption{The triangular grid}
    \label{fig:triangular-grid}
\end{figure}

Cellular automata on the triangular grid (\textit{Figure \ref{fig:triangular-grid}}), or \textbf{Triangular Automata} (TA) for short, have already been studied in a few papers \cite{
gerlingClassificationTriangularHoneycomb1990,
baysCellularAutomataTriangular1994,
imaiComputationuniversalTwodimensional8state2000,
naumovGeneralizedCoordinatesCellular2003,
linApplicationUnstructuredCellular2009,
baysCellularAutomataTriangular2009,
baysGameLifeNonsquare2010,
zawidzkiApplicationSemitotalistic2D2011,
brecklingCellularAutomataEcological2011,
saadatCellularAutomataTriangular2016,
ortigozaACFUEGOSUnstructuredTriangular2016,
uguzStructureReversibility2D2017,
saadatCellularAutomataApproach2018,
wainerIntroductionCellularAutomata2019,
pavlovaUsingCellularAutomata2020,
saadatGeneratingPatternsTriangular2021,
saadatCopyMachinesSelfreproduction2023}. This paper focuses on a natural subset of TA called \textbf{Elementary TA} (ETA). 

\begin{figure}[H]
     \centering
     \hfill
     \begin{subfigure}[b]{0.1\textwidth}
         \centering
         \includegraphics[width=\textwidth]{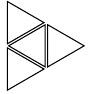}
         \caption*{0}
     \end{subfigure}
     \hfill
     \begin{subfigure}[b]{0.1\textwidth}
         \centering
         \includegraphics[width=\textwidth]{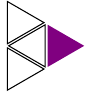}
         \caption*{1}
     \end{subfigure}
     \hfill
     \begin{subfigure}[b]{0.1\textwidth}
         \centering
         \includegraphics[width=\textwidth]{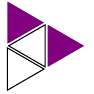}
         \caption*{2}
     \end{subfigure}
     \hfill
     \begin{subfigure}[b]{0.1\textwidth}
         \centering
         \includegraphics[width=\textwidth]{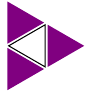}
         \caption*{3}
     \end{subfigure}
     \hfill
     \begin{subfigure}[b]{0.1\textwidth}
         \centering
         \includegraphics[width=\textwidth]{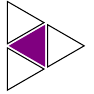}
         \caption*{4}
     \end{subfigure}
     \hfill
     \begin{subfigure}[b]{0.1\textwidth}
         \centering
         \includegraphics[width=\textwidth]{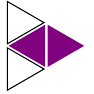}
         \caption*{5}
     \end{subfigure}
     \hfill
     \begin{subfigure}[b]{0.1\textwidth}
         \centering
         \includegraphics[width=\textwidth]{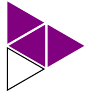}
         \caption*{6}
     \end{subfigure}
     \hfill
     \begin{subfigure}[b]{0.1\textwidth}
         \centering
         \includegraphics[width=\textwidth]{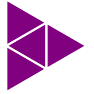}
         \caption*{7}
     \end{subfigure}
     \hfill
     \caption{All possible local configurations}
        \label{fig:configurations}
\end{figure}

\noindent ETA cells hold only \textbf{binary states}; each cell can thus be either:
\begin{enumerate}
    \item[•] ``alive" and colored purple \raisebox{-0.2\height}{\includegraphics[height=3mm]{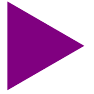}}, with a state $s=1$ 
    \item[•] ``dead" and colored white \raisebox{-0.2\height}{\includegraphics[height=3mm]{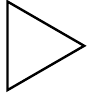}},  with a state $s=0$
\end{enumerate}
\  \\
ETA \textbf{rules} determine the future state of a cell based on its current state and the states of its neighbors, regardless of their orientation. This results in only 8 possible local configurations, as shown in \textit{Figure \ref{fig:configurations}}.

\begin{figure}[H]
     \centering
     \begin{subfigure}[b]{0.40\textwidth}
         \centering
         \includegraphics[width=\textwidth]{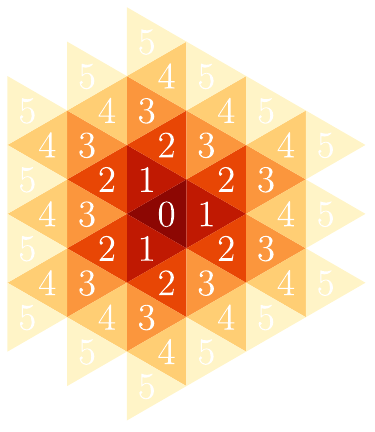}
         \caption{Grid and layers}
         \label{fig:layers}
     \end{subfigure}
     \hspace{0.05\textwidth}
     \begin{subfigure}[b]{0.35\textwidth}
         \centering
         \raisebox{4mm}{\includegraphics[width=\textwidth]{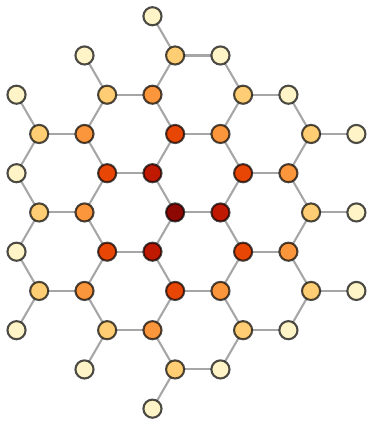}}
         \caption{Corresponding graph}
         \label{fig:graph}
     \end{subfigure}
        \caption{Structure of the triangular grid}
        \label{fig:structure-of-the-triangular-grid}
\end{figure}

This paper uses a graph-theoretical framework developed in a previous work on Graph-Rewriting Automata \cite{cousin2022organic}. The triangular grid will here be considered as a graph (\textit{Figure \ref{fig:structure-of-the-triangular-grid}}). This graph must be expanded at each time step to simulate an infinite grid. The \textbf{region of influence} of a single cell grows in hexagonal layers (\textit{Figure \ref{fig:layers}}). This is therefore the most efficient way to expand the graph as well. How to do this in practice is detailed in \textit{Section \ref{growing-the-grid}}.

 It is useful to see the triangular grid as a graph because computing the evolution of an ETA is made relatively straightforward by properties of its \textbf{adjacency matrix} $\mathcal{A}$\linebreak and \textbf{state vector} $\mathcal{S}$. Every \textbf{vertex} $v$ of this graph will hold a \textbf{state} $s(v)$. The \textbf{neighborhood} $N(v)$ of a vertex is defined as the set of its adjacent vertices.

\begin{equation}
\begin{array}{ccc}
    \mathcal{A}_{ij}=
    \begin{cases}
        1 & \text{ if } v_i\in N(v_j) \\
        0 & \text{ otherwise}
    \end{cases}
&\hspace{1cm}&
    \mathcal{S}_i=s(v_i)\in\{0,1\}
\end{array}
\end{equation}

The \textbf{configuration} $c(v)$ of a vertex is a number that, when indexed as in \textit{Figure \ref{fig:configurations}},\linebreak can be expressed as:
\begin{equation}
c(v)=4\times s(v)+\sum_{i\in N(v)} s(i)
\end{equation}

The space of possible ETA rules is finite. For each of the 8 configurations, a rule must specify whether the vertex will be dead or alive at $t+1$. Consequently, there are only $2^8=256$ possible rules. For this reason, ETA can be seen as the two-dimensional counterpart of Wolfram’s 256 Elementary Cellular Automata \cite{wolfram2002new, weisstein2002elementary}. In fact, there is a direct equivalence between 64 of each (see \textit{Section \ref{elementary-cellular-automata}}). Furthermore, the triangle is the regular polygon that tiles the 2D plane with the smallest number of neighbors per cell. ETA are thus the most basic two-dimensional cellular automata and are fundamental in this regard. \\

\noindent A \textbf{rule} $R$ is a map from \textbf{configuration space} to \textbf{state space}.
\begin{equation}
\begin{aligned}
& R: \{0,1,2,3,4,5,6,7\}\rightarrow\{0,1\}\\  
& R\big(c_t(v)\big)=s_{t+1}\big(v\big)
\end{aligned}
\end{equation}

\noindent Each rule can be labeled by a unique \textbf{rule number} $n$: 

\begin{equation} \label{eq:rule-number}
    n=\sum_{i=0}^7 2^i R(i)
\end{equation}

\begin{figure}[H]
    \centering
    \includegraphics[width=.8\textwidth]{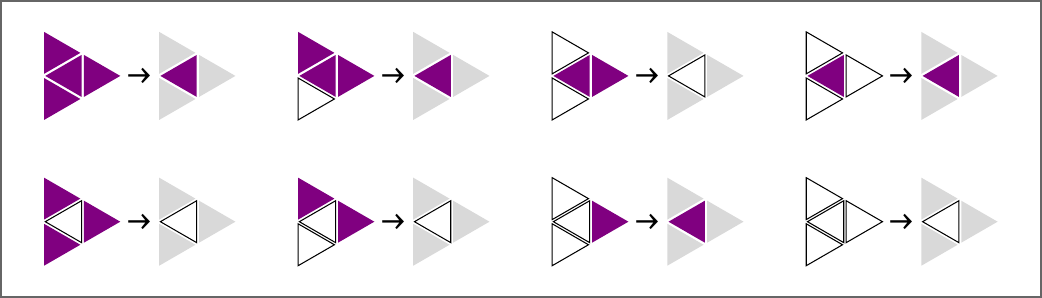}
    \caption{Rule $210=1101\,0010_2$}
    \label{fig:rule-plot-210}
\end{figure}

This labeling system was proposed independently in \cite{zawidzkiApplicationSemitotalistic2D2011} and \cite{cousin2022organic} and will be used in this paper, since it is fairly natural and has useful properties. This system, inspired by the Wolfram code \cite{wolfram2002new}, is such that a rule number in its binary form displays the behavior of the rule. Starting from the right, its digits indicate the future state for each configuration as they have been ordered previously. \textit{Figure \ref{fig:rule-plot-210}} shows the example of rule 210.

\pagebreak
\section{Behavior} \label{behavior}
Before going into the details of how to compute these automata, we can take a look at how they behave. In this section, a preliminary study of ETA is presented to motivate a future, more in-depth analysis. It is of particular interest to see what happens to a single living cell under different ETA rules, so, unless otherwise noted, the following figures come from this starting point.

\subsection{Beauty} \label{beauty}
One of the most striking aspects of these automata is their aesthetic quality, which cannot be better illustrated than by a few selected examples.

\begin{figure}[H]
    \centering
    \begin{subfigure}[b]{0.45\textwidth}
        \centering
        \includegraphics[width=\textwidth]{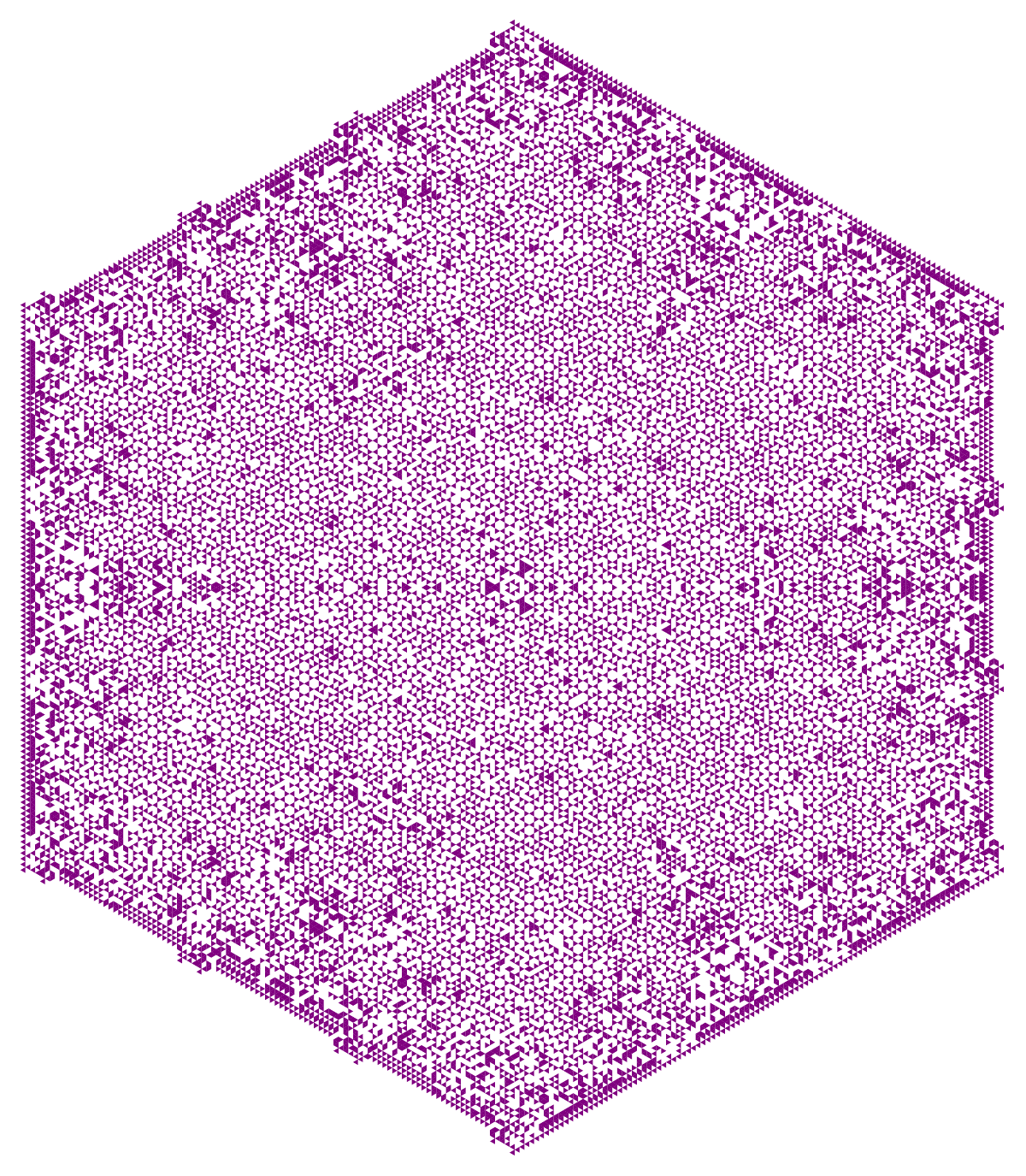}
        \caption{Rule 89 at $t=200$}
        \label{fig:rule-89-time-200-OneAlive}
    \end{subfigure}
    \begin{subfigure}[b]{0.45\textwidth}
        \centering
        \raisebox{0.08\textwidth}{
            \hspace{-0.035\textwidth}
            \includegraphics[width=\textwidth]{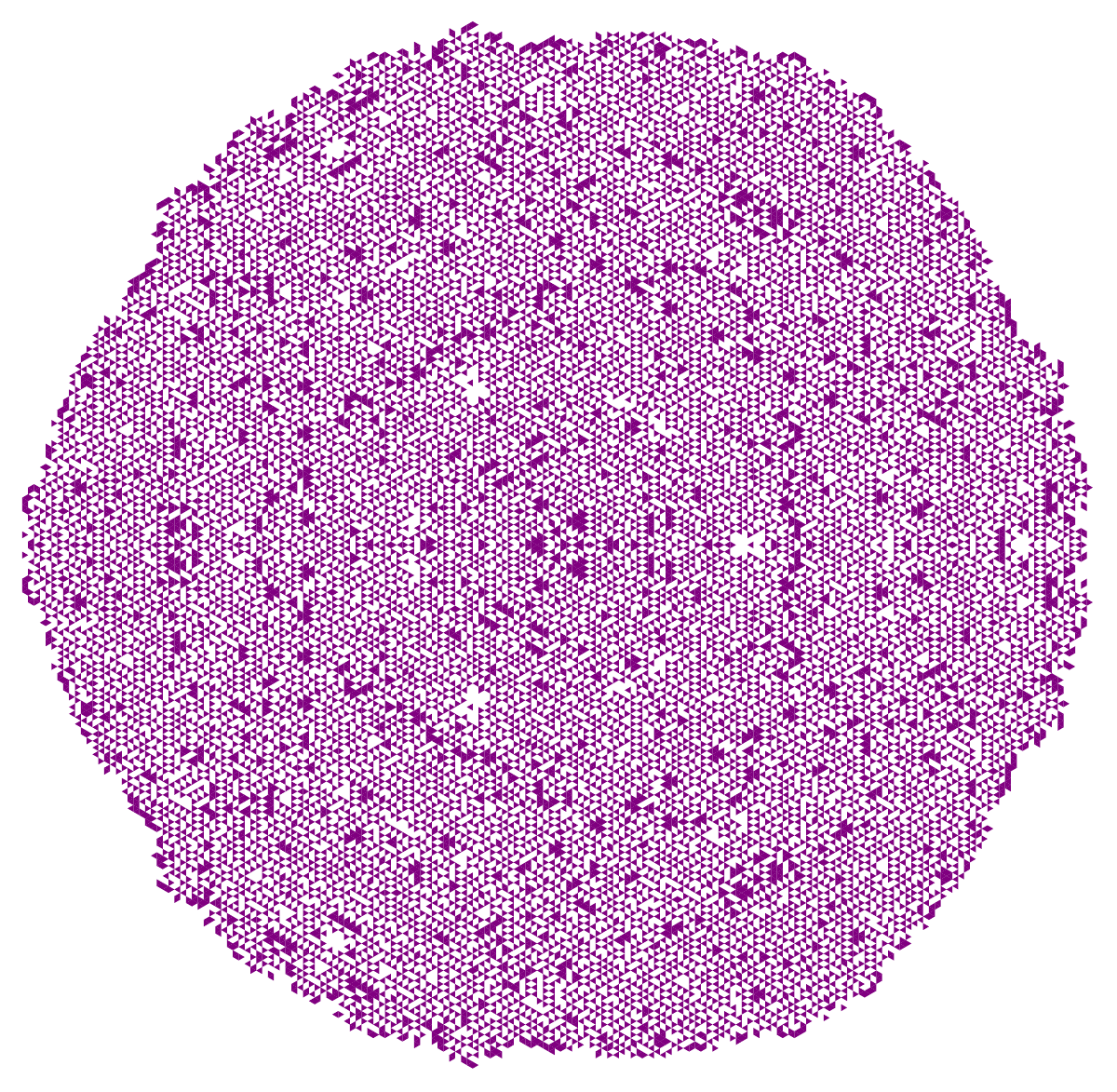}
        }
        \caption{Rule 57 at $t=320$}
        \label{fig:rule-57-time-320-OneAlive}
    \end{subfigure}
    \break
    \begin{subfigure}[b]{0.45\textwidth}
        \centering
        \includegraphics[width=\textwidth]{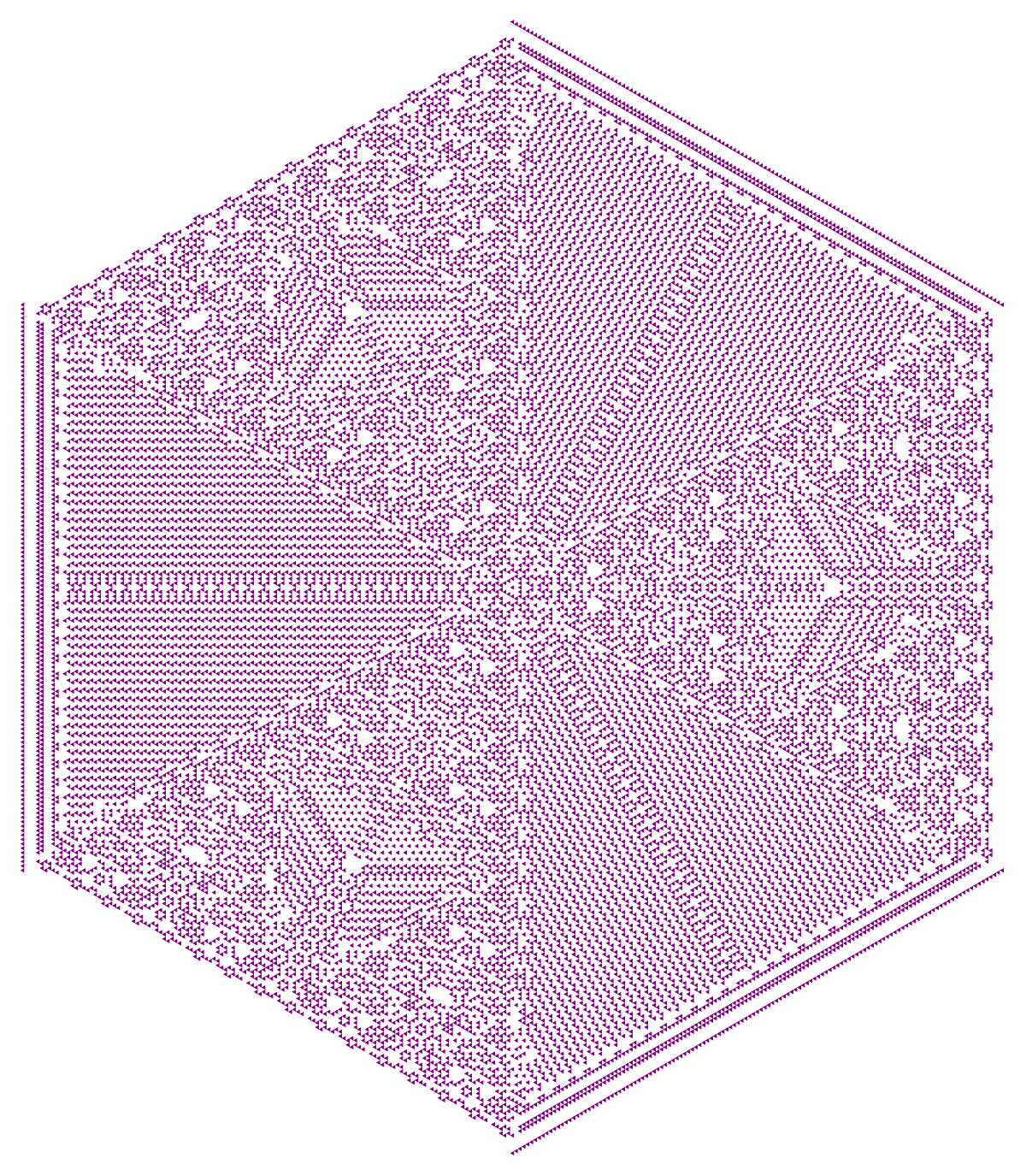}
        \caption{Rule 73 at $t=256$}
        \label{fig:rule-73-time-256-OneAlive}
    \end{subfigure}
    \begin{subfigure}[b]{0.45\textwidth}
        \centering
        \includegraphics[width=\textwidth]{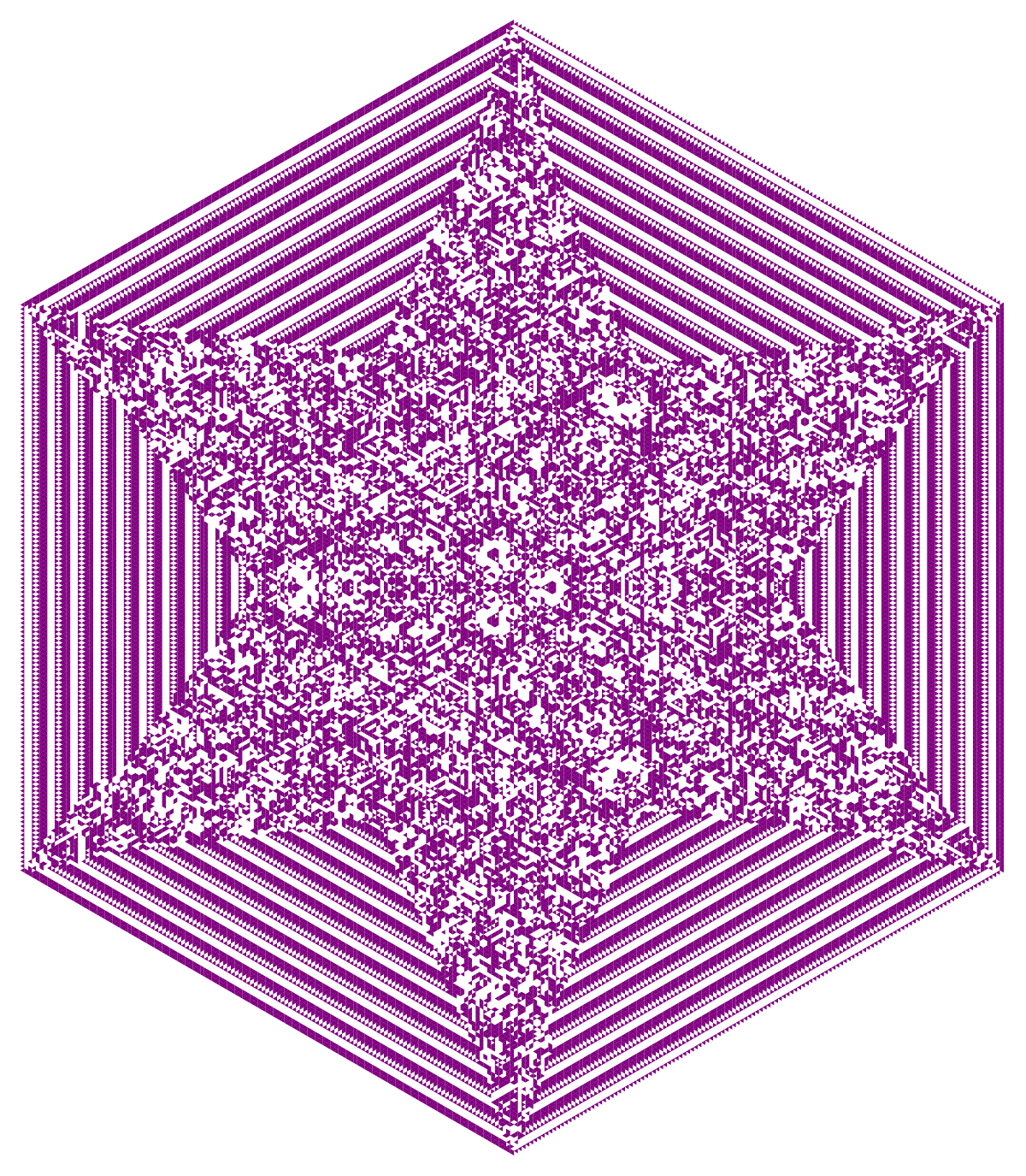}
        \caption{Rule 62 at $t=256$}
        \label{fig:rule-62-time-256-OneAlive}
    \end{subfigure}
    \caption{The beauty of ETA}
    \label{fig:beauty}
\end{figure}

\pagebreak
\subsection{Fractals} \label{fractals}
Some ETA rules produce remarkable scale-free structures. 

\begin{figure}[H]
    \centering
    \begin{subfigure}[b]{0.68\textwidth}
        \centering
        \includegraphics[width=\textwidth]{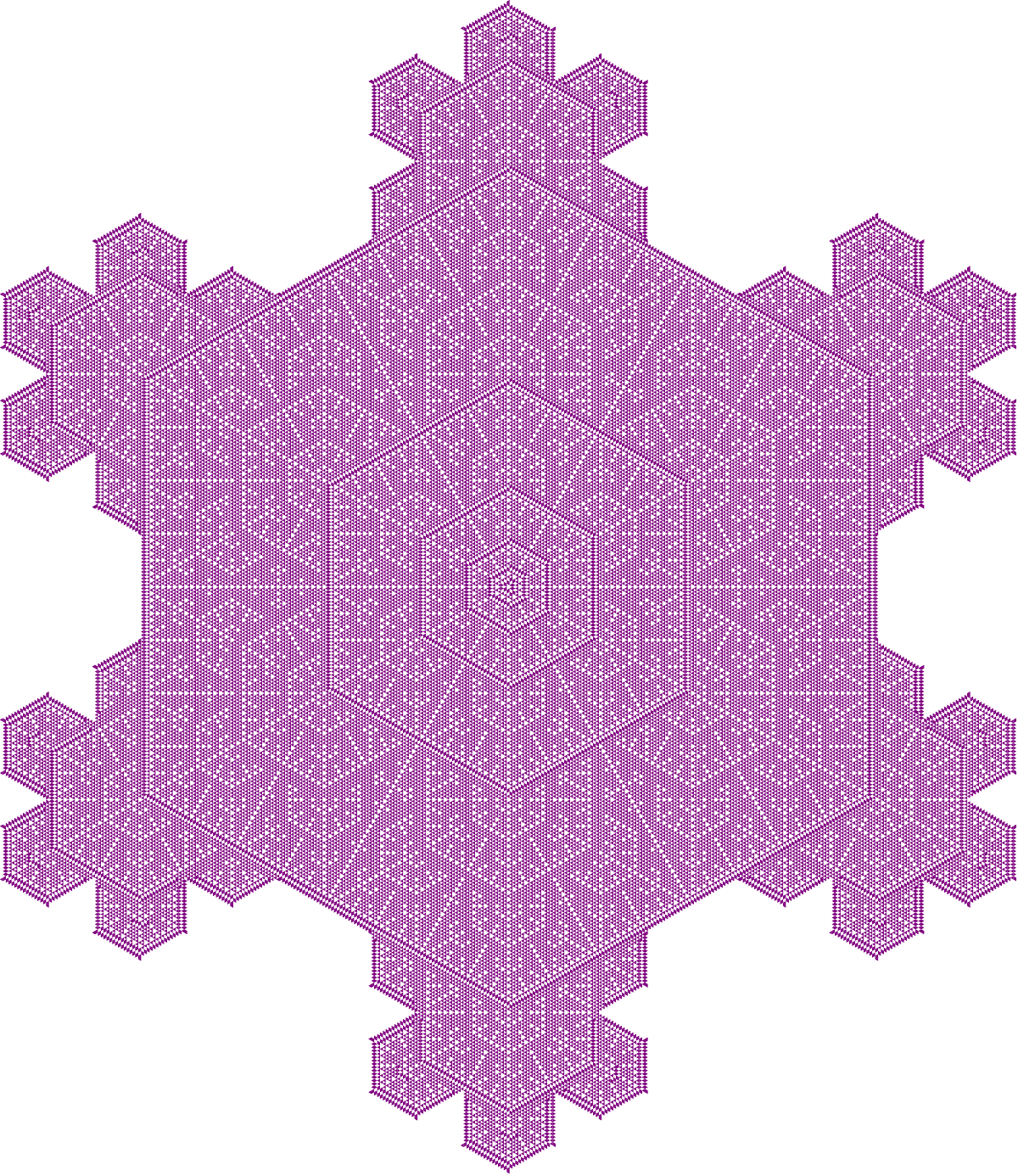}
        \caption{Rule 50 at $t=352$}
        \label{fig:rule-50-time-352-OneAlive}
    \end{subfigure}
    \break
    \begin{subfigure}[b]{0.46\textwidth}
        \centering
        \includegraphics[width=\textwidth]{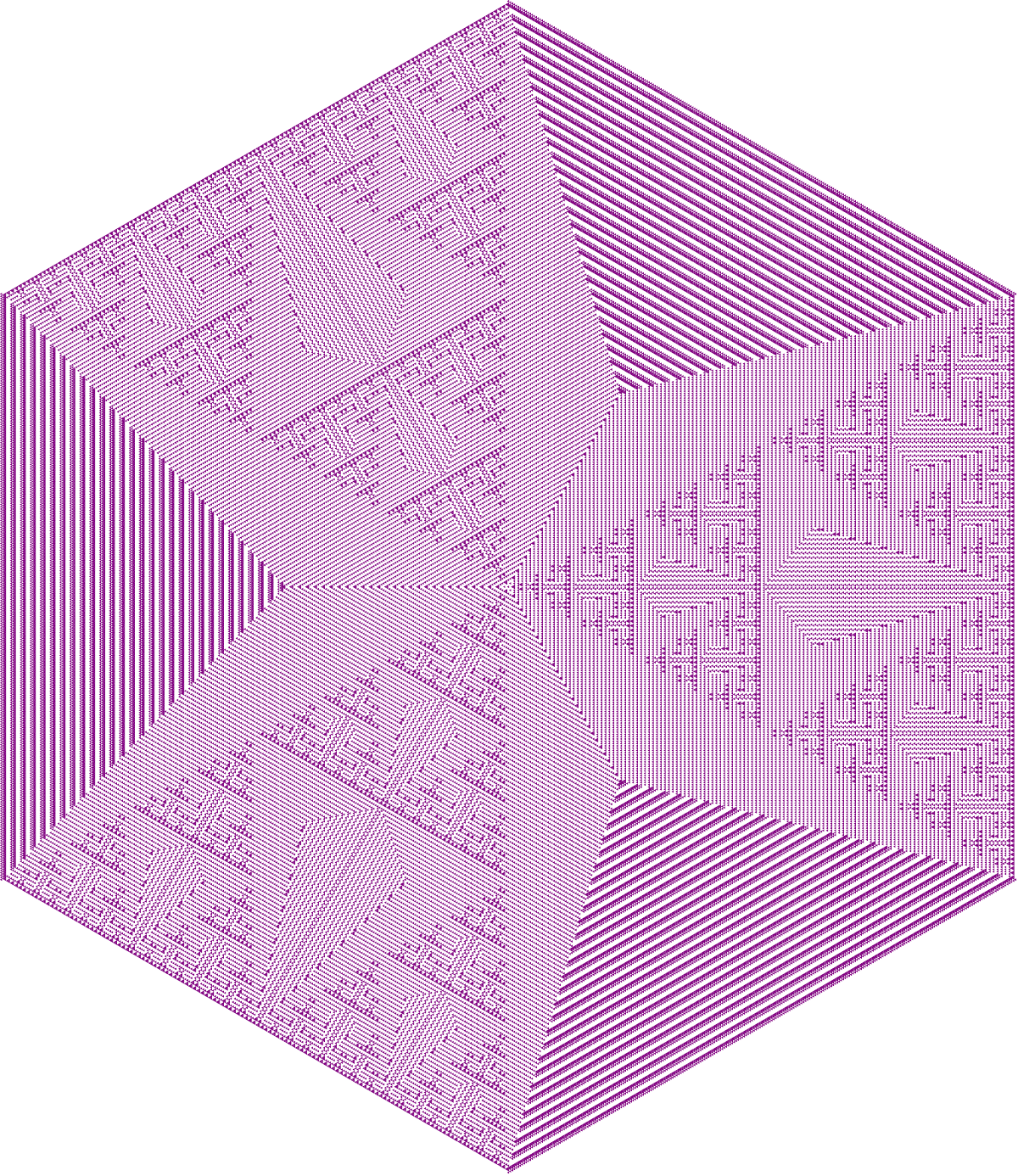}
        \caption{Rule 65 at $t=512$}
        \label{fig:rule-65-time-512-OneAlive}
    \end{subfigure}
    \hspace{0.06\textwidth}
    \begin{subfigure}[b]{0.46\textwidth}
        \centering
        \includegraphics[width=\textwidth]{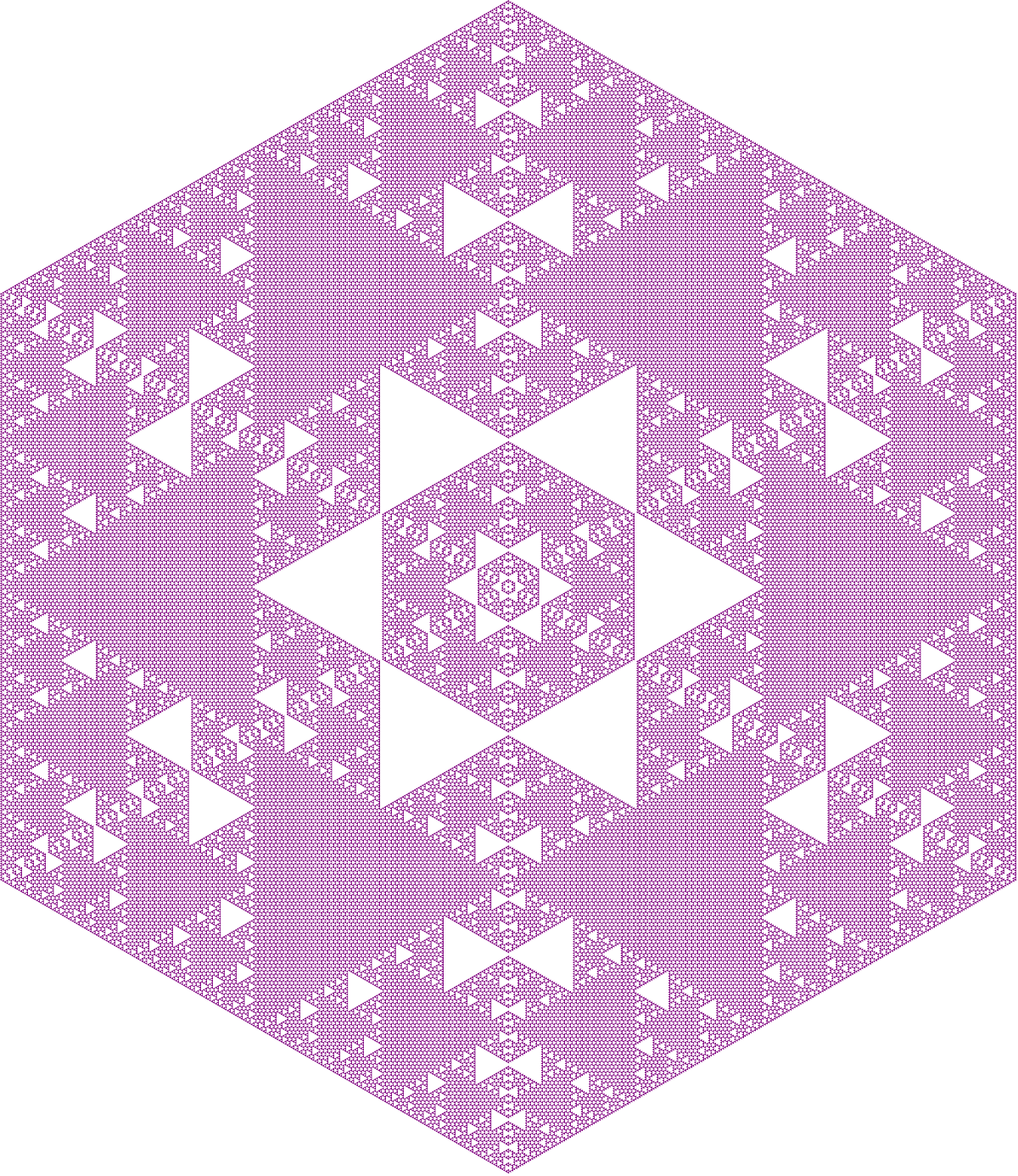}
        \caption{Rule 106 at $t=510$}
        \label{fig:rule-106-time-510-OneAlive}
    \end{subfigure}
    \caption{Fractals}
    \label{fig:fractals}
\end{figure}

\pagebreak
\subsection{Chaos} \label{chaos}
Given the existing literature on cellular automata, it is not surprising that some rules behave chaotically. The example of rule 53 confirms it. Starting from two randomly generated 64-layer-wide grids that are completely similar except for the central cell which is alive in (1) and dead in (2), the trajectories strongly diverge.

\begin{figure}[H]
     \centering
     \begin{subfigure}[b]{0.23\textwidth}
         \centering
         \includegraphics[width=\textwidth]{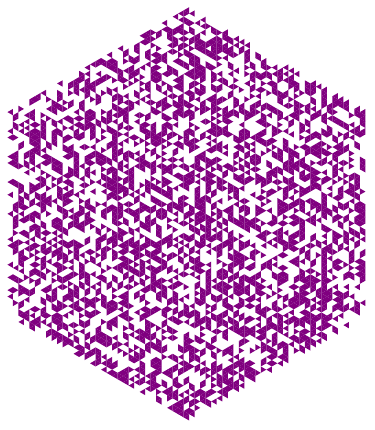}
         \caption*{(1) at $t=0$}
     \end{subfigure}
     \hspace{0.17\textwidth}
     \begin{subfigure}[b]{0.23\textwidth}
         \centering
         \includegraphics[width=\textwidth]{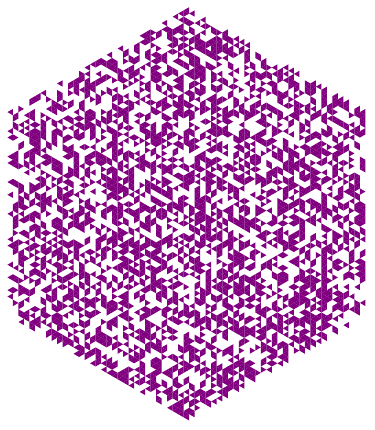}
         \caption*{(2) at $t=0$}
     \end{subfigure}
    \break\vspace{5mm}
     \begin{subfigure}[b]{0.48\textwidth}
         \centering
         \includegraphics[width=\textwidth]{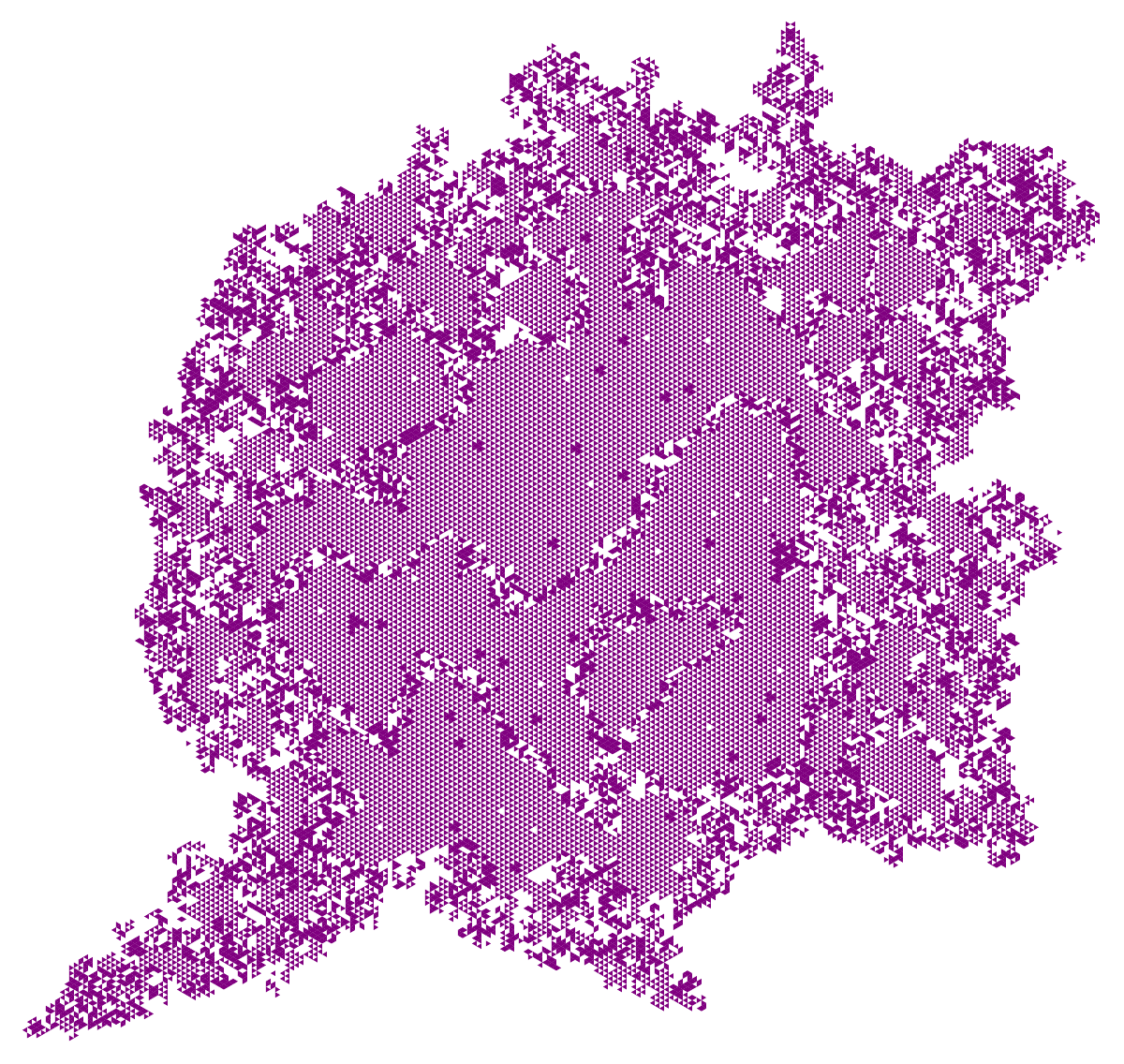}
         \caption*{(1) at $t=512$}
     \end{subfigure}
     \begin{subfigure}[b]{0.48\textwidth}
         \centering
         \includegraphics[width=\textwidth]{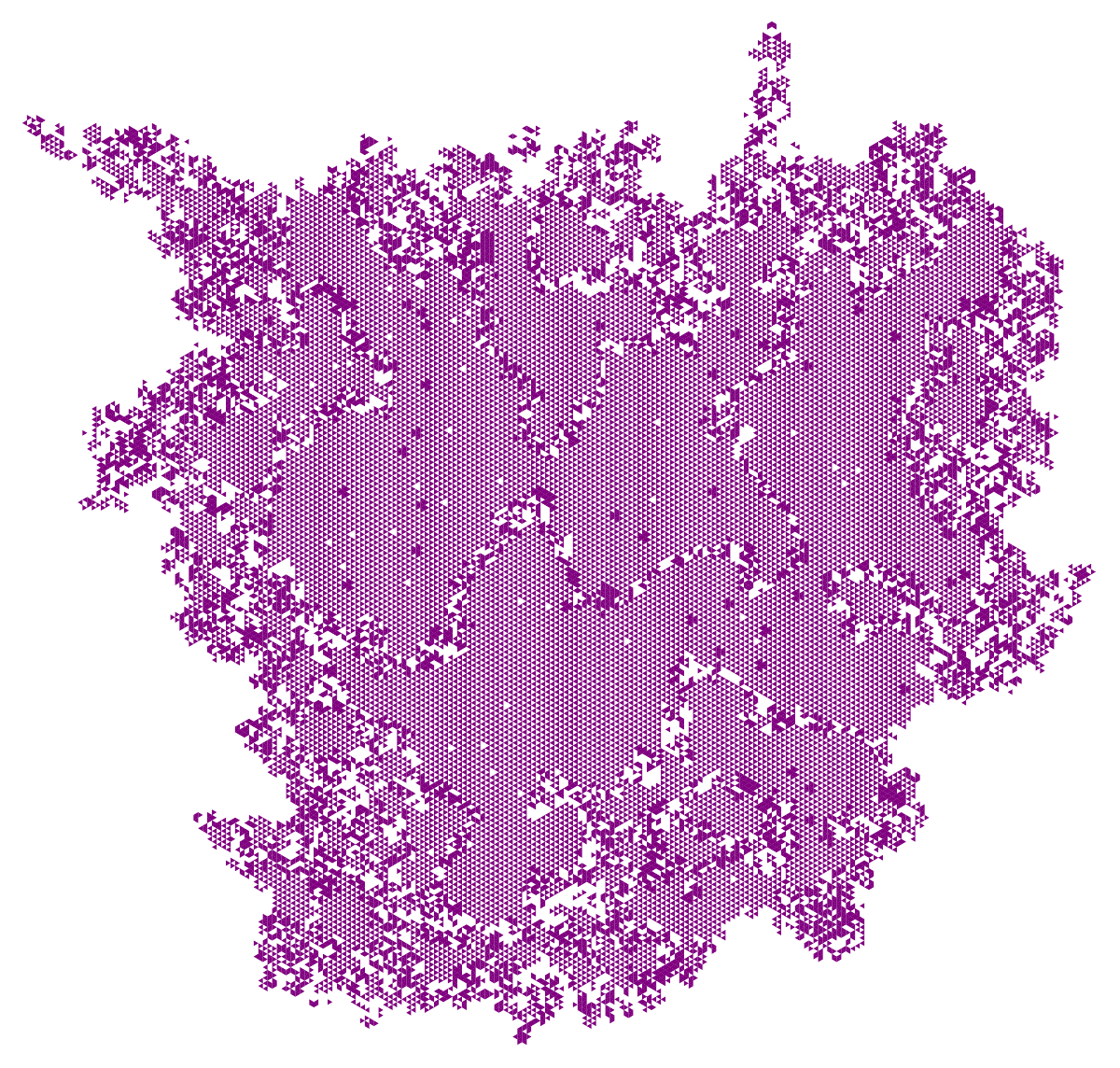}
         \caption*{(2) at $t=512$}
     \end{subfigure}
     \vspace{-5mm}
     \caption{Chaotic behavior of rule 53}
        \label{fig:chaos}
\end{figure}

\subsection{Gliders} \label{gliders}
\begin{figure}[H]
    \centering
        \includegraphics[width=.5555555\textwidth]{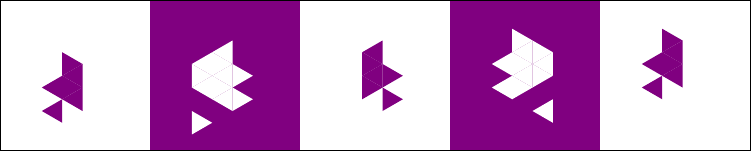}
        \vspace{-2mm}\caption{A glider of rule 21}
    \label{fig:glider21}
\end{figure}

\begin{figure}[H]
    \centering
        \includegraphics[width=1\textwidth]{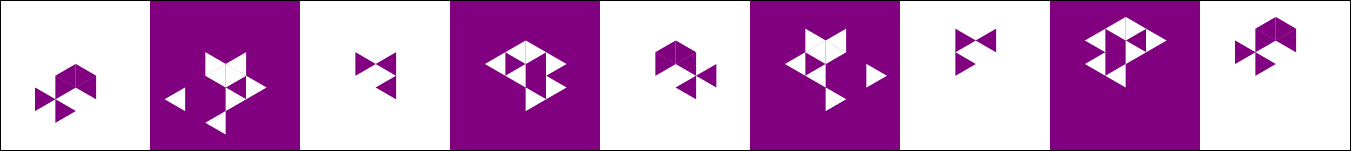}
        \vspace{-6mm}
    \caption{A glider of rule 53}
    \label{fig:glider53}
\end{figure}

\textit{Figures \ref{fig:glider21}} and \textit{\ref{fig:glider53}} show two small gliders.

\vspace{-2mm}

\pagebreak
\subsection{Self-Reproduction} \label{self-reproduction}

As mentioned in \cite{saadatCopyMachinesSelfreproduction2023}, one of the original motivations for the development of cellular automata was to create a mathematical model of self-reproduction. Interestingly, 4 of\linebreak the 256 ETA rules naturally replicate any finite pattern given as initial conditions: rules 85, 90, 165 and 170. A proof of self-reproduction based on path counting already exists for rule 170 \cite{saadatCopyMachinesSelfreproduction2023}. Similar proofs could probably be proposed for the others.

\vspace{.6cm}

\begin{table}[H]
    \centering
    \begin{tblr}{|Q[c,m]|Q[c,h] Q[c,h] Q[c,h] Q[c,h]|}
        \hline
            \textit{t} & rule 85 & rule 90 & rule 165 & rule 170 \\
        \hline
            0 &
            \raisebox{-.05\height}{\includegraphics[width=0.19\textwidth]{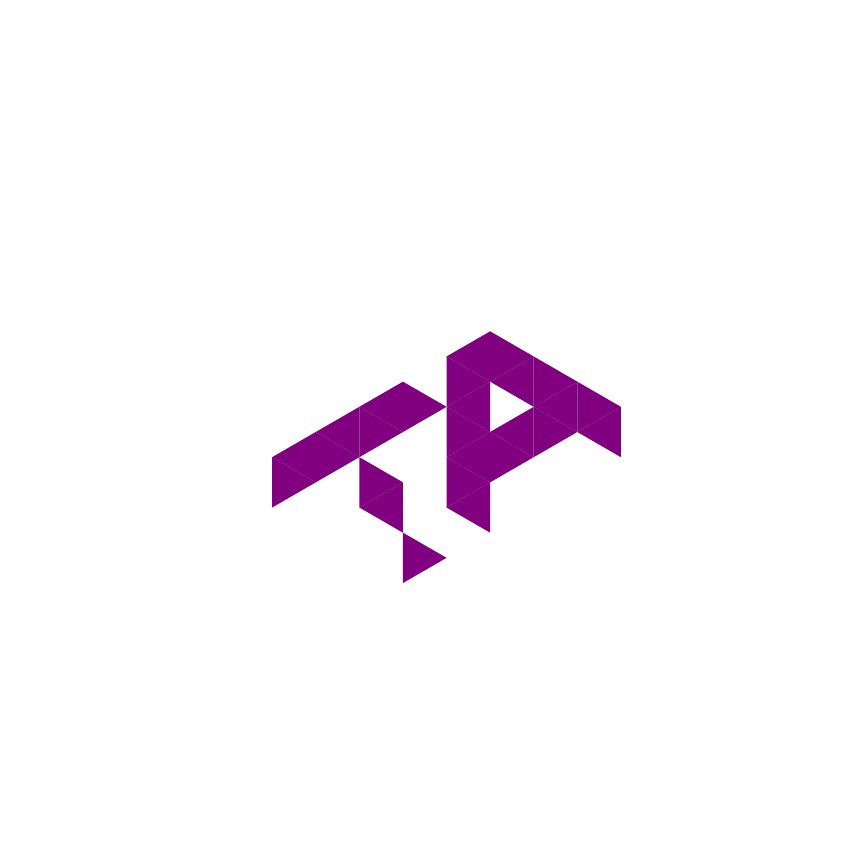}} &
            \raisebox{-.05\height}{\includegraphics[width=0.19\textwidth]{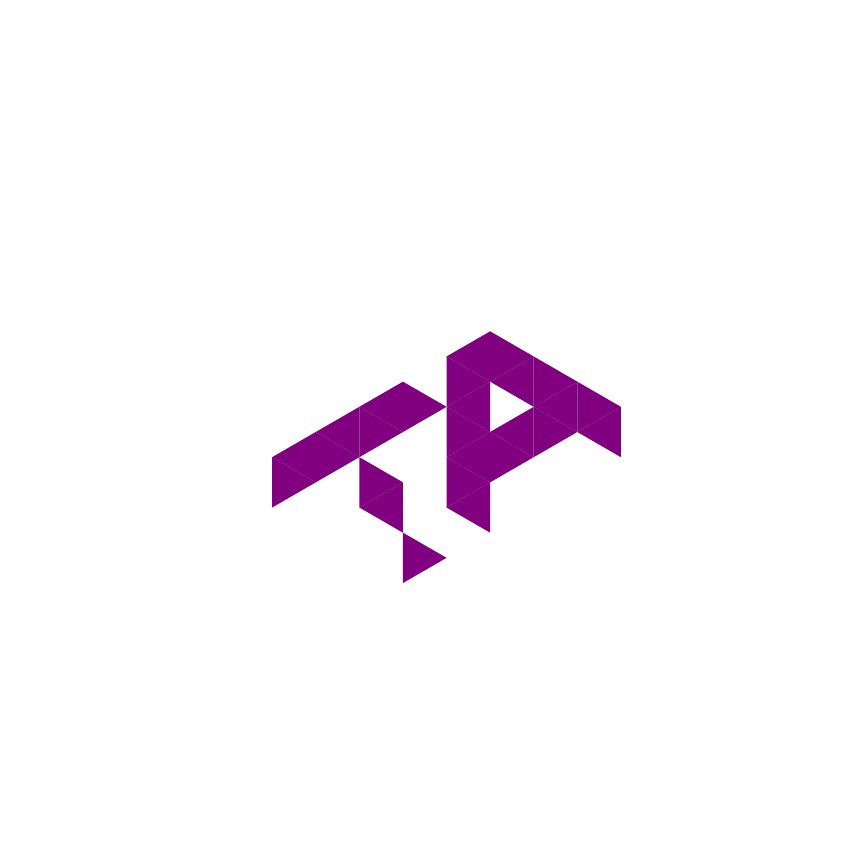}} &
            \raisebox{-.05\height}{\includegraphics[width=0.19\textwidth]{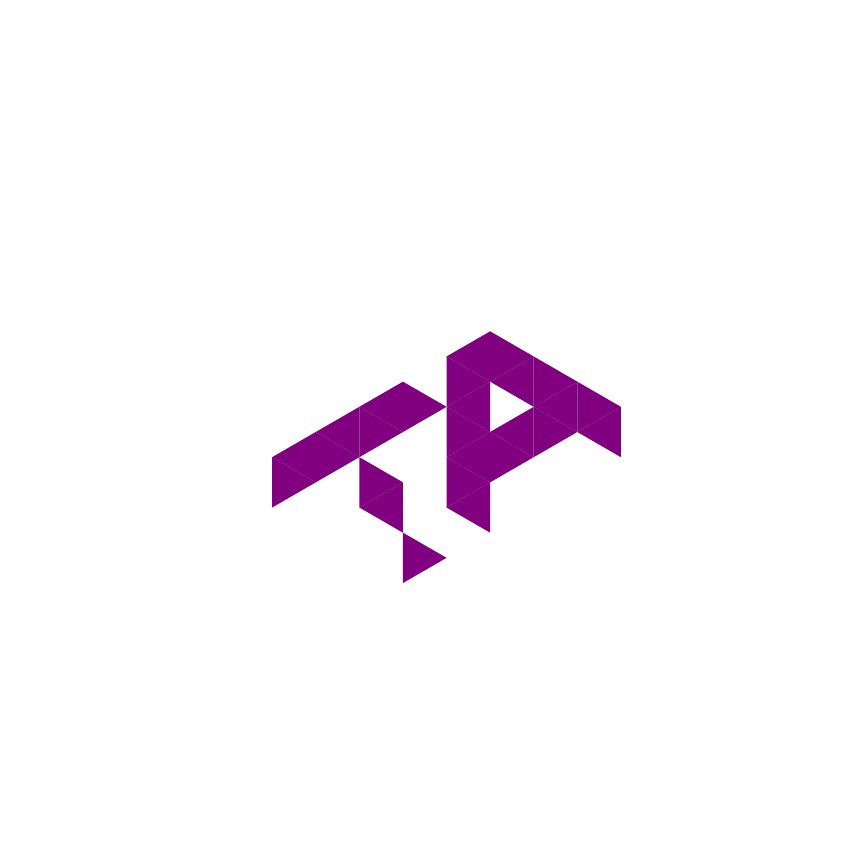}} &
            \raisebox{-.05\height}{\includegraphics[width=0.19\textwidth]{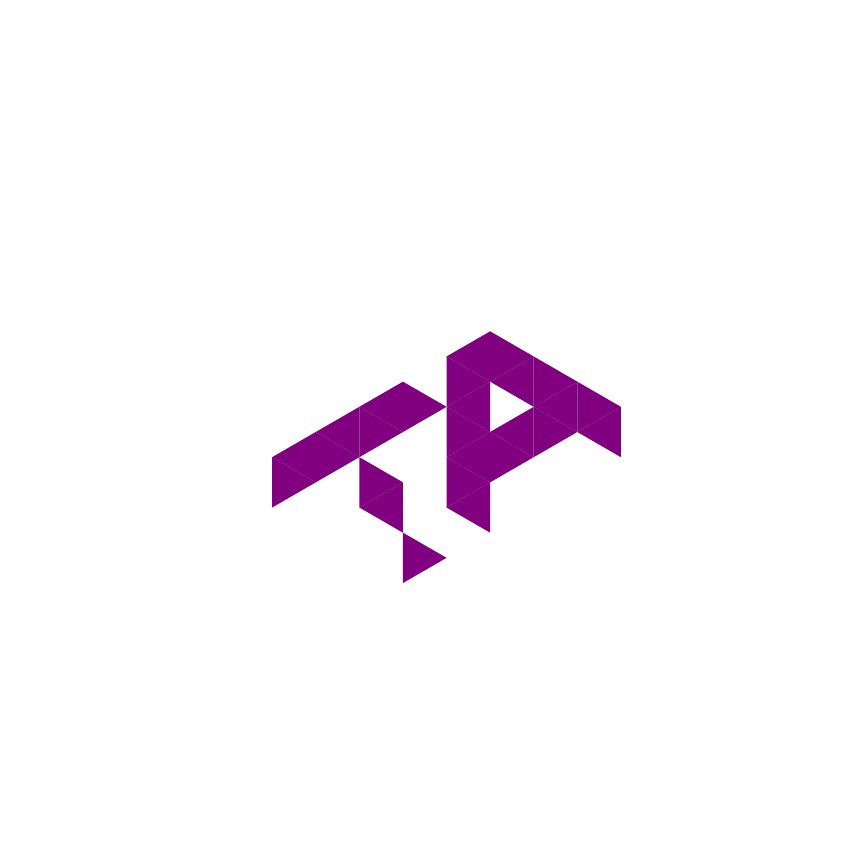}} \\
            16 &
            \raisebox{-.05\height}{\includegraphics[width=0.19\textwidth]{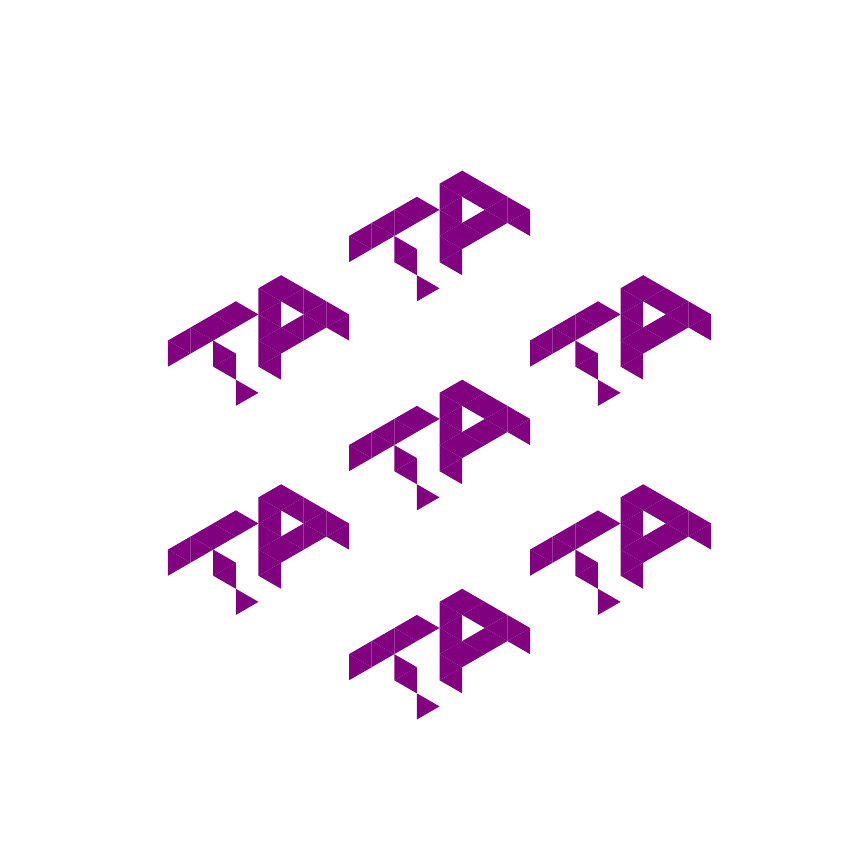}} &
            \raisebox{-.05\height}{\includegraphics[width=0.19\textwidth]{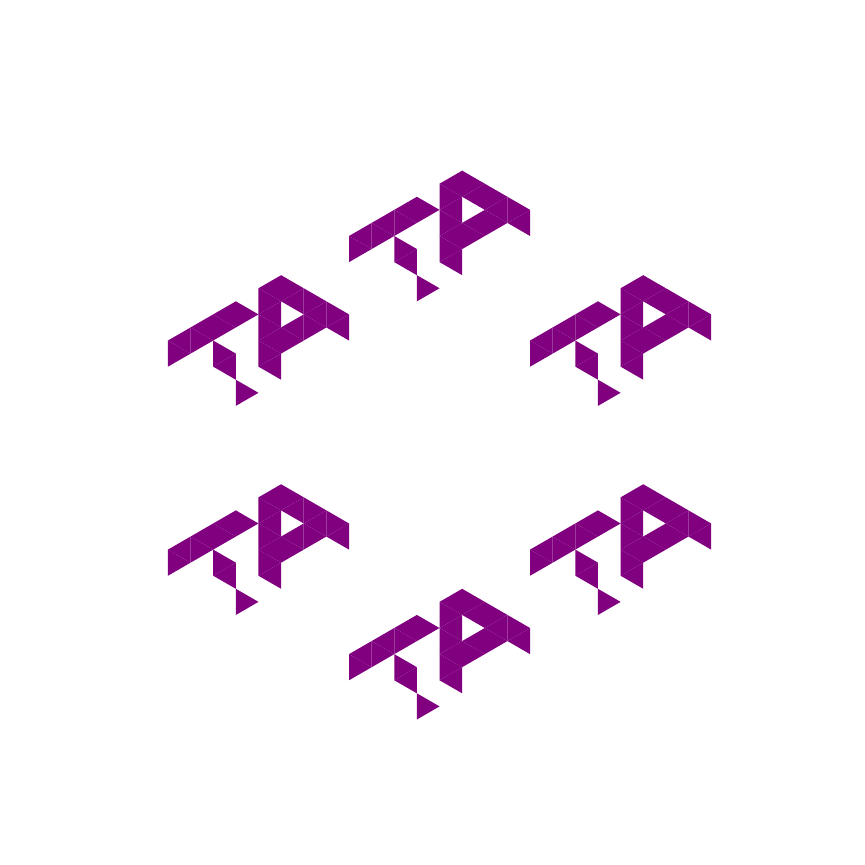}} &
            \raisebox{-.05\height}{\includegraphics[width=0.19\textwidth]{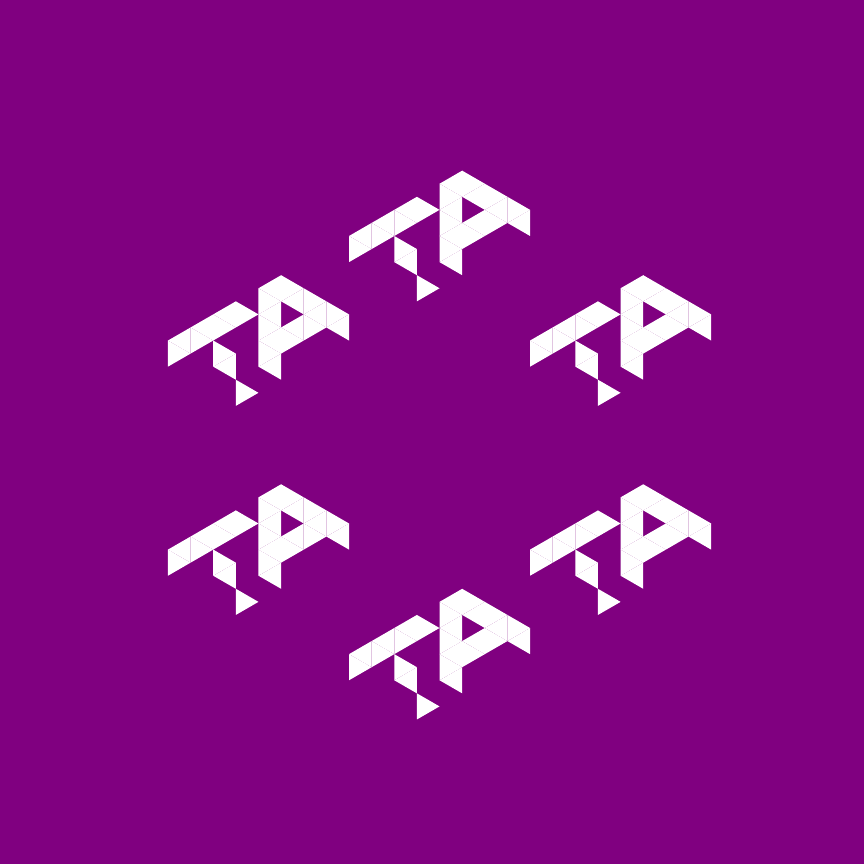}} &
            \raisebox{-.05\height}{\includegraphics[width=0.19\textwidth]{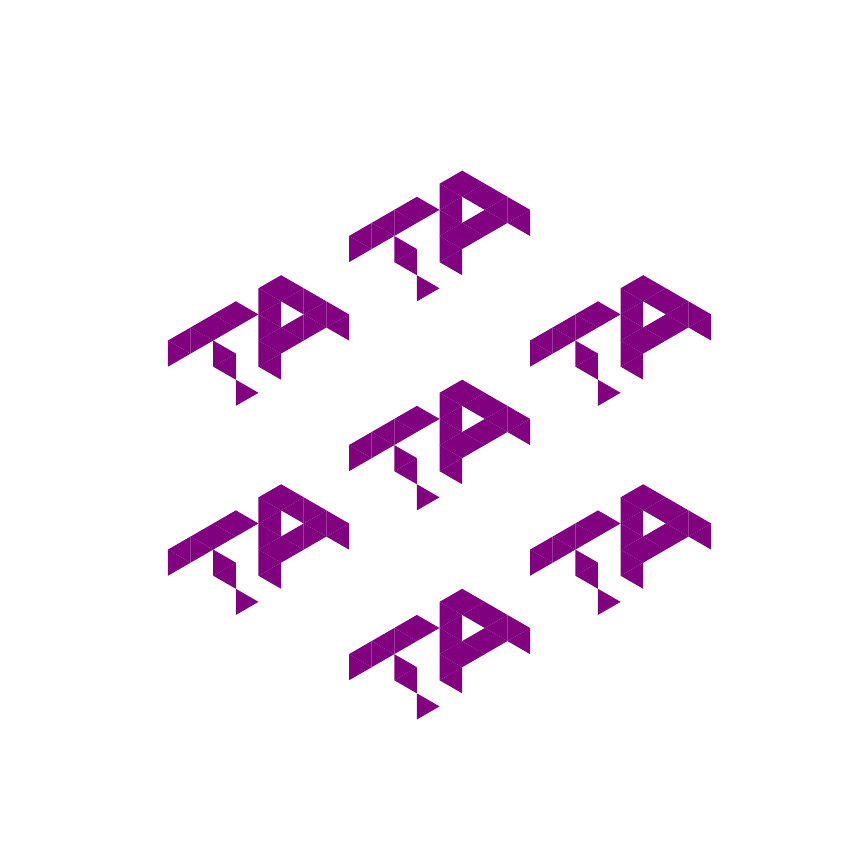}} \\
            32 &
            \raisebox{-.05\height}{\includegraphics[width=0.19\textwidth]{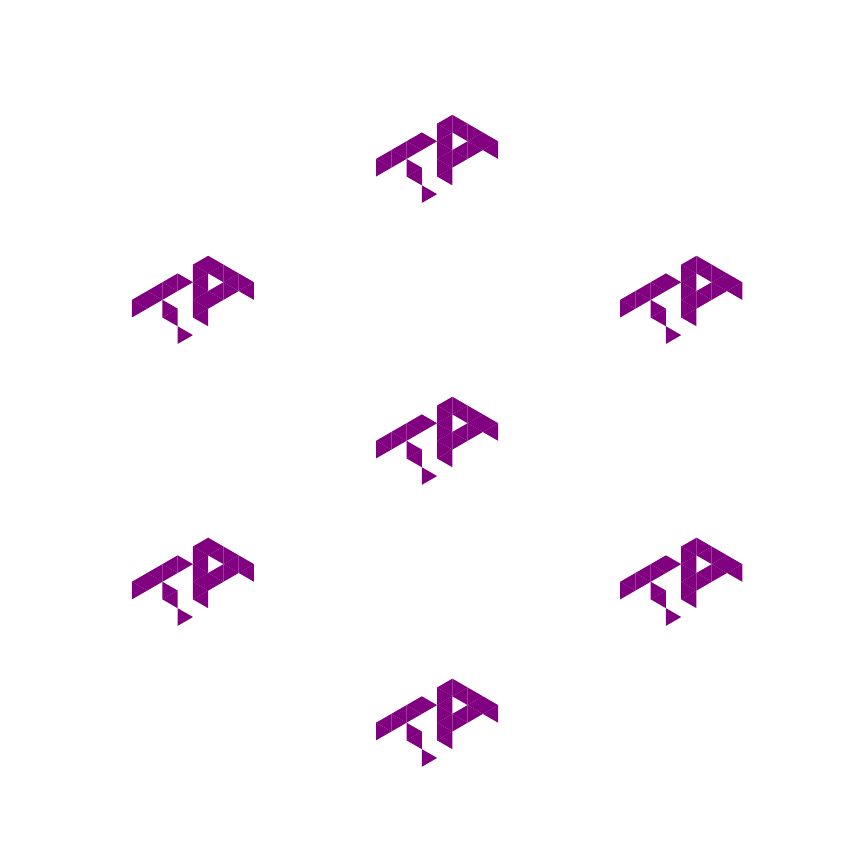}} &
            \raisebox{-.05\height}{\includegraphics[width=0.19\textwidth]{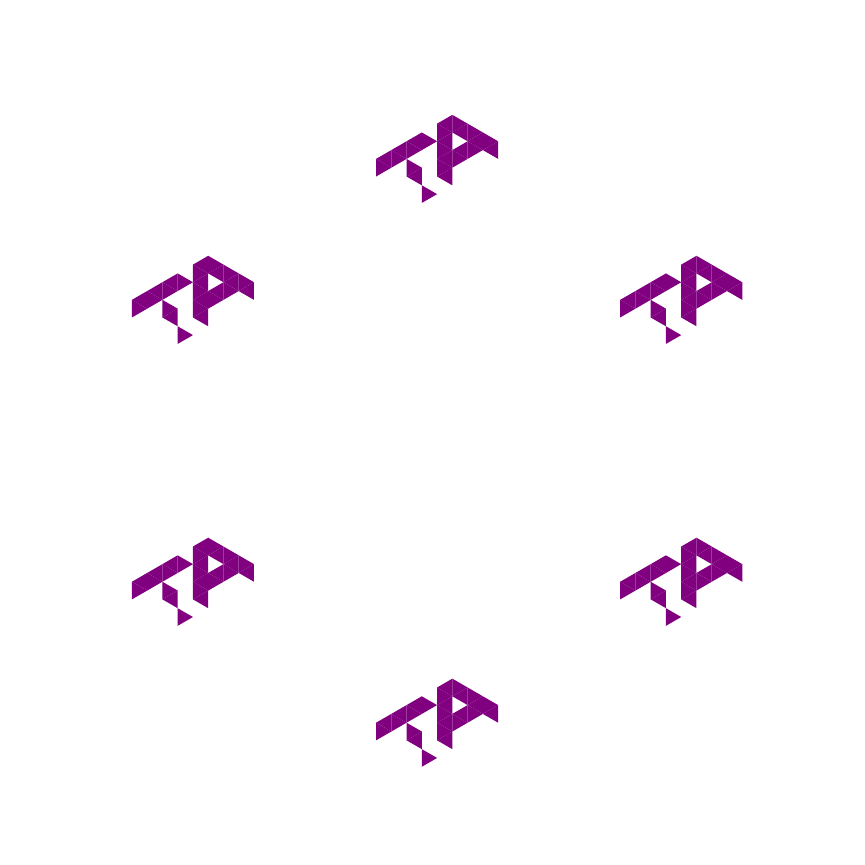}} &
            \raisebox{-.05\height}{\includegraphics[width=0.19\textwidth]{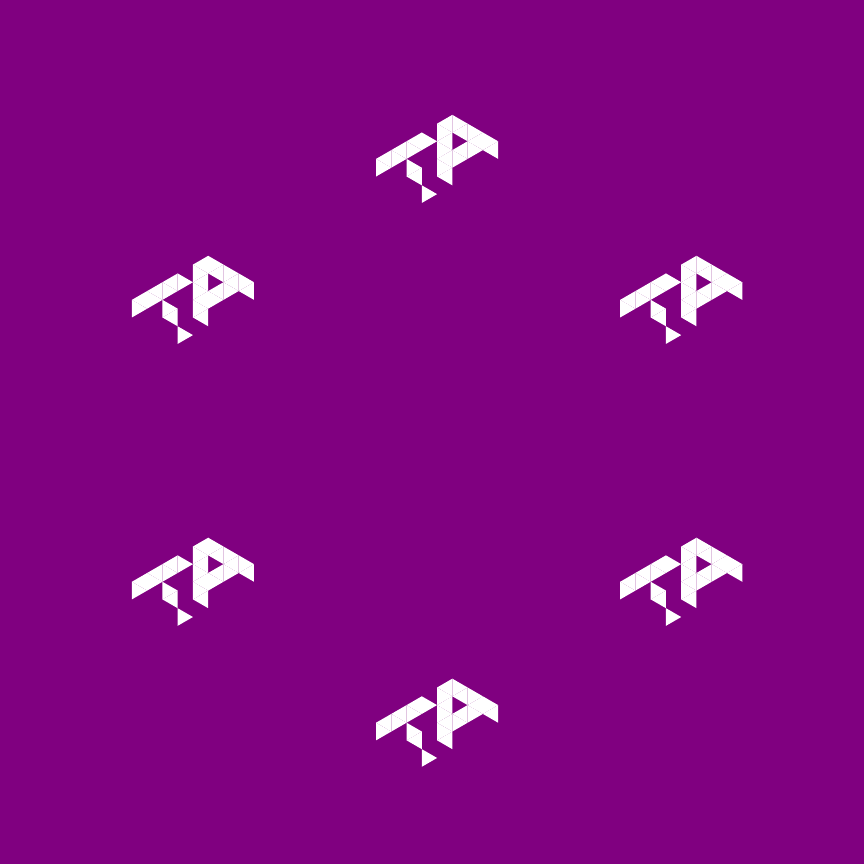}} &
            \raisebox{-.05\height}{\includegraphics[width=0.19\textwidth]{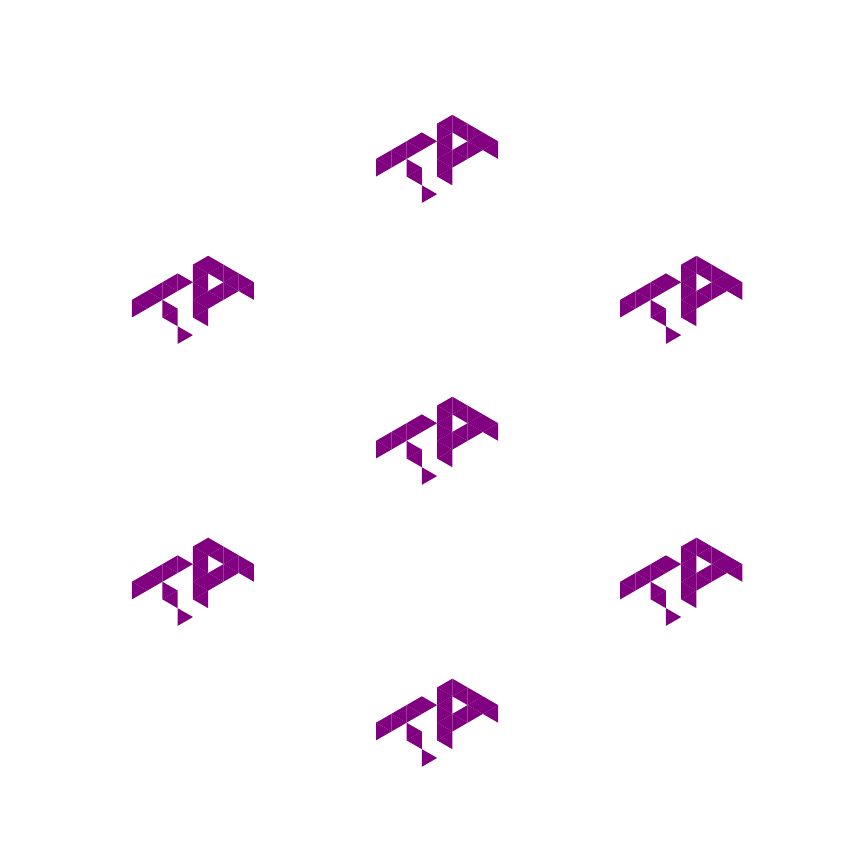}} \\
            48 &
            \raisebox{-.05\height}{\includegraphics[width=0.19\textwidth]{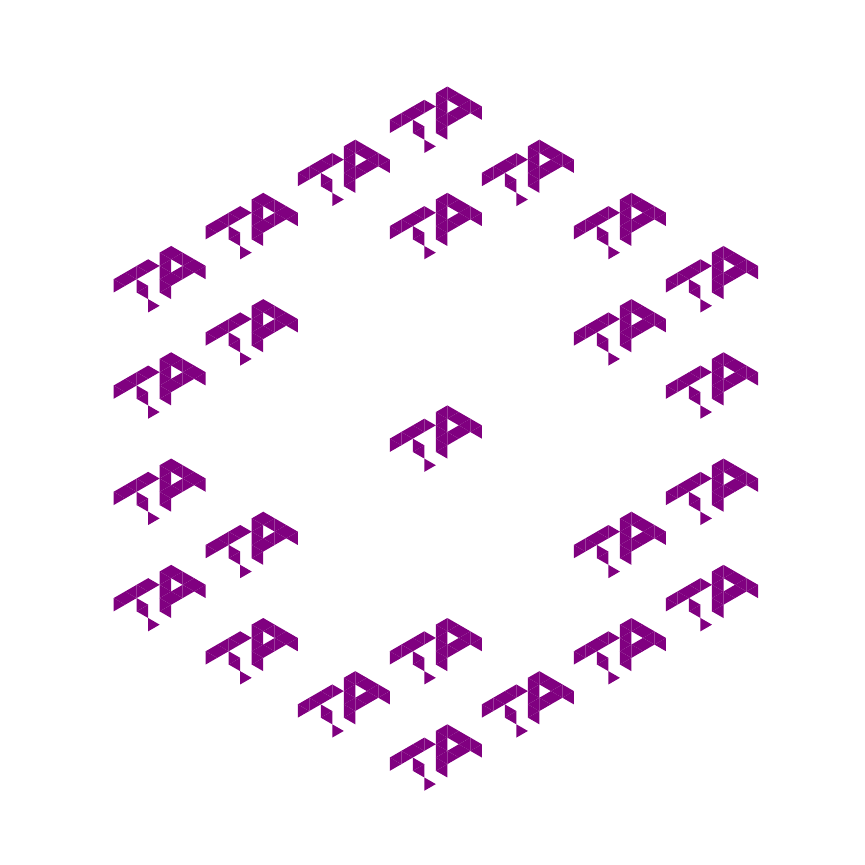}} &
            \raisebox{-.05\height}{\includegraphics[width=0.19\textwidth]{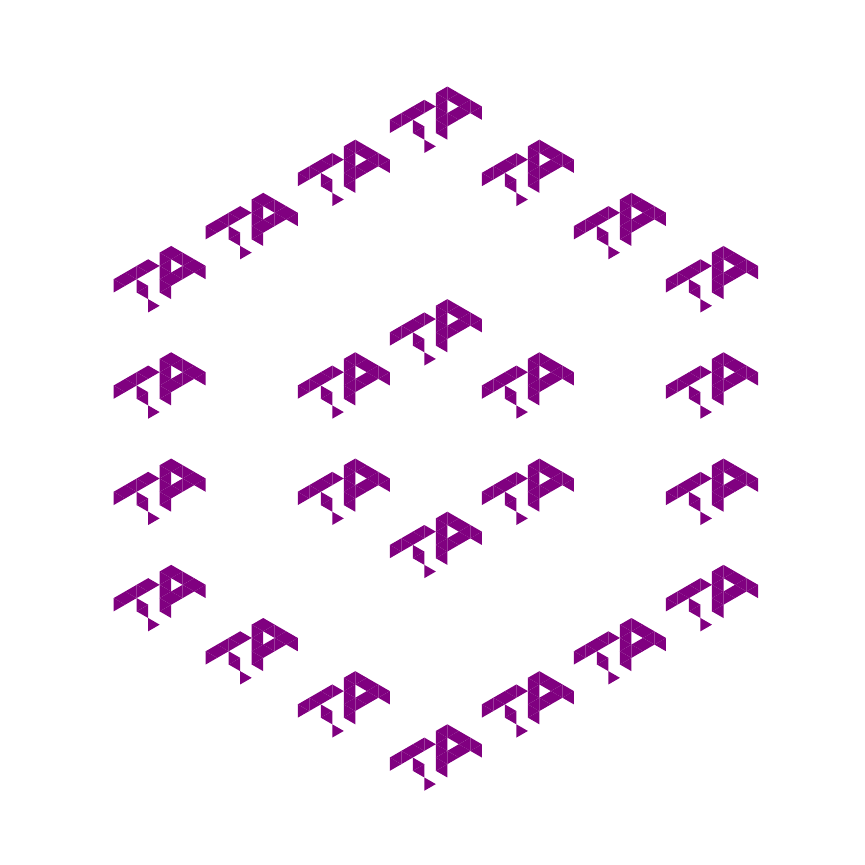}} &
            \raisebox{-.05\height}{\includegraphics[width=0.19\textwidth]{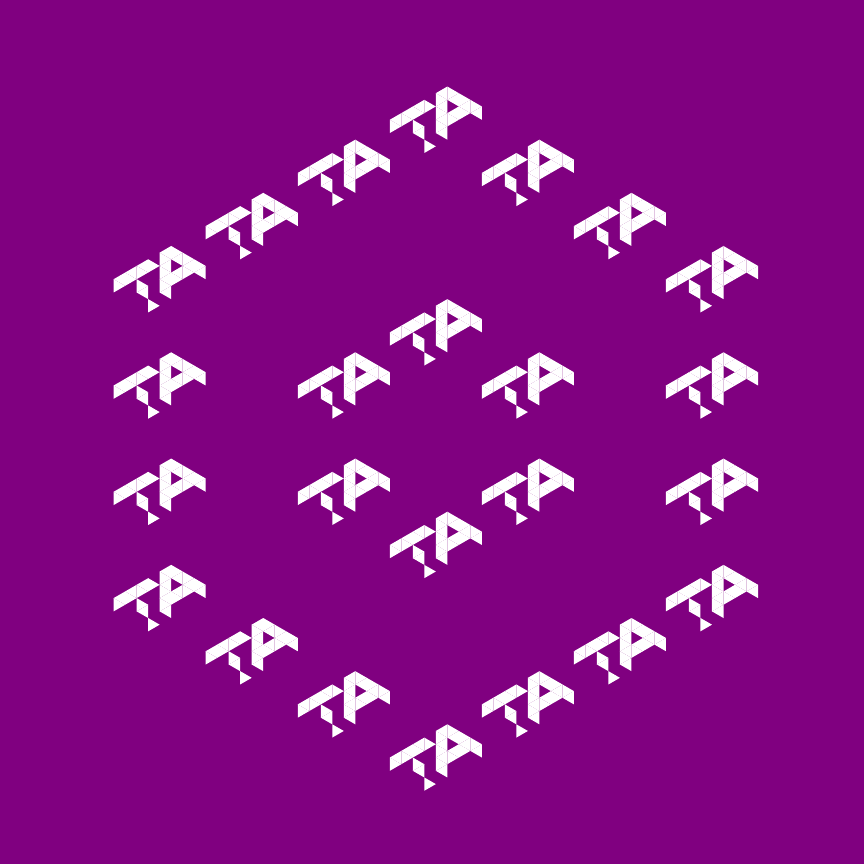}} &
            \raisebox{-.05\height}{\includegraphics[width=0.19\textwidth]{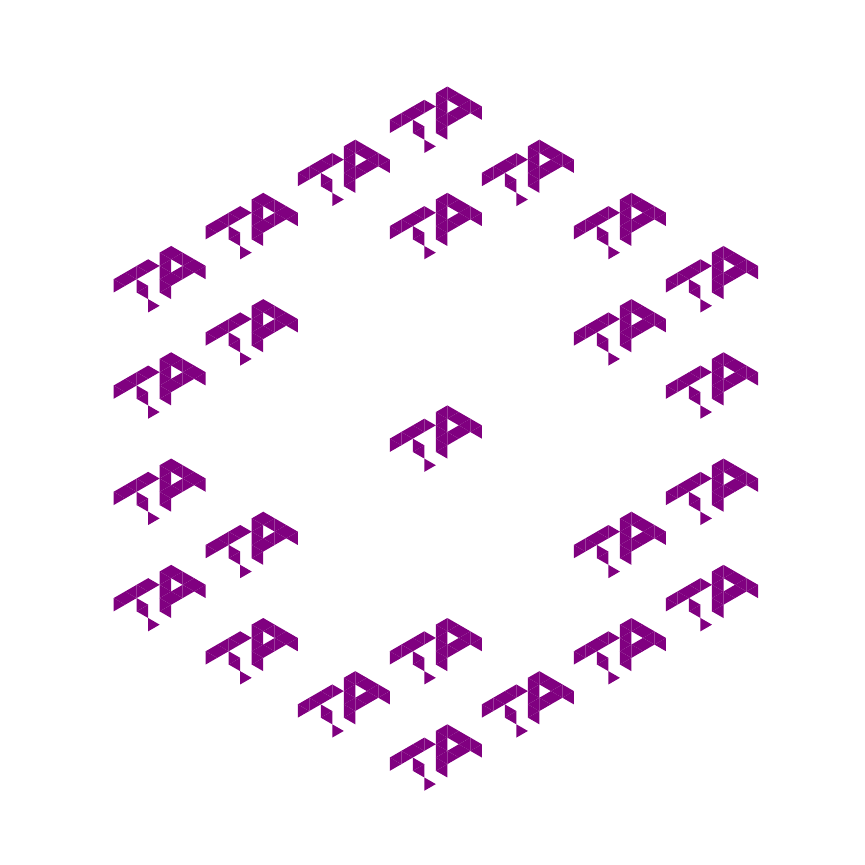}} \\
            112 &
            \raisebox{-.05\height}{\includegraphics[width=0.19\textwidth]{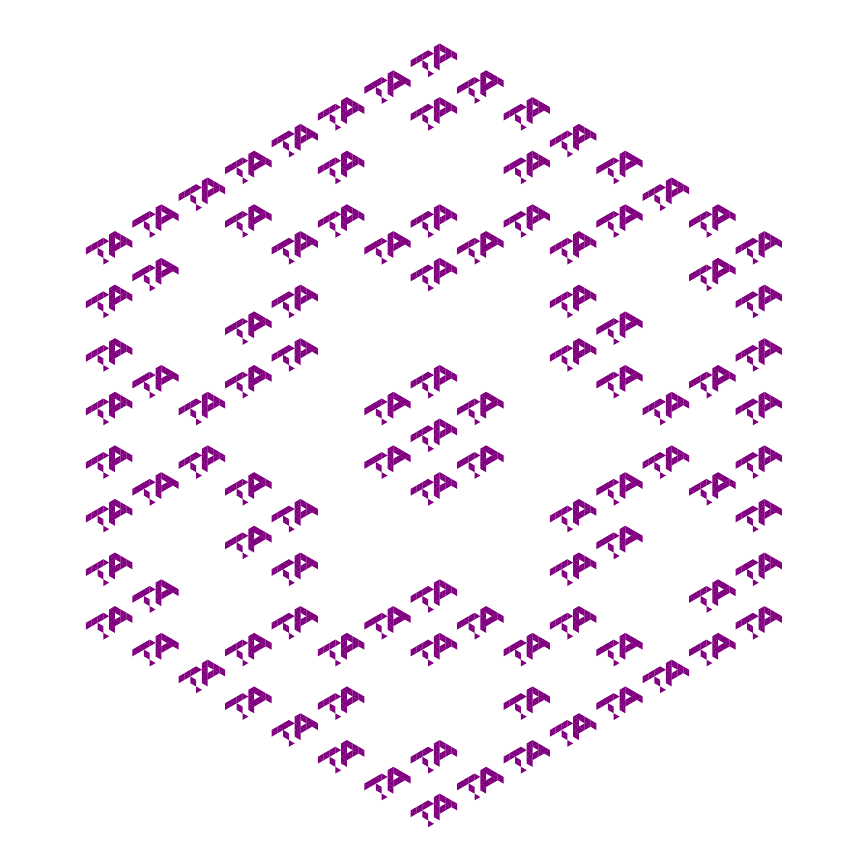}} &
            \raisebox{-.05\height}{\includegraphics[width=0.19\textwidth]{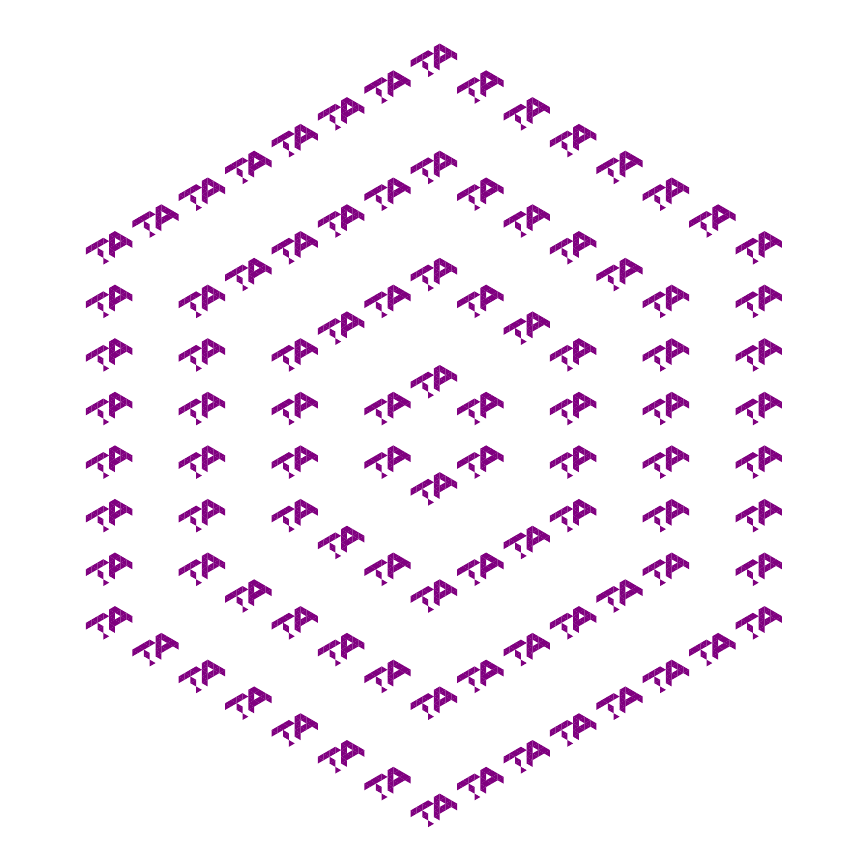}} &
            \raisebox{-.05\height}{\includegraphics[width=0.19\textwidth]{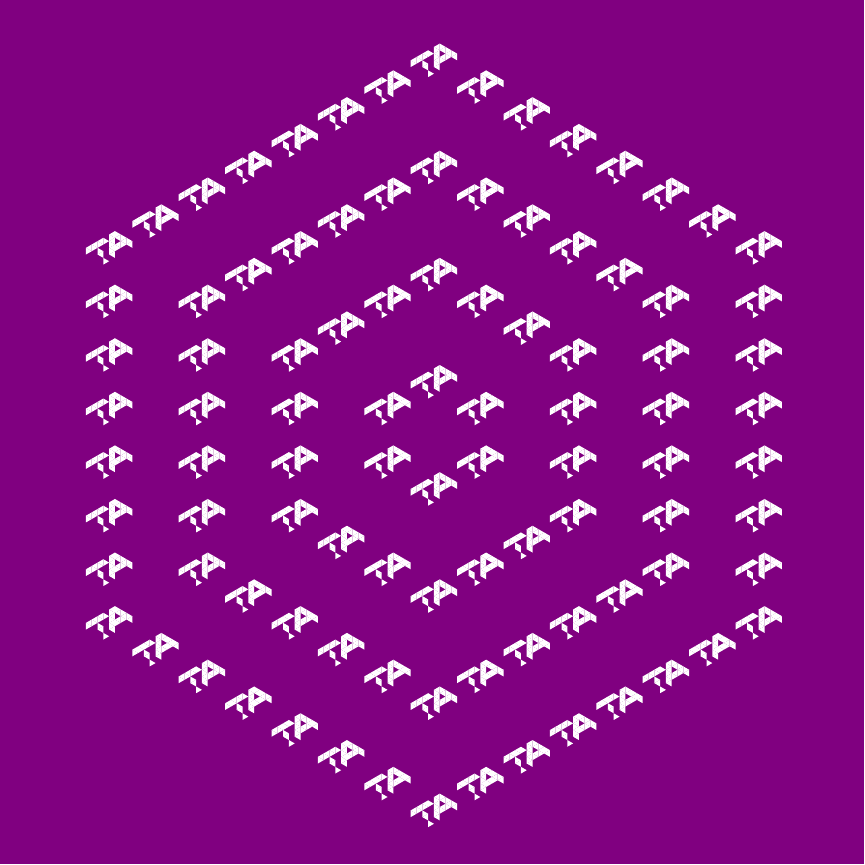}} &
            \raisebox{-.05\height}{\includegraphics[width=0.19\textwidth]{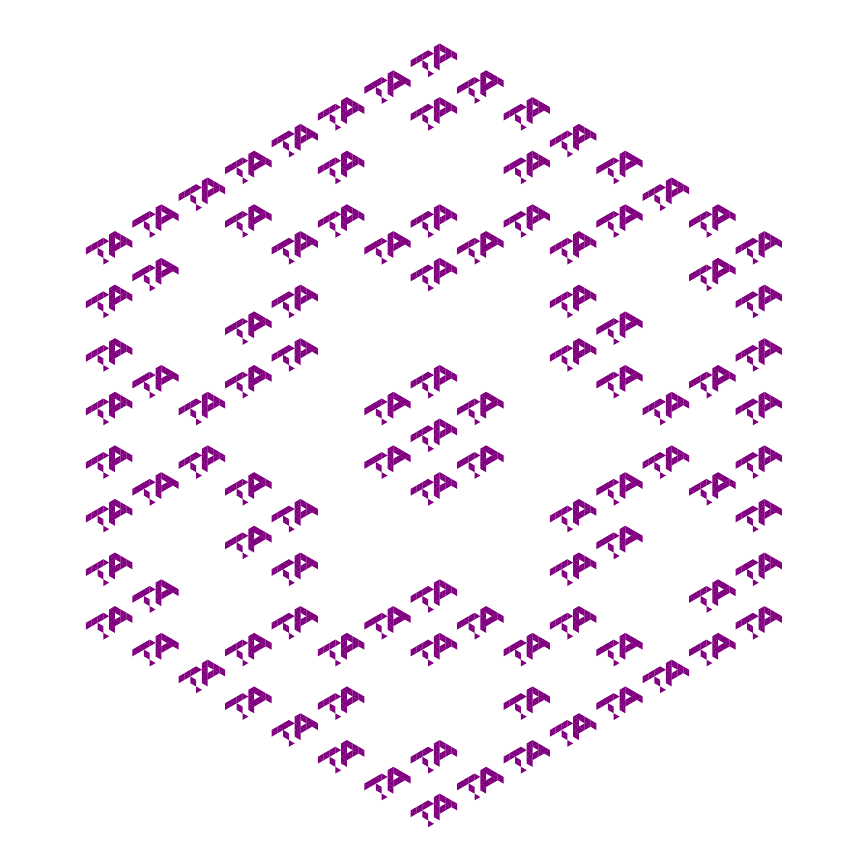}} \vspace{0pt} \\ 
        \hline
    \end{tblr}
    \caption{Pattern replication from a recognizable starting point}
    \label{tab:self-reproduction}
\end{table}

\pagebreak
\subsection{Space-Time} \label{space-time}
\begin{figure}[H]
    \centering
    \includegraphics[width=\textwidth]{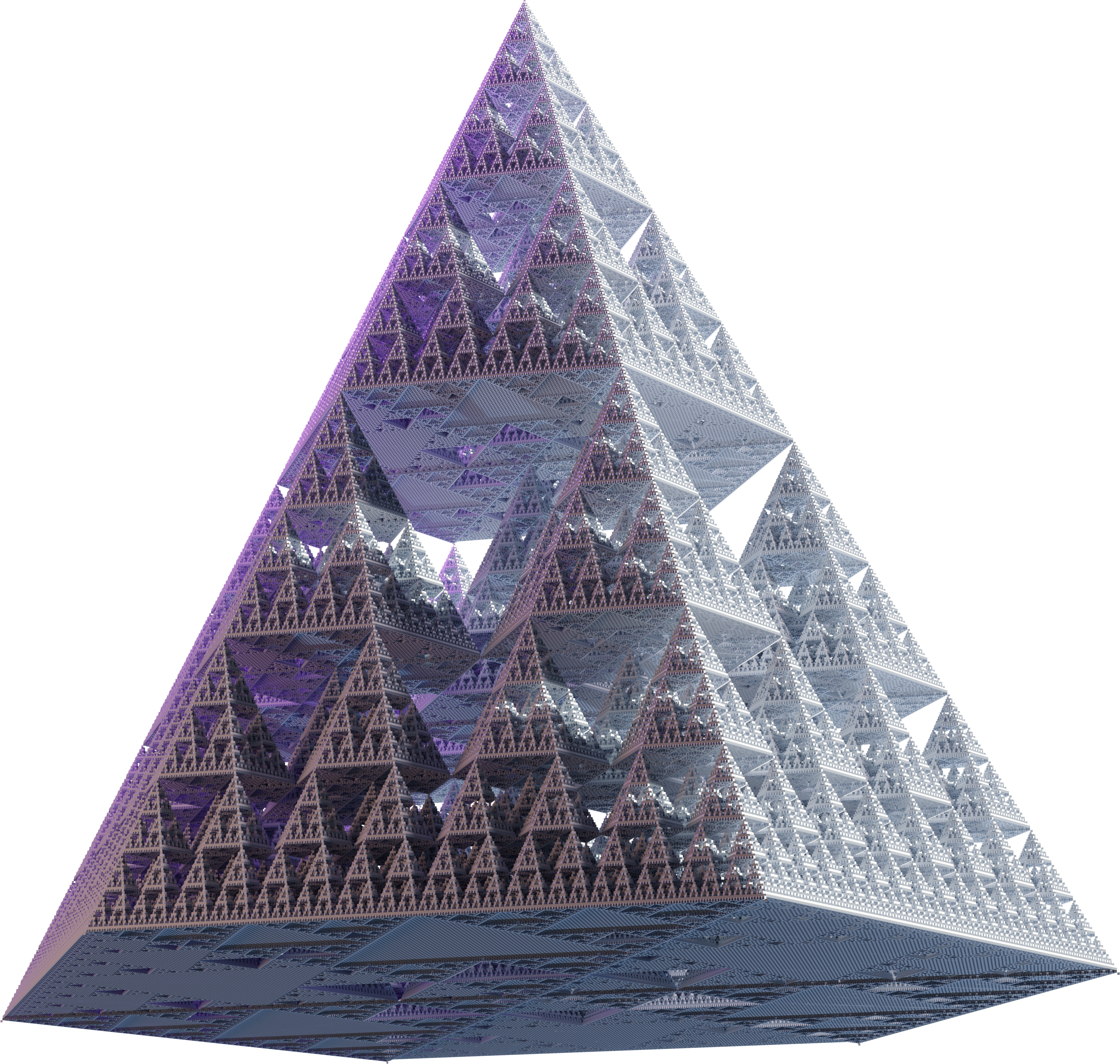}
    \caption{Space-time plot of rule 10 up to $t=512$}
    \label{fig:rule-10-perspective}
\end{figure}

Similar to the way Elementary Cellular Automata \cite{wolfram2002new, weisstein2002elementary} are most often represented, the evolution of an ETA can be displayed in a single plot. Here, an instant is two-dimensional, so adding the dimension of time creates a 3D structure. In these \textbf{space-time plots}, time flows downward. The successive grids are stacked beneath each other, starting from the initial conditions at the top. When the background state is 1,\linebreak we can display the dead cells to avoid the infinite planes. A lot of information is therefore lost. We do not see most of the internal structure and we cannot know the state of the universe. Nevertheless, this representation helps visualize some properties of ETA that are difficult to notice otherwise. For instance, certain rules create 3D space-time fractals, as in \textit{Figure \ref{fig:rule-10-perspective}}.

\vspace{1cm}

\begin{figure}[H]
    \centering
    \includegraphics[width=.5\textwidth]{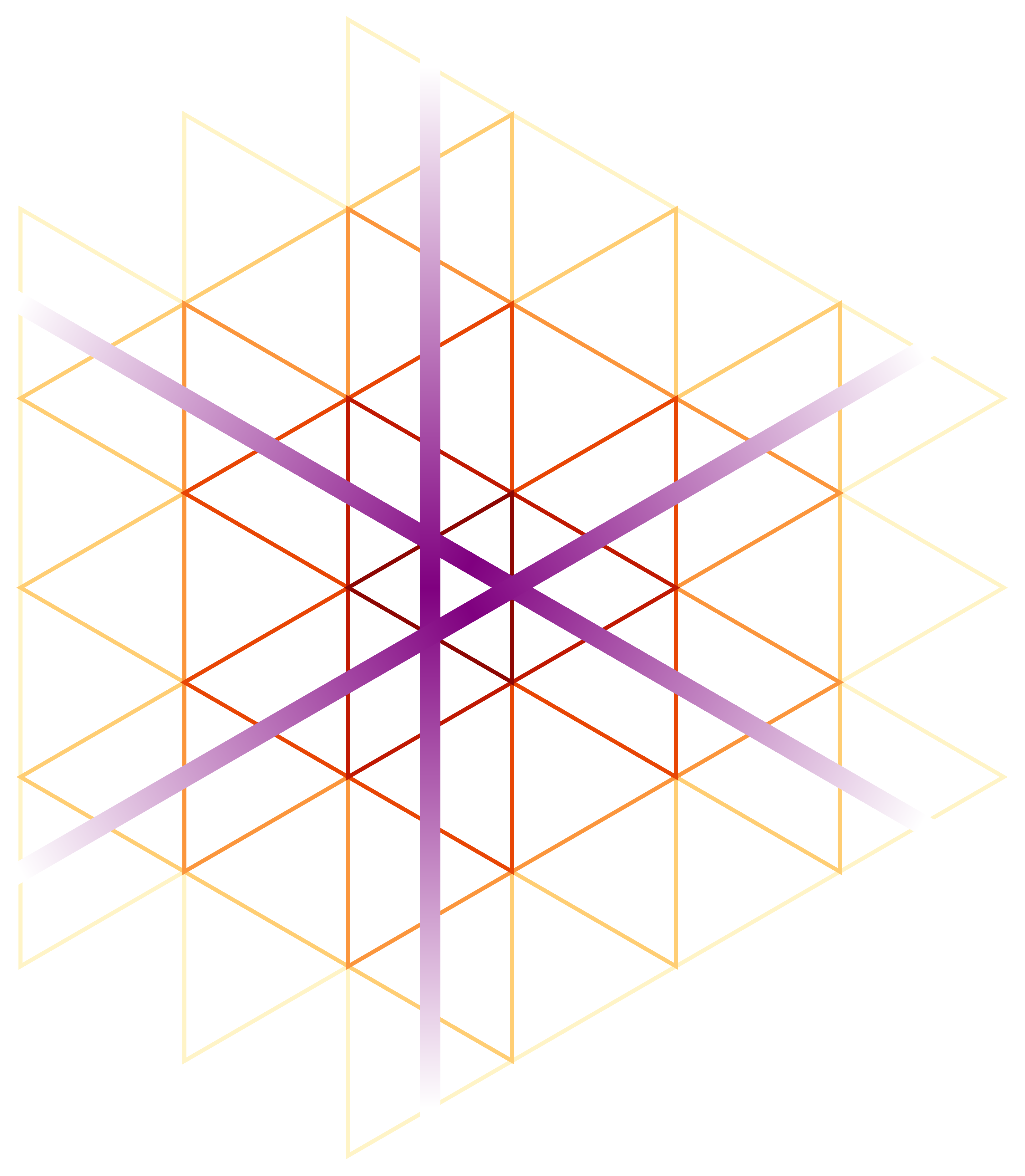}
    \caption{Natural slicing lines}
    \label{fig:slice-lines}
\end{figure}

\pagebreak
Another way to see the time evolution of an ETA in a single plot is to look at just one line of the grid, that is, to look at a sliced space-time plot from the side. There is a most natural way to slice: the lines parallel to the sides of the triangles preserve the neighborhood relations (\textit{Figure \ref{fig:slice-lines}}).

\begin{figure}[H]
    \centering
    \includegraphics[width=\textwidth]{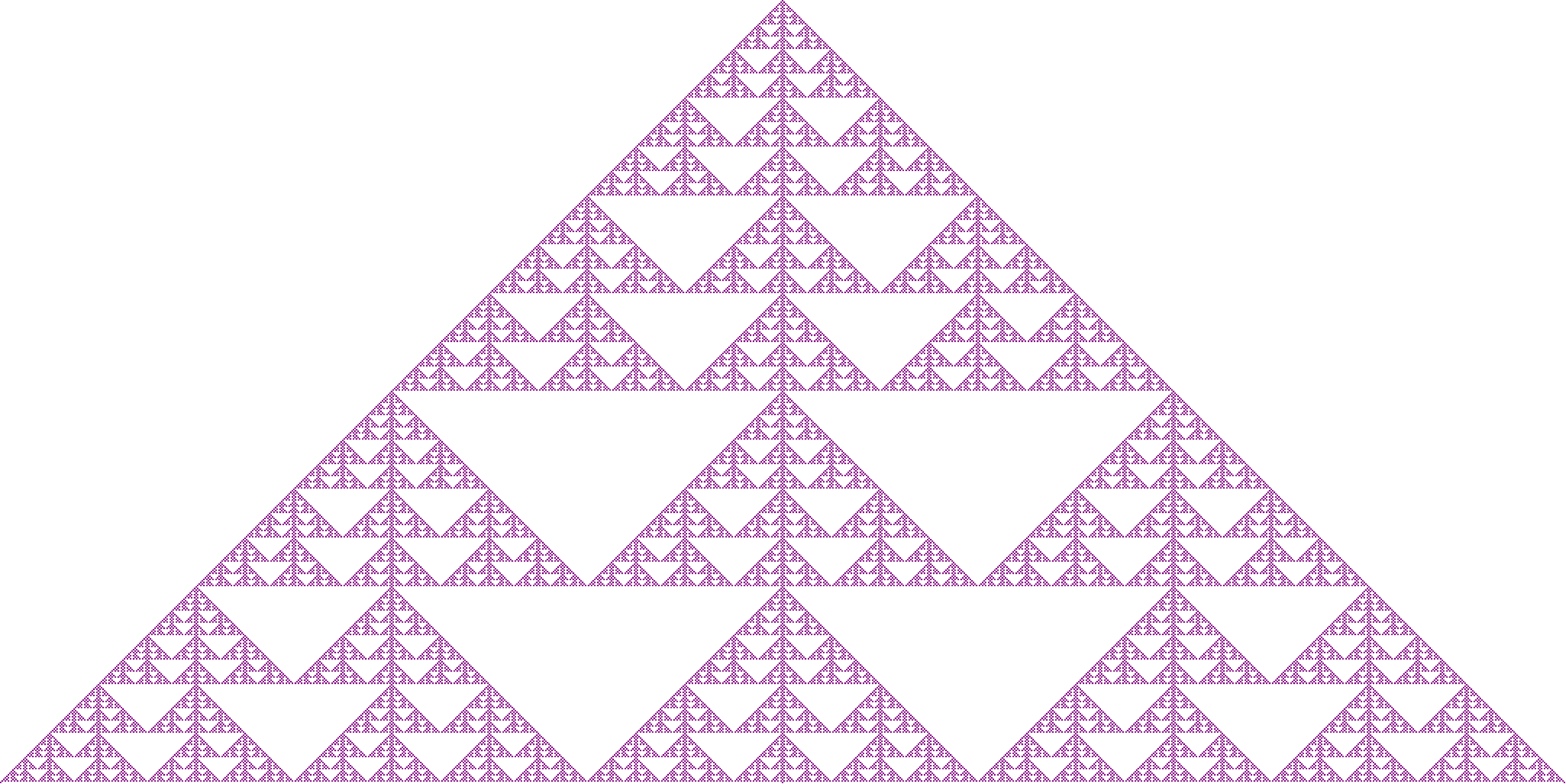}
    \caption{Slice plot of rule 10 up to $t=512$}
    \label{fig:rule-10-slice-plot-512}
\end{figure}

The states along these lines can be plotted beneath each other to create Wolfram-style plots, here called \textbf{slice plots} (\textit{Figure \ref{fig:rule-10-slice-plot-512}}).  

\pagebreak
\subsection{Elementary Cellular Automata} \label{elementary-cellular-automata}
Under a certain condition, ETA rules maintain the uniformity of layers. This is the case when the second and third digits of the binary rule number are the same, as well as the sixth and seventh.

\begin{equation}
    \Big[R(1)=R(2)\Big] \land \Big[R(5)=R(6)\Big]
\end{equation}

\noindent 64 ETA rules satisfy this criterion. When the initial conditions have uniform layers (e.g., a single living cell or infinite stripes), these rules are each strictly equivalent to a unique symmetric \textbf{Elementary Cellular Automaton} (ECA) \cite{wolfram2002new}.

\begin{figure}[H]
    \centering
    \boxed{\includegraphics[width=.5\textwidth]{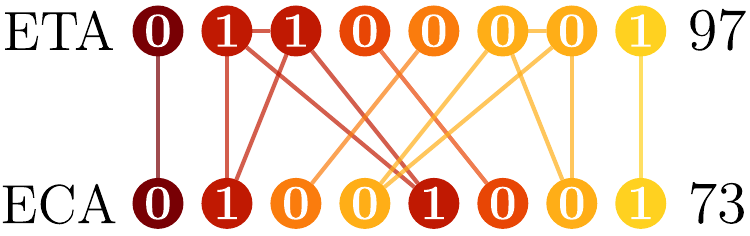}}
    \caption{Conversion between ETA and ECA rules}
    \label{fig:ETA-to-ECA}
\end{figure}

\noindent In this case, taking a slice plot of the ETA gives the classic ECA plot.

\begin{figure}[H]
    \centering
    \includegraphics[width=.6\textwidth]{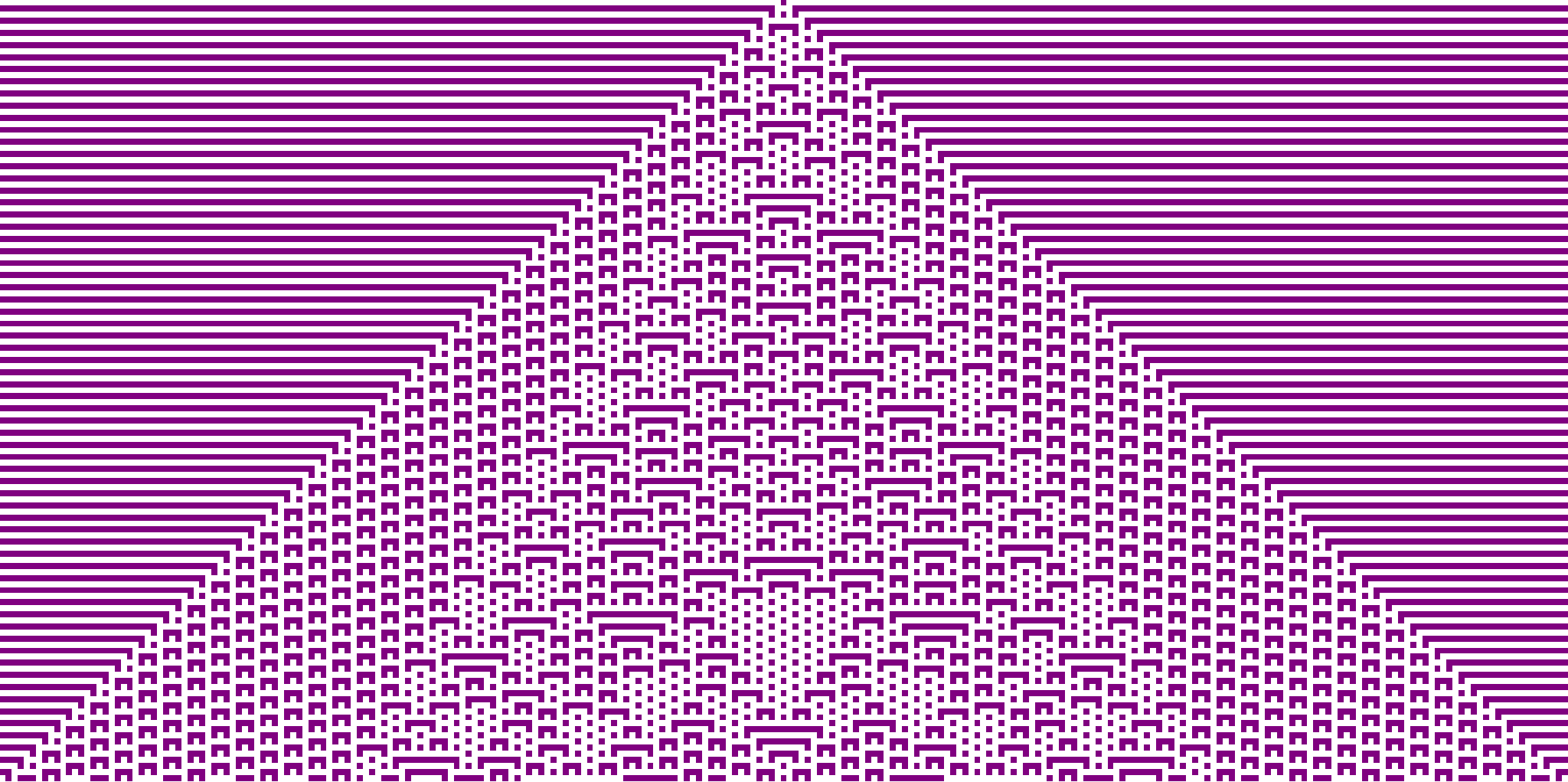}
    \caption{Slice plot of rule 97 (ECA rule 73) up to $t=128$}
    \label{fig:rule-97-slice-plot-128}
\end{figure}

\vspace{-10pt}

\begin{figure}[H]
    \centering
    \includegraphics[width=.4\textwidth]{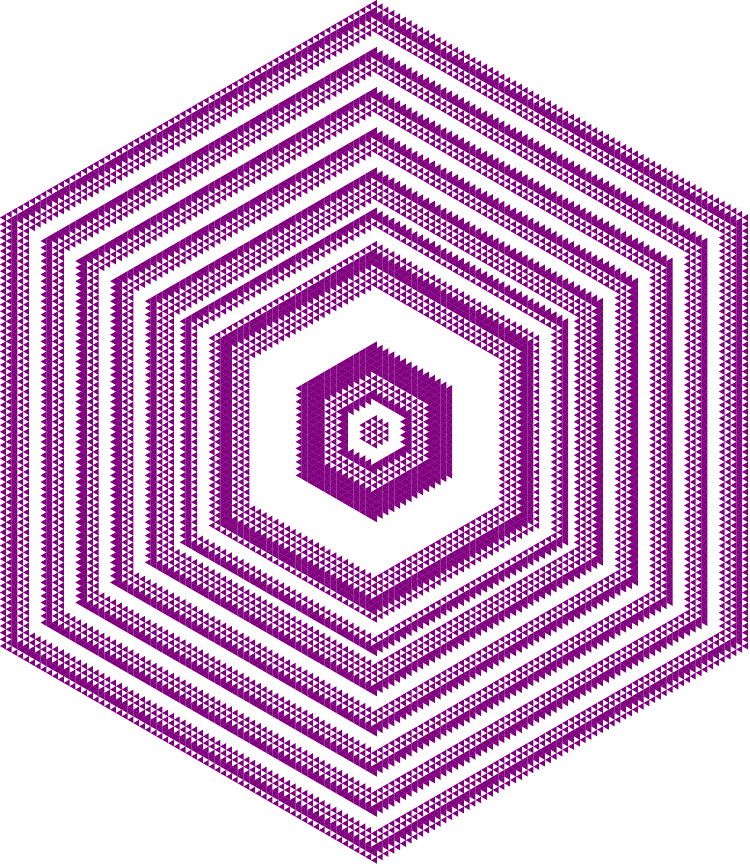}
    \caption{Rule 97 at $t=128$}
    \label{fig:rule-97-time-128}
\end{figure}

\pagebreak
\subsection{Center Column} \label{center-column}
\noindent There are a few rules where the state of the first living cell evolves in an interesting way. This sequence of states forms the center column of the space-time plots.

\begin{figure}[H]
    \centering
    \begin{subfigure}[b]{0.24\textwidth}
        \centering
        \includegraphics[width=.9\textwidth]{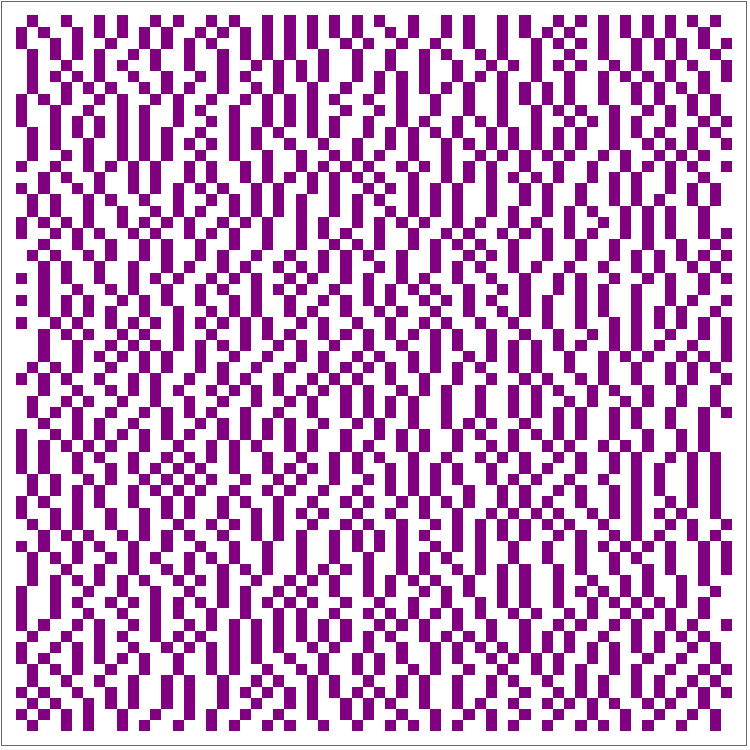}
        \caption*{Rule 37\\(OEIS \href{https://oeis.org/A371844}{A371844})}
        \label{fig:rule-37-center_column_plot-4096}
    \end{subfigure}
    \begin{subfigure}[b]{0.24\textwidth}
        \centering
        \includegraphics[width=.9\textwidth]{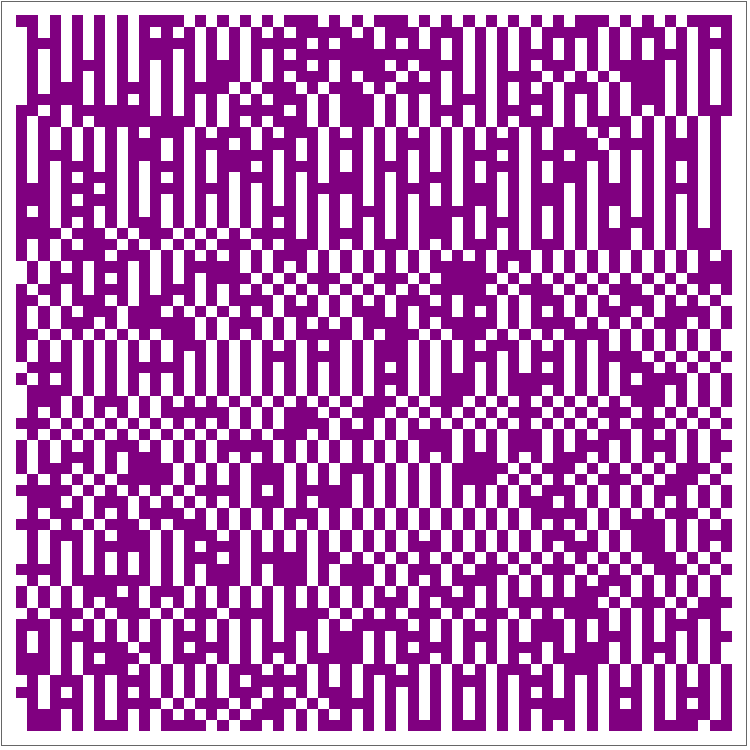}
        \caption*{Rule 61\\(OEIS \href{https://oeis.org/A372553}{A372553})}
        \label{fig:rule-61-center_column_plot-4096}
    \end{subfigure}
    \begin{subfigure}[b]{0.24\textwidth}
        \centering
        \includegraphics[width=.9\textwidth]{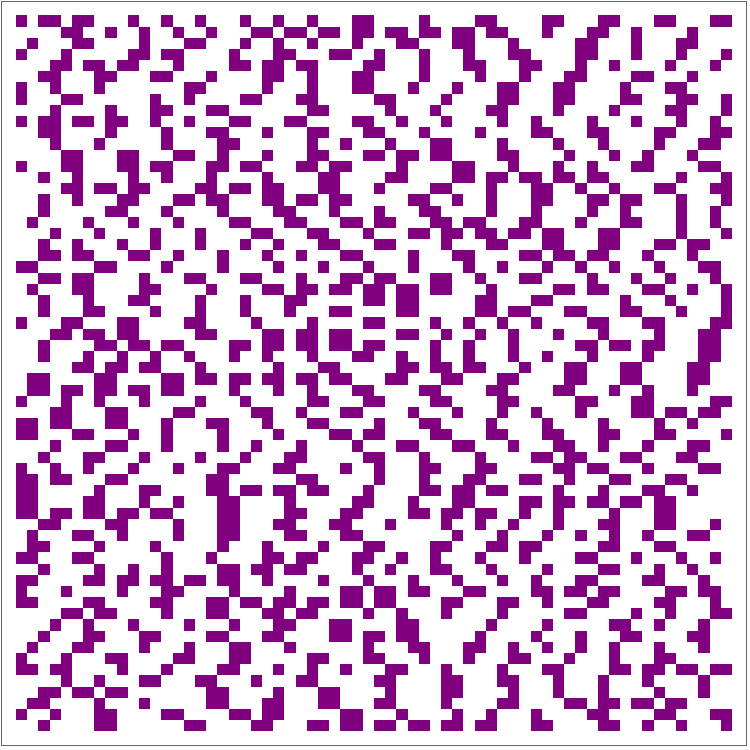}
        \caption*{Rule 62\\(OEIS \href{https://oeis.org/A371931}{A371931})}
        \label{fig:rule-62-center_column_plot-4096}
    \end{subfigure}
    \begin{subfigure}[b]{0.24\textwidth}
        \centering
        \includegraphics[width=.9\textwidth]{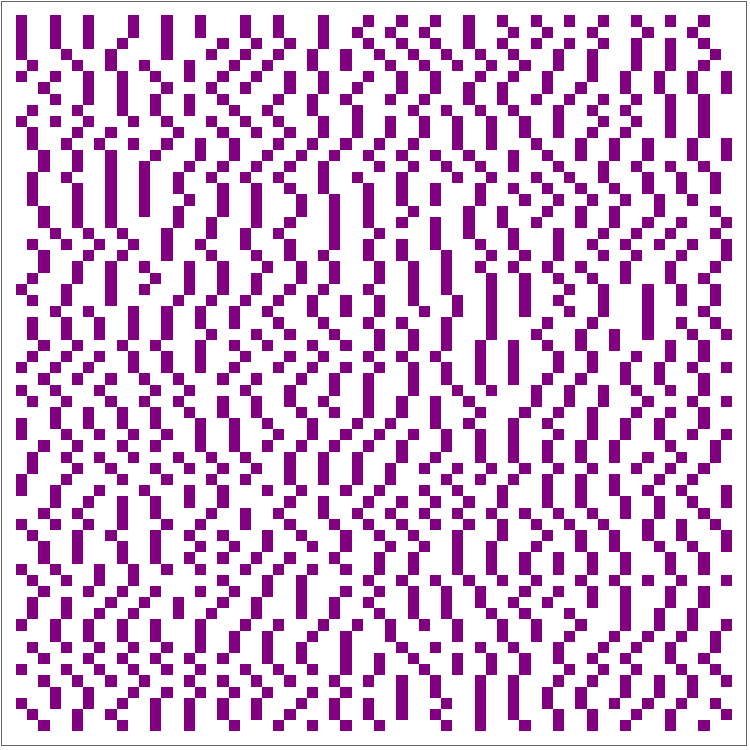}
        \caption*{Rule 94\\(OEIS \href{https://oeis.org/A372552}{A372552})}
        \label{fig:rule-94-center_column_plot-4096}
    \end{subfigure}
       \caption{Center column for $1\leq t \leq 4096$ (read left-to-right, top-to-bottom)}
       \label{fig:center-column}
\end{figure}

\begin{figure}[H]
    \centering
    \includegraphics[width=.2\textwidth]{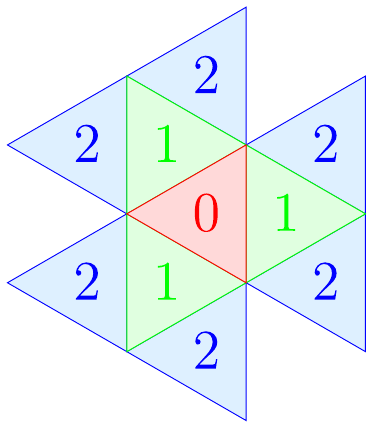}
    \caption{RGB map}
    \label{fig:rgb-map}
\end{figure}

\begin{figure}[H]
    \centering
    \begin{subfigure}[b]{0.1\textwidth}
        \centering
        \includegraphics[width=.9\textwidth]{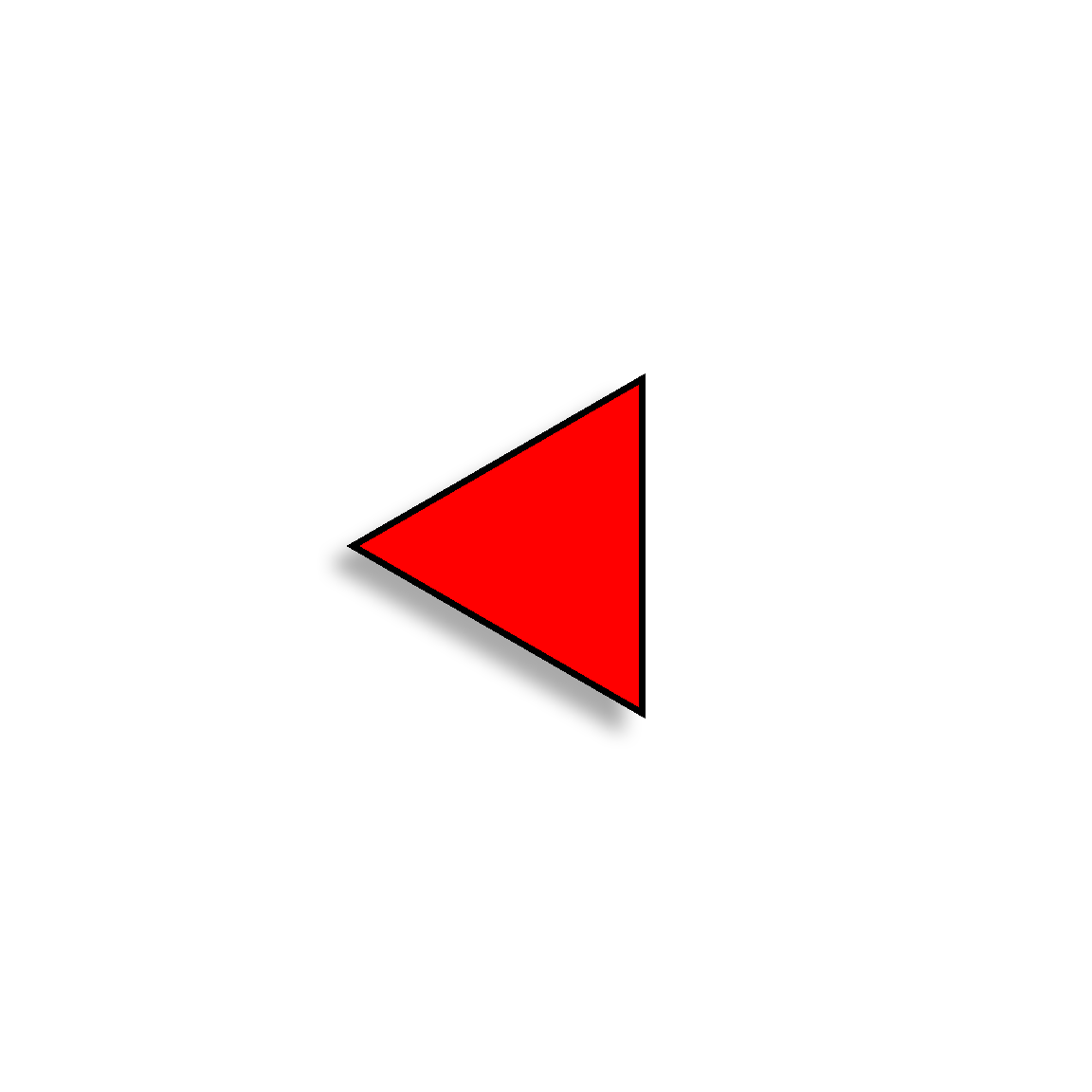}
    \end{subfigure}
    \begin{subfigure}[b]{0.1\textwidth}
        \centering
        \includegraphics[width=.9\textwidth]{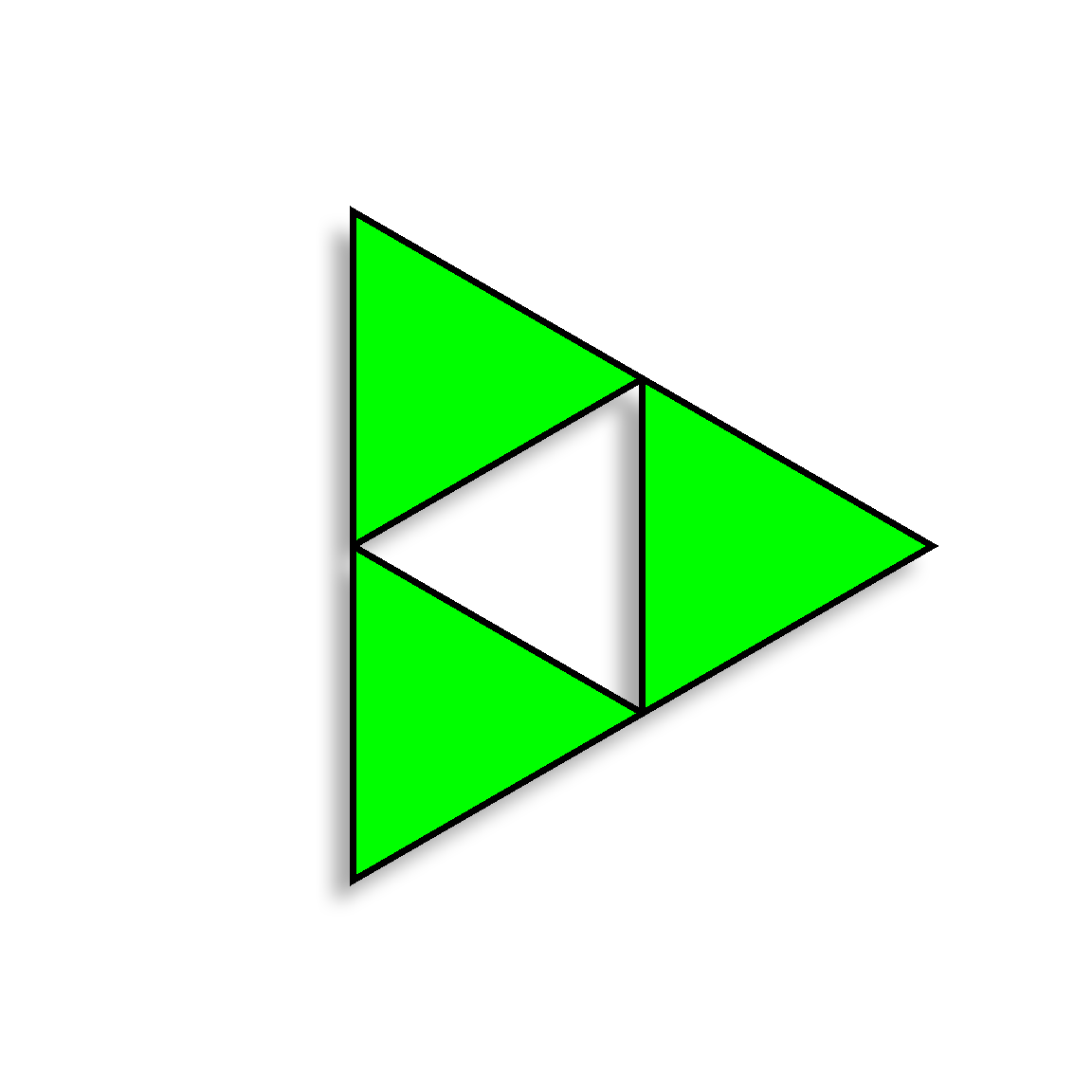}
    \end{subfigure}
    \begin{subfigure}[b]{0.1\textwidth}
        \centering
        \includegraphics[width=.9\textwidth]{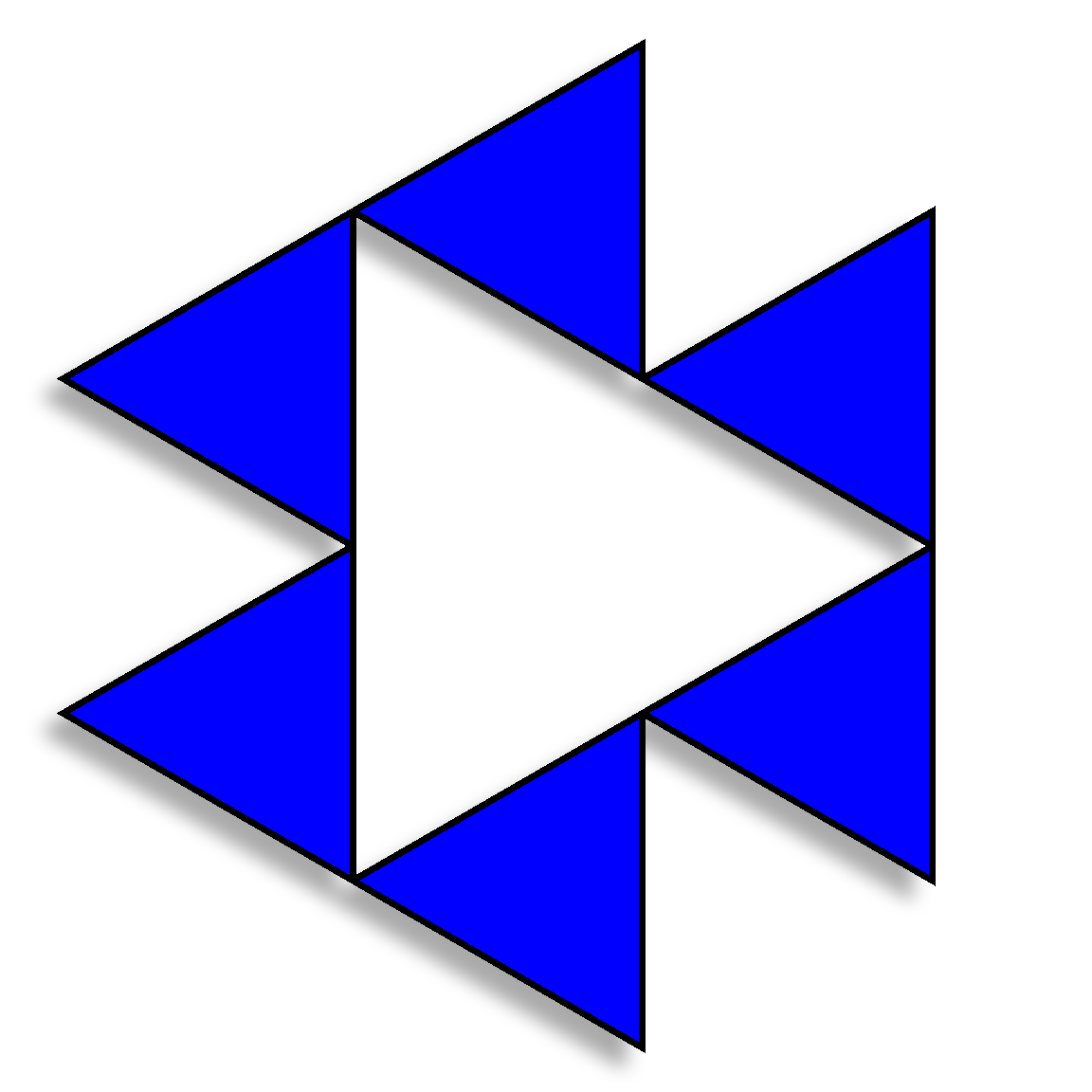}
    \end{subfigure}
    \begin{subfigure}[b]{0.1\textwidth}
        \centering
        \includegraphics[width=.9\textwidth]{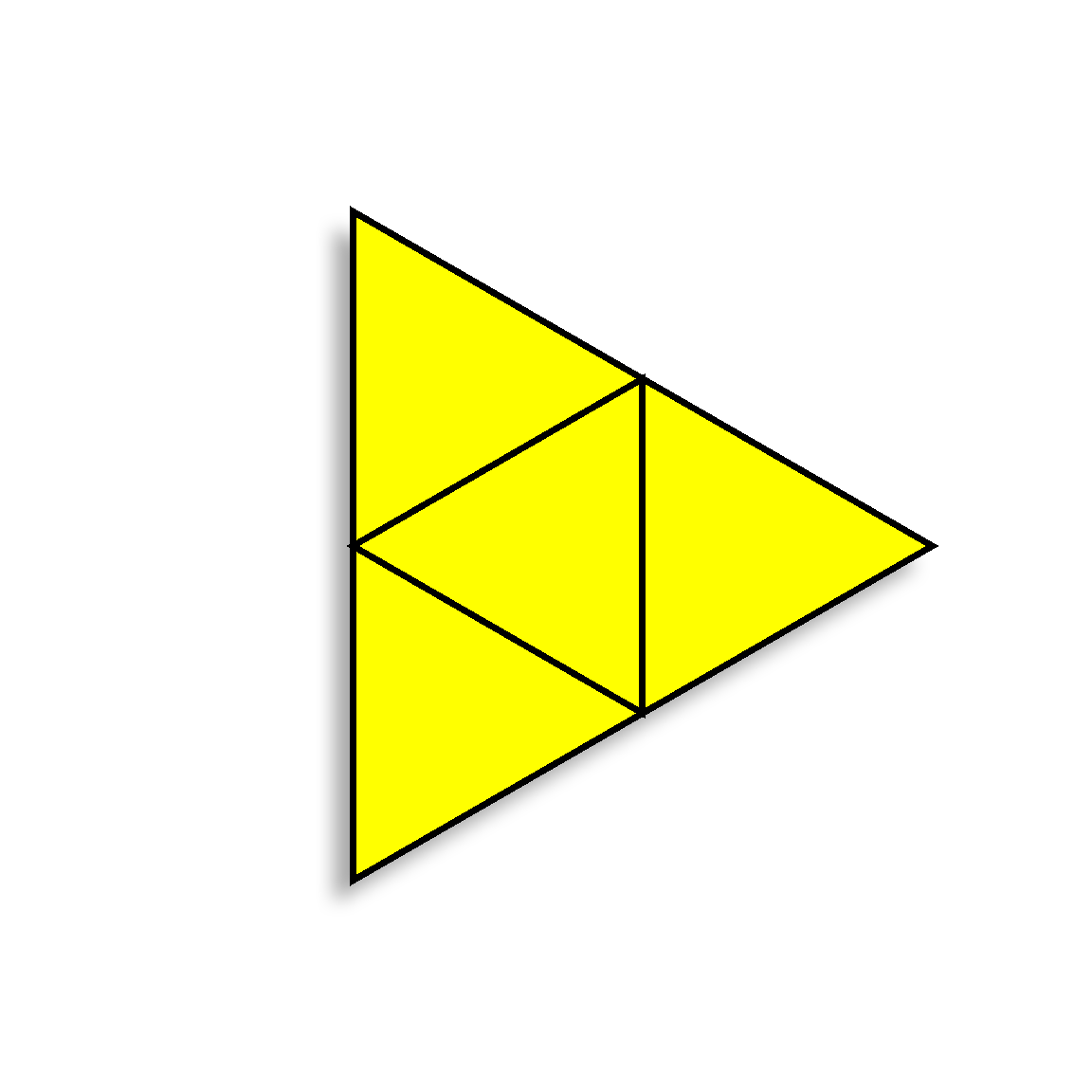}
    \end{subfigure}
    \begin{subfigure}[b]{0.1\textwidth}
        \centering
        \includegraphics[width=.9\textwidth]{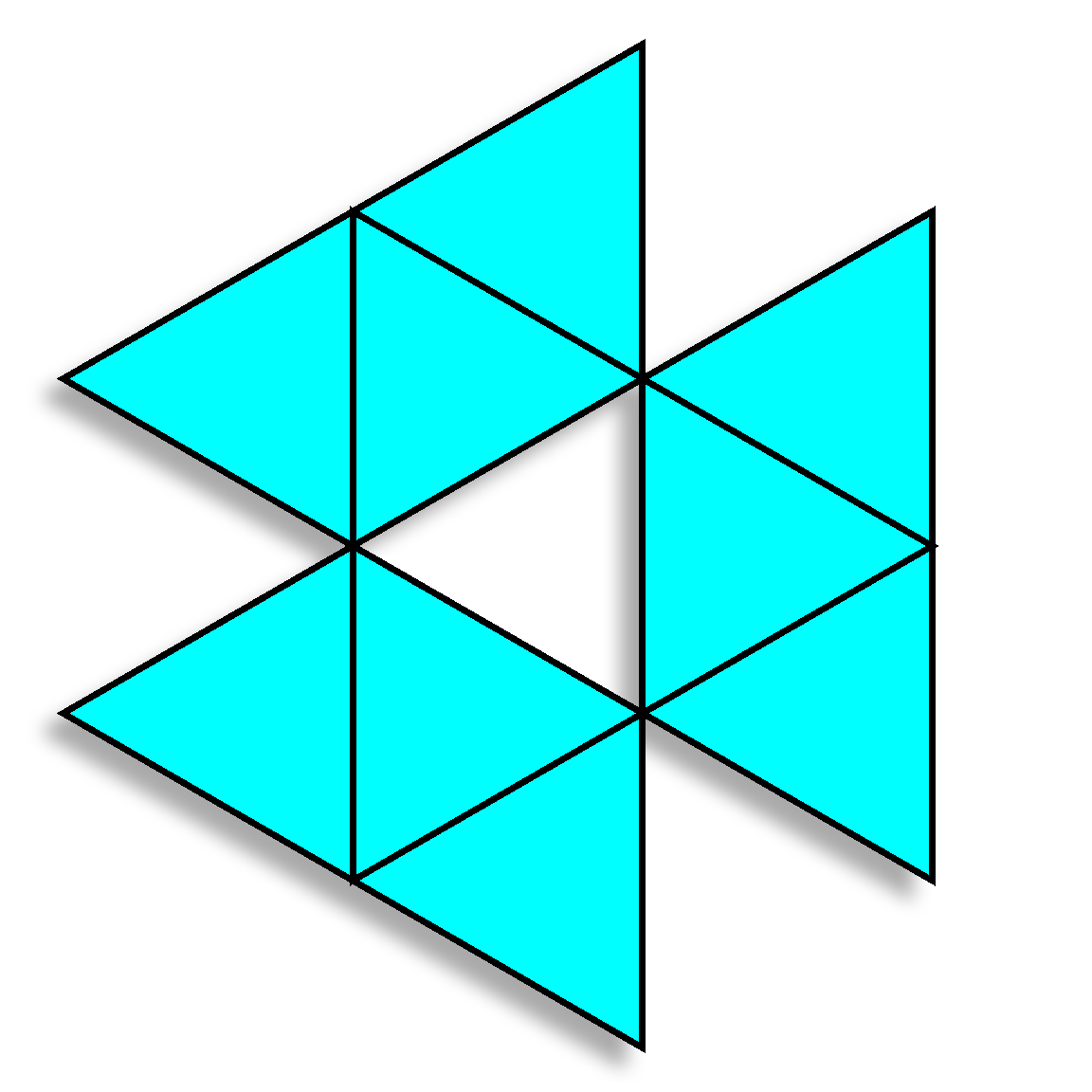}
    \end{subfigure}
    \begin{subfigure}[b]{0.1\textwidth}
        \centering
        \includegraphics[width=.9\textwidth]{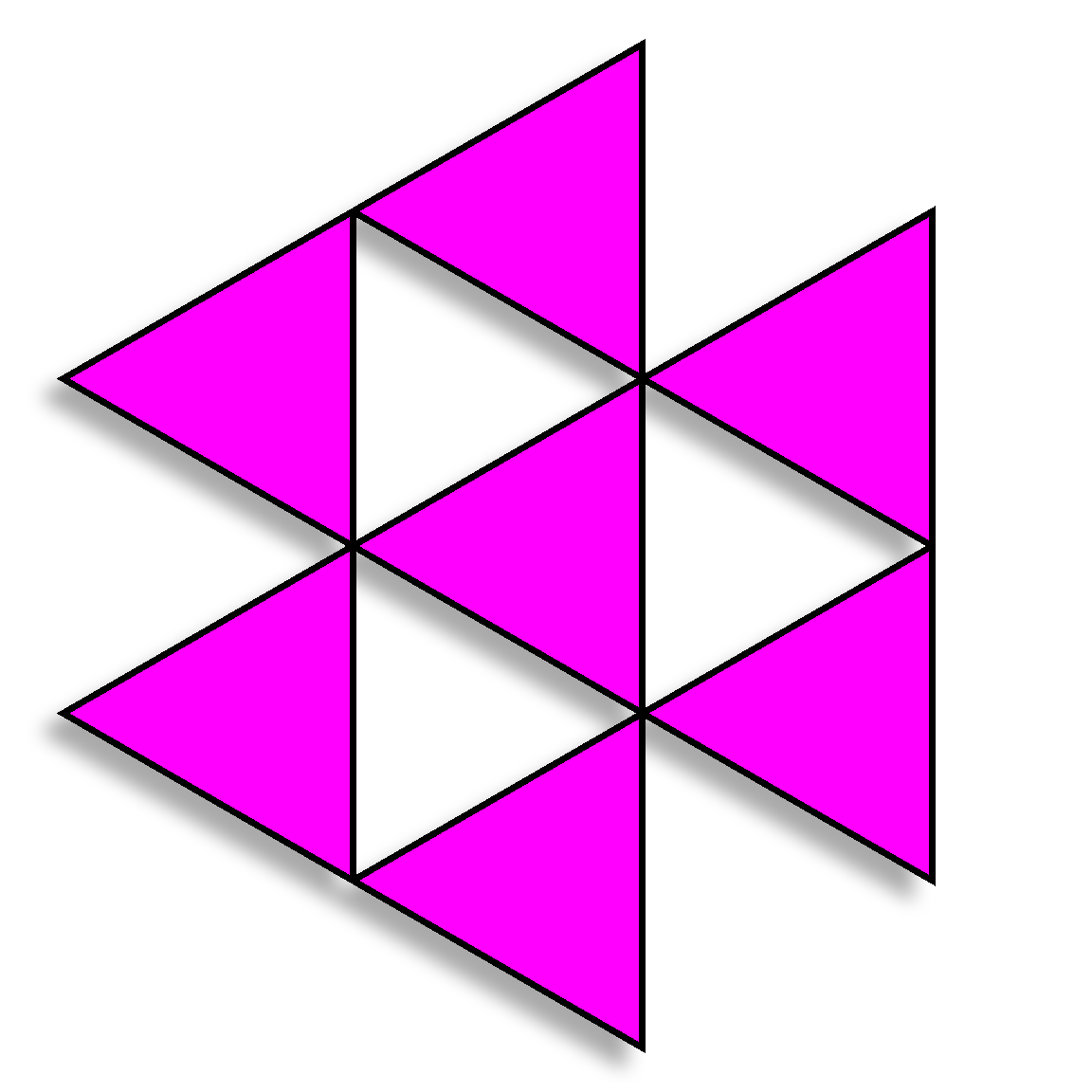}
    \end{subfigure}
    \begin{subfigure}[b]{0.1\textwidth}
        \centering
        \includegraphics[width=.9\textwidth]{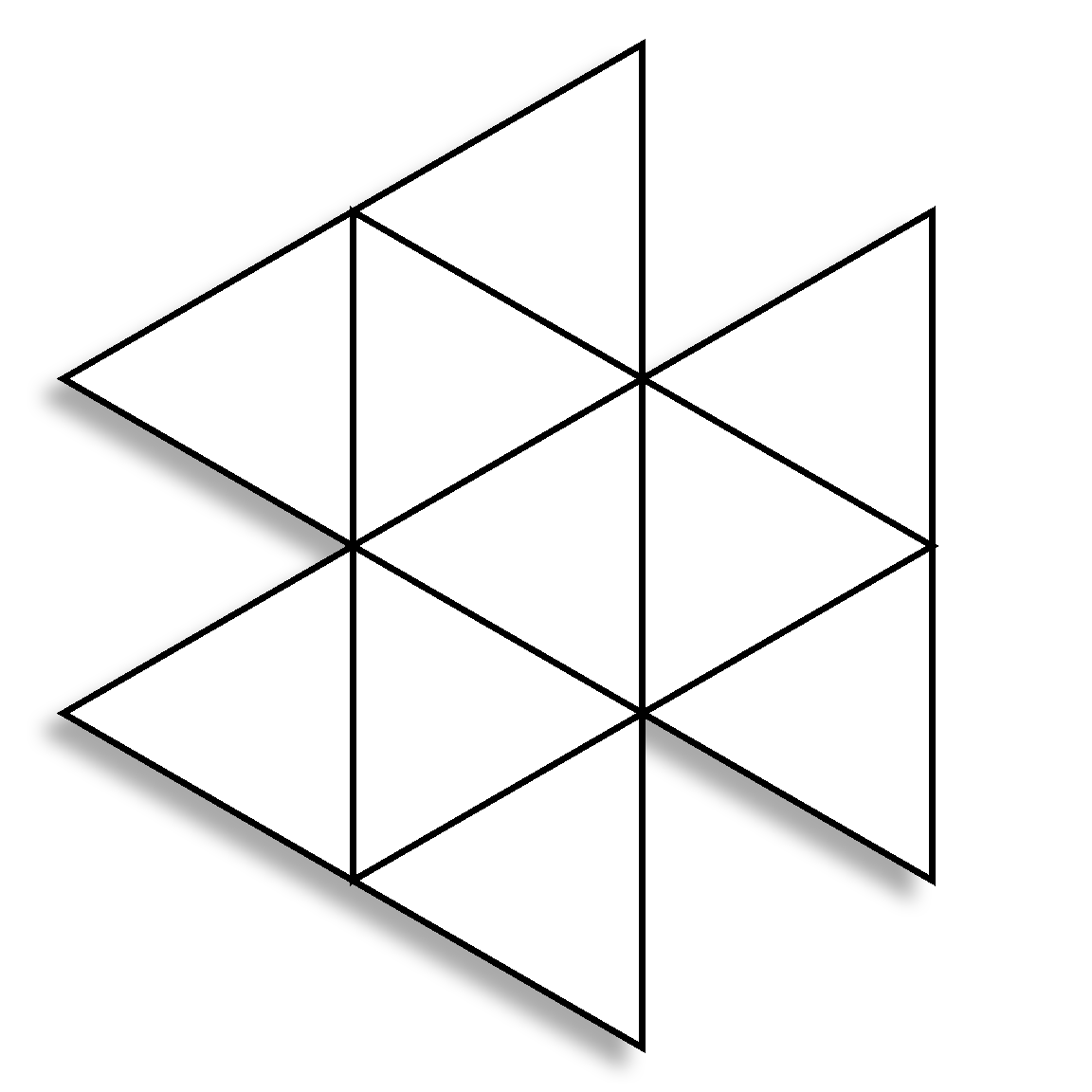}
    \end{subfigure}
       \caption{Living cells with corresponding colors}
       \label{fig:colors}
\end{figure}

The symmetries of the triangle impose that, starting from a single living cell, layers 1 and 2 remain uniform. We therefore have a three-layer-wide pillar, which is fully described by a sequence of bit triplets. These can be mapped to RGB colors (\textit{Figures \ref{fig:rgb-map}}\linebreak and \textit{\ref{fig:colors}}) to extend the previous plots.

\vspace{-8pt}

\vspace{-4pt}

\begin{figure}[H]
    \centering
    \begin{subfigure}[b]{0.135\textwidth}
        \centering
        \includegraphics[width=\textwidth]{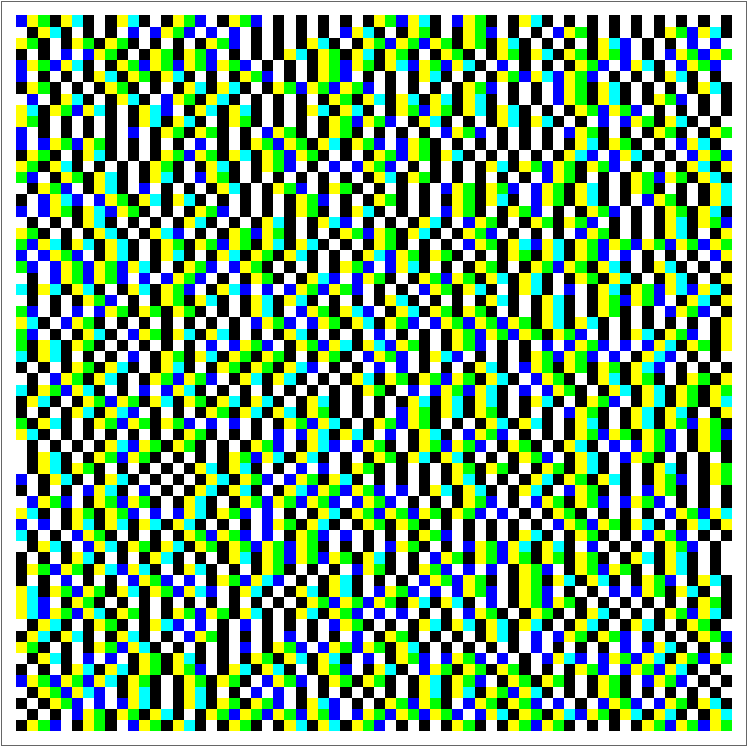}
        \vspace{-16pt}\caption*{Rule 37}
        \label{fig:rule-37-center_columns-4096}
    \end{subfigure}
    \begin{subfigure}[b]{0.135\textwidth}
        \centering
        \includegraphics[width=\textwidth]{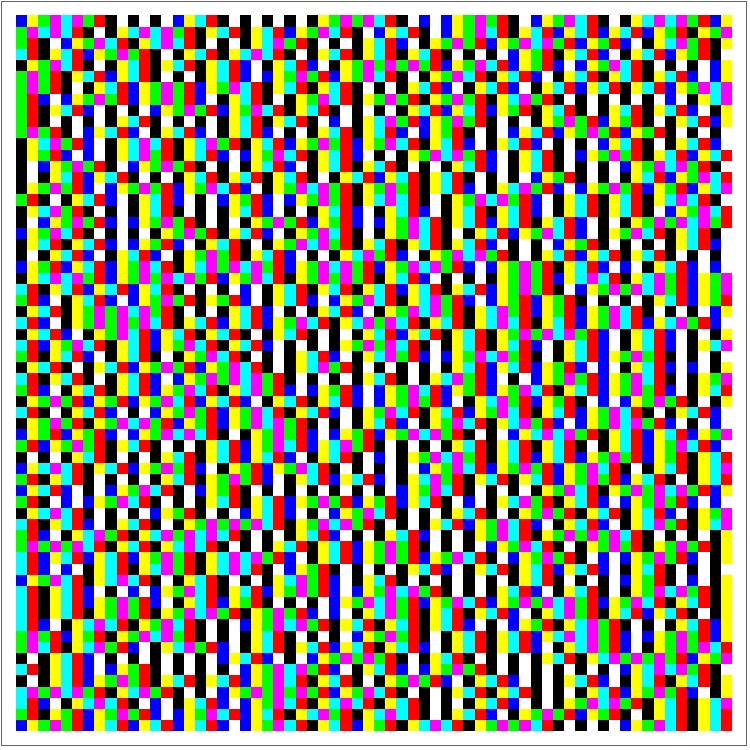}
        \vspace{-16pt}\caption*{Rule 45}
        \label{fig:rule-45-center_columns-4096}
    \end{subfigure}
    \begin{subfigure}[b]{0.135\textwidth}
        \centering
        \includegraphics[width=\textwidth]{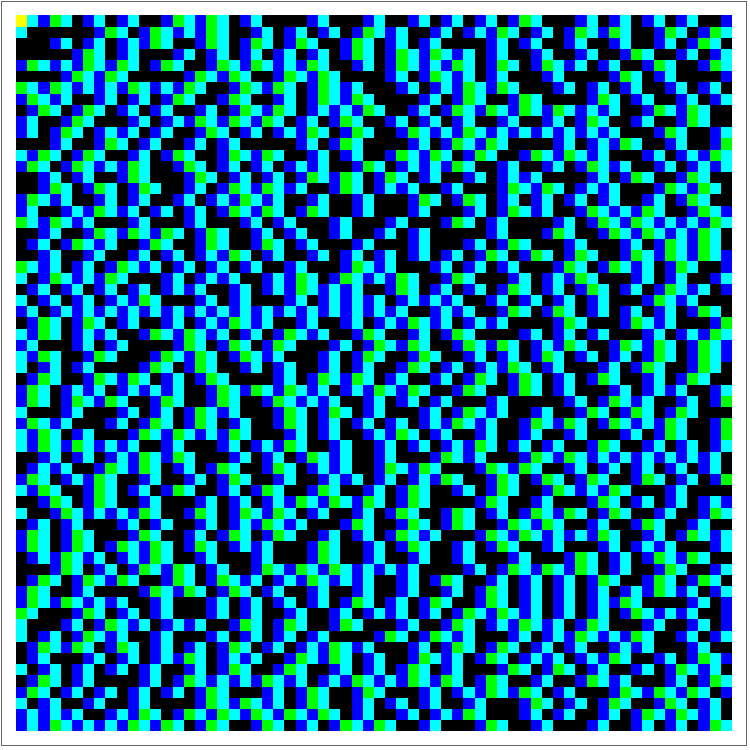}
        \vspace{-16pt}\caption*{Rule 54}
        \label{fig:rule-54-center_columns-4096}
    \end{subfigure}
    \begin{subfigure}[b]{0.135\textwidth}
        \centering
        \includegraphics[width=\textwidth]{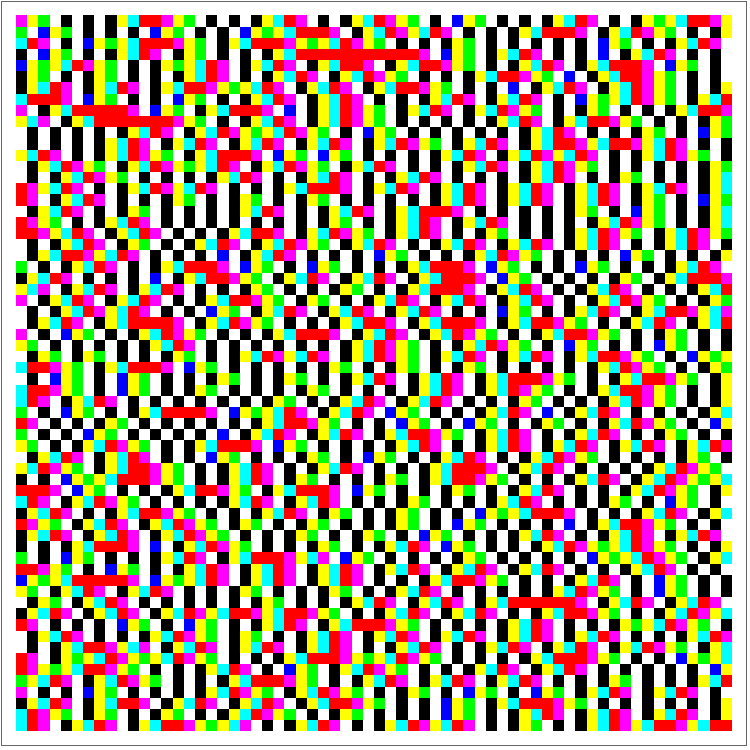}
        \vspace{-16pt}\caption*{Rule 61}
        \label{fig:rule-61-center_columns-4096}
    \end{subfigure}
    \begin{subfigure}[b]{0.135\textwidth}
        \centering
        \includegraphics[width=\textwidth]{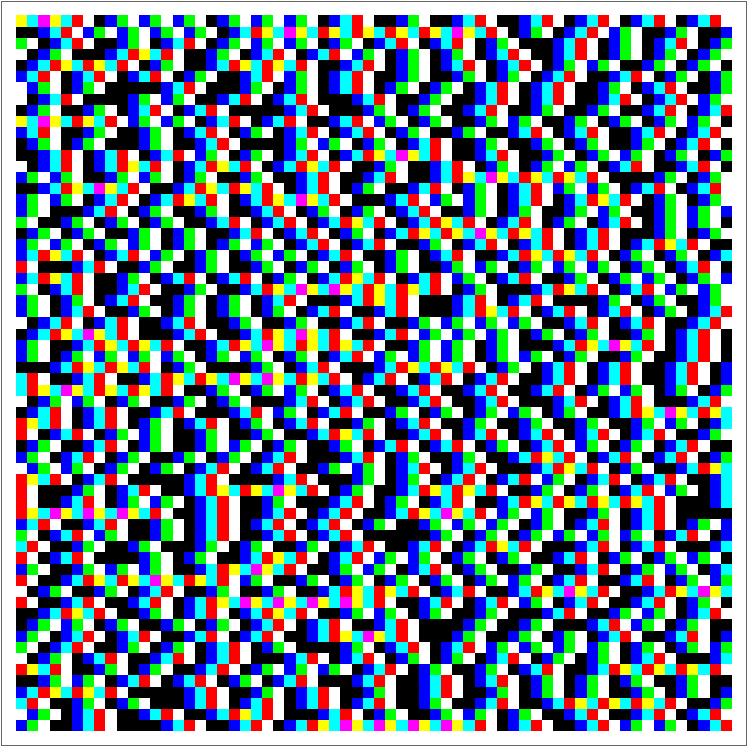}
        \vspace{-16pt}\caption*{Rule 62}
        \label{fig:rule-62-center_columns-4096}
    \end{subfigure}
    \begin{subfigure}[b]{0.135\textwidth}
        \centering
        \includegraphics[width=\textwidth]{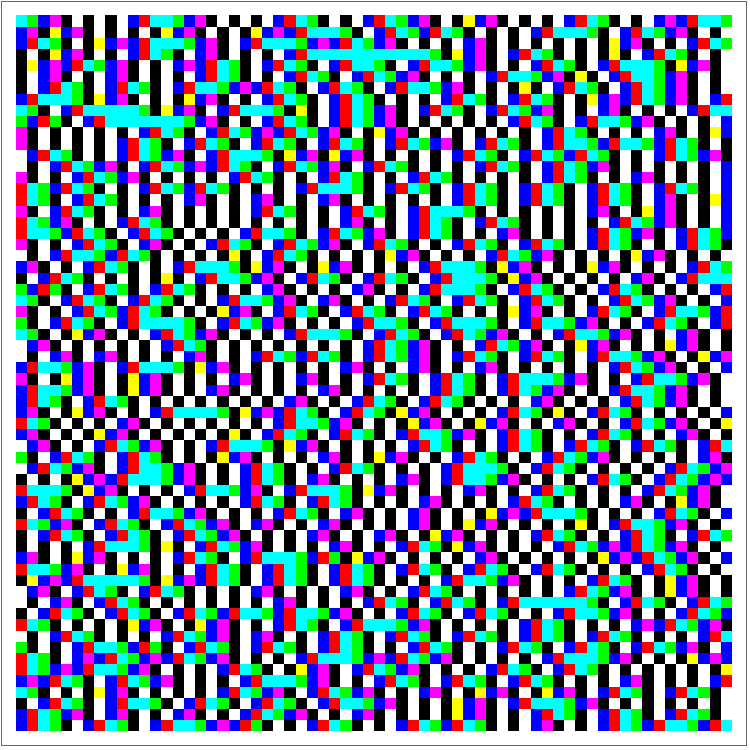}
        \vspace{-16pt}\caption*{Rule 67}
        \label{fig:rule-67-center_columns-4096}
    \end{subfigure}
    \begin{subfigure}[b]{0.135\textwidth}
        \centering
        \includegraphics[width=\textwidth]{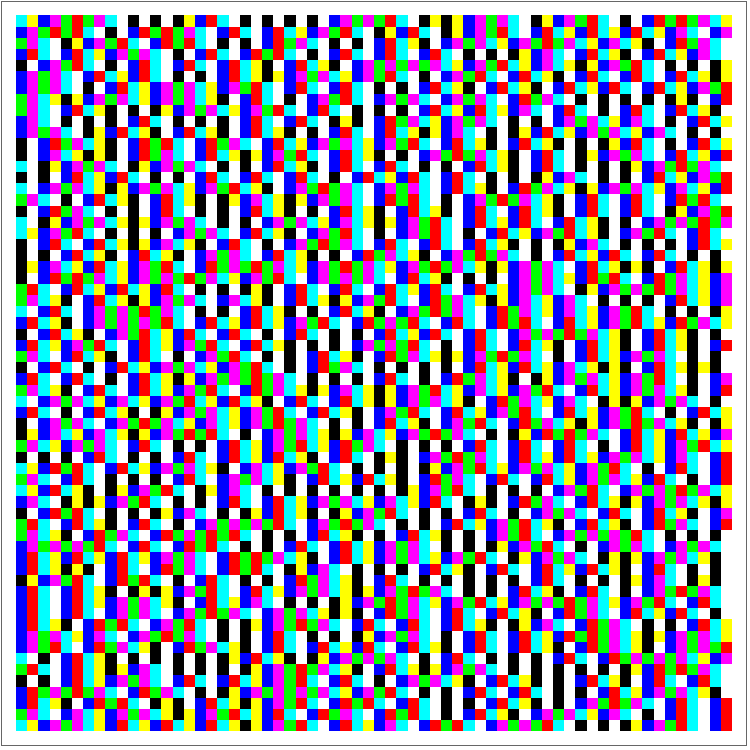}
        \vspace{-16pt}\caption*{Rule 75}
        \label{fig:rule-75-center_columns-4096}
    \end{subfigure}
    \begin{subfigure}[b]{0.135\textwidth}
        \centering
        \includegraphics[width=\textwidth]{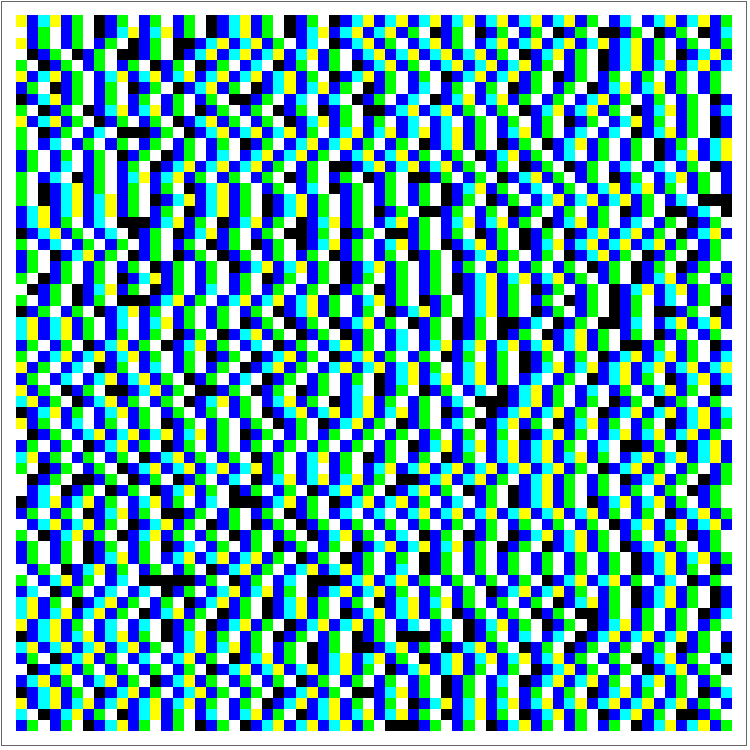}
        \vspace{-16pt}\caption*{Rule 94}
        \label{fig:rule-94-center_columns-4096}
    \end{subfigure}
    \begin{subfigure}[b]{0.135\textwidth}
        \centering
        \includegraphics[width=\textwidth]{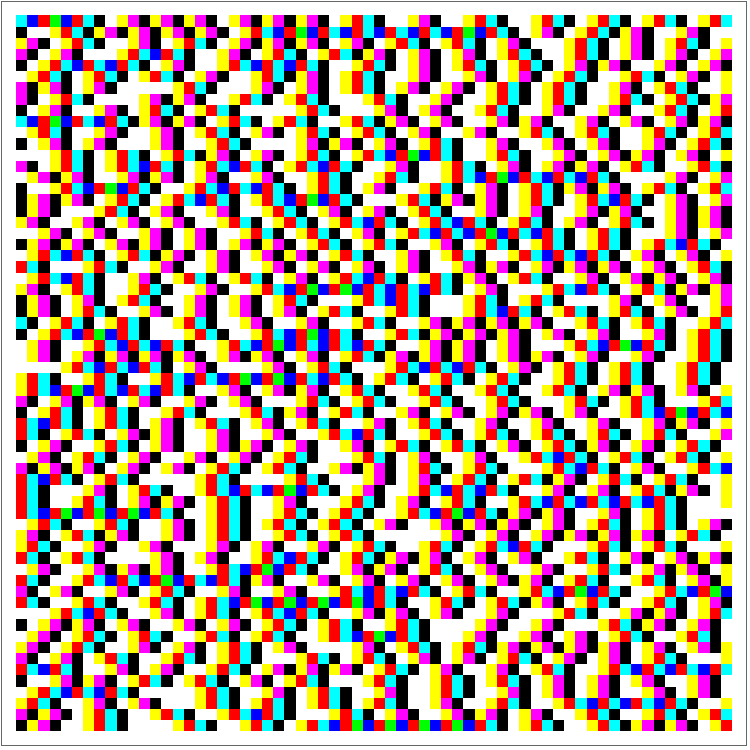}
        \vspace{-16pt}\caption*{Rule 131}
        \label{fig:rule-131-center_columns-4096}
    \end{subfigure}
    \begin{subfigure}[b]{0.135\textwidth}
        \centering
        \includegraphics[width=\textwidth]{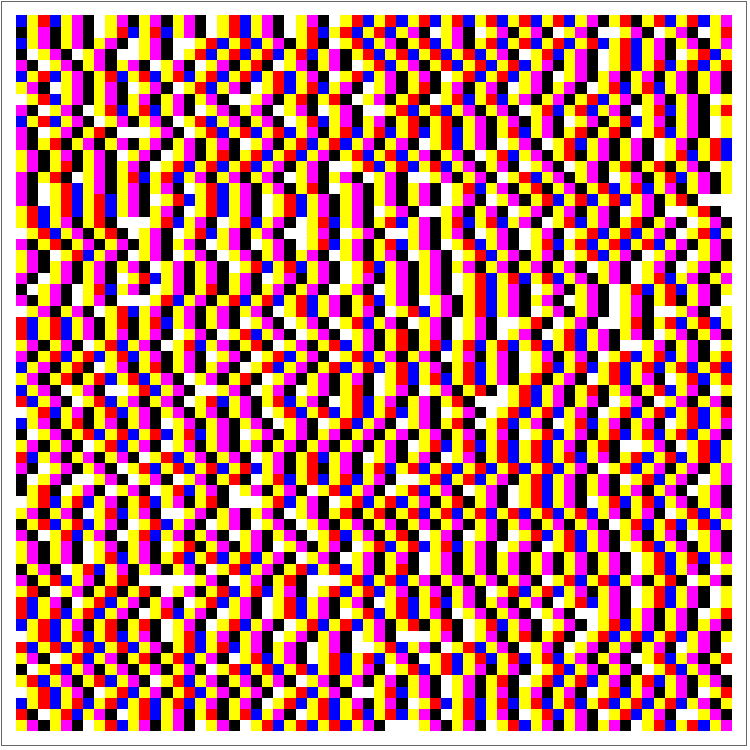}
        \vspace{-16pt}\caption*{Rule 133}
        \label{fig:rule-133-center_columns-4096}
    \end{subfigure}
    \begin{subfigure}[b]{0.135\textwidth}
        \centering
        \includegraphics[width=\textwidth]{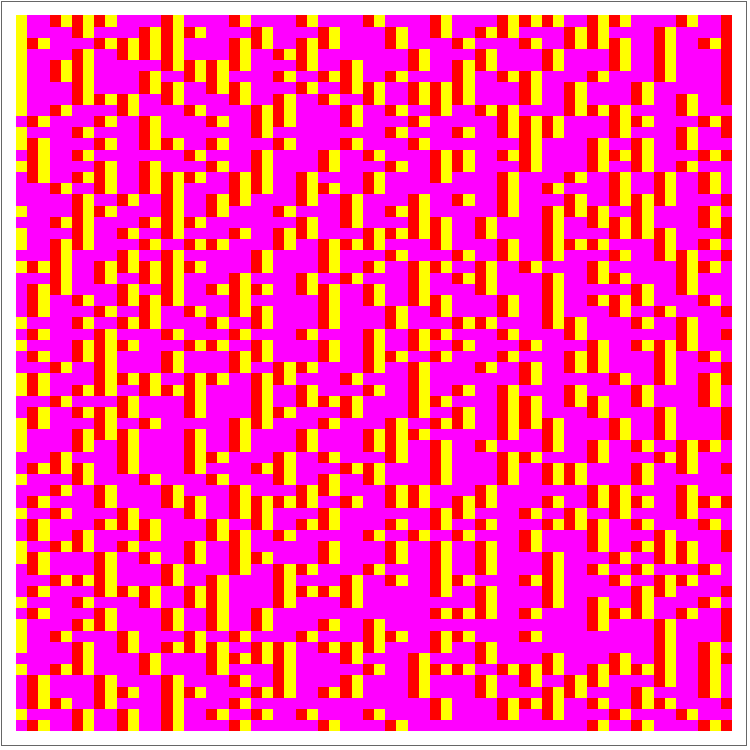}
        \vspace{-16pt}\caption*{Rule 146}
        \label{fig:rule-146-center_columns-4096}
    \end{subfigure}
    \begin{subfigure}[b]{0.135\textwidth}
        \centering
        \includegraphics[width=\textwidth]{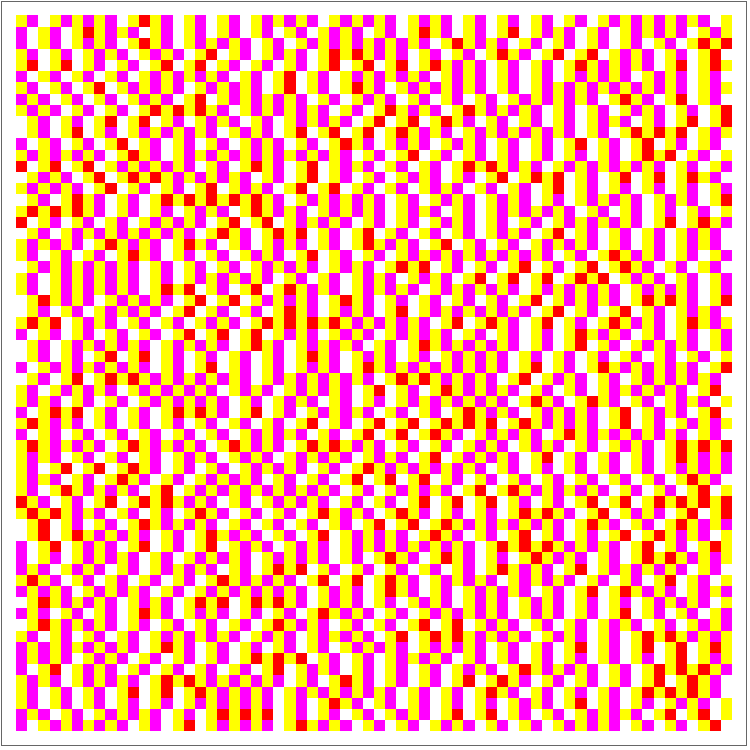}
        \vspace{-16pt}\caption*{Rule 154}
        \label{fig:rule-154-center_columns-4096}
    \end{subfigure}
    \begin{subfigure}[b]{0.135\textwidth}
        \centering
        \includegraphics[width=\textwidth]{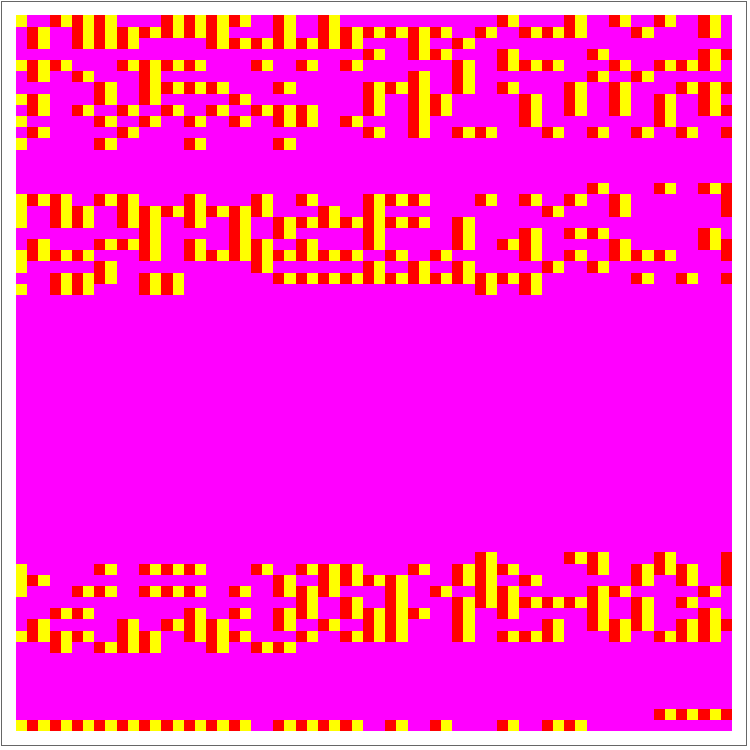}
        \vspace{-16pt}\caption*{Rule 210}
        \label{fig:rule-210-center_columns-4096}
    \end{subfigure}
       \caption{Three center columns, $1\leq t \leq 4096$}
       \label{fig:center-columns}
\end{figure}

\pagebreak
\subsection{Rule 210} \label{rule-210}
Rule 210 is easy to describe: ``cells with one living neighbor change state". It is special for several reasons: it creates larger structures as time goes on, and a new type of structure with very large features appears from around $t=1024$.

\begin{figure}[H]
    \centering
    \begin{tblr}{c c c c c}
            \raisebox{-.05\height}{\includegraphics[width=0.165\textwidth]{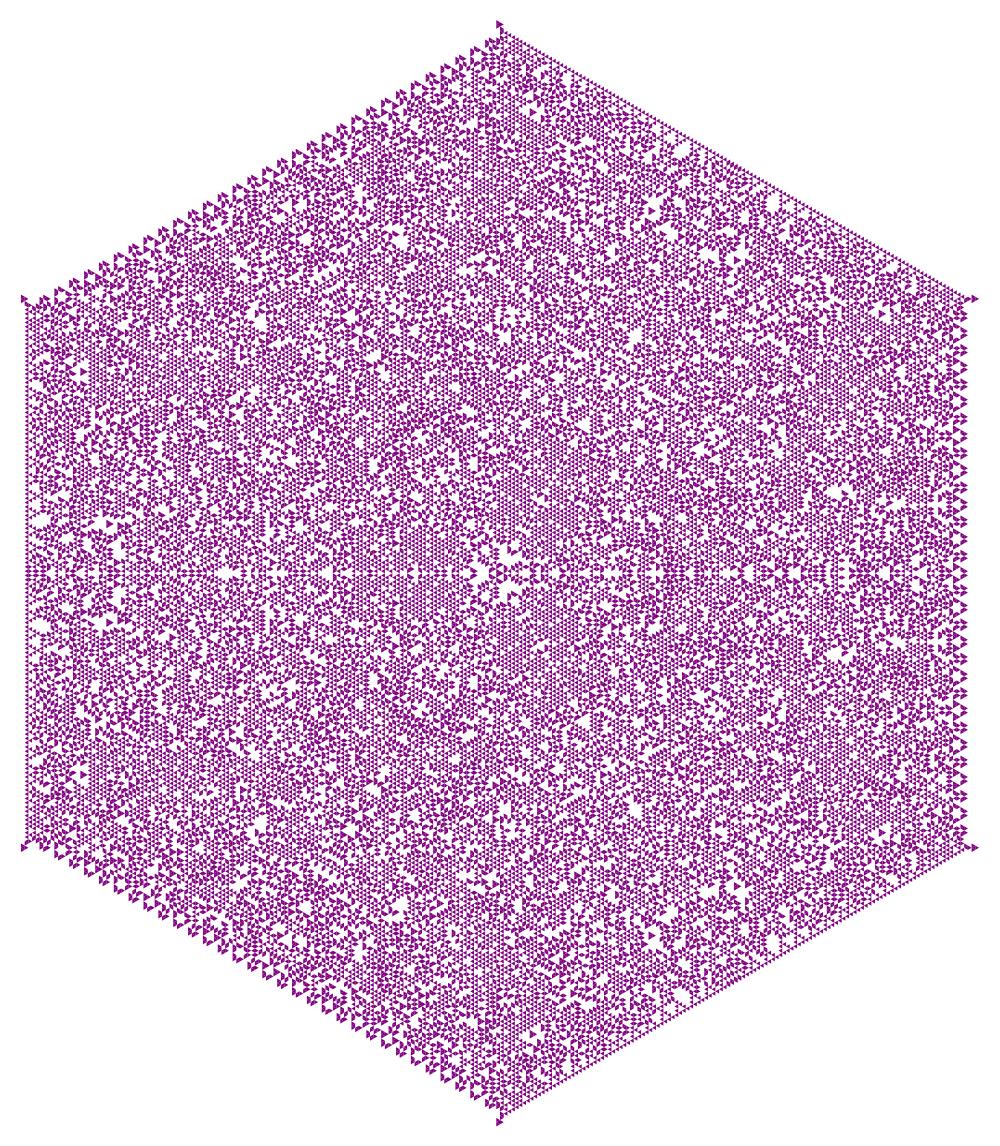}} &
            \raisebox{-.05\height}{\includegraphics[width=0.165\textwidth]{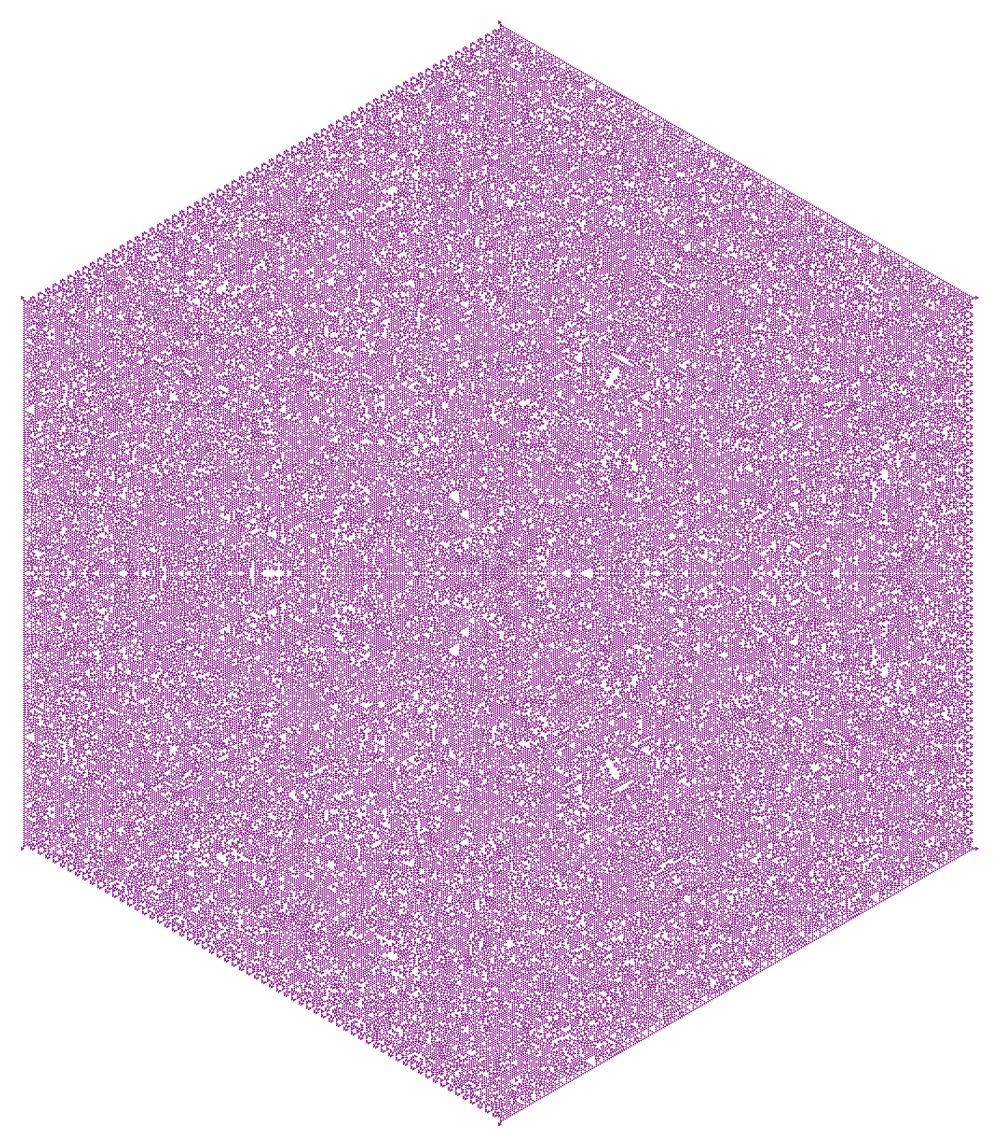}} &
            \raisebox{-.05\height}{\includegraphics[width=0.165\textwidth]{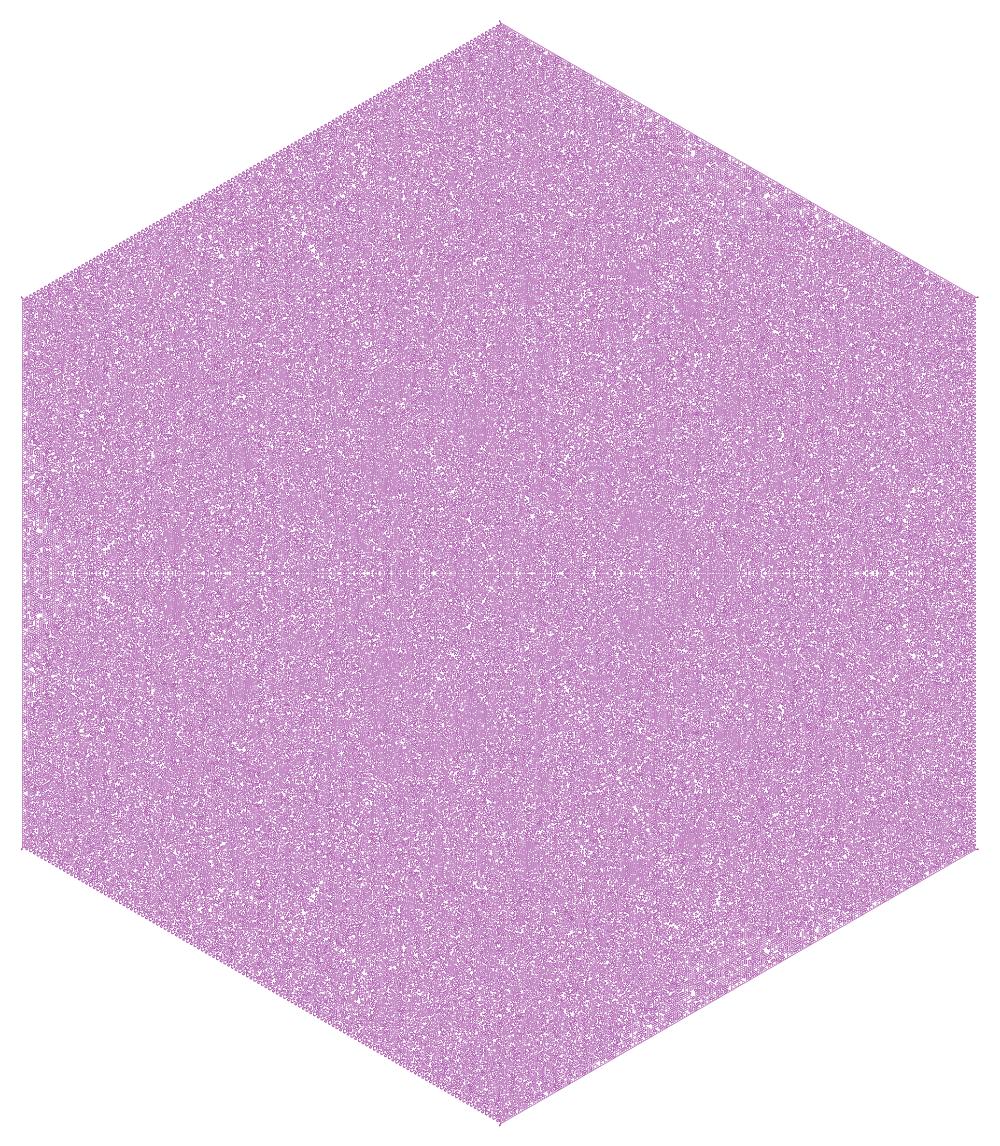}} &
            \raisebox{-.05\height}{\includegraphics[width=0.165\textwidth]{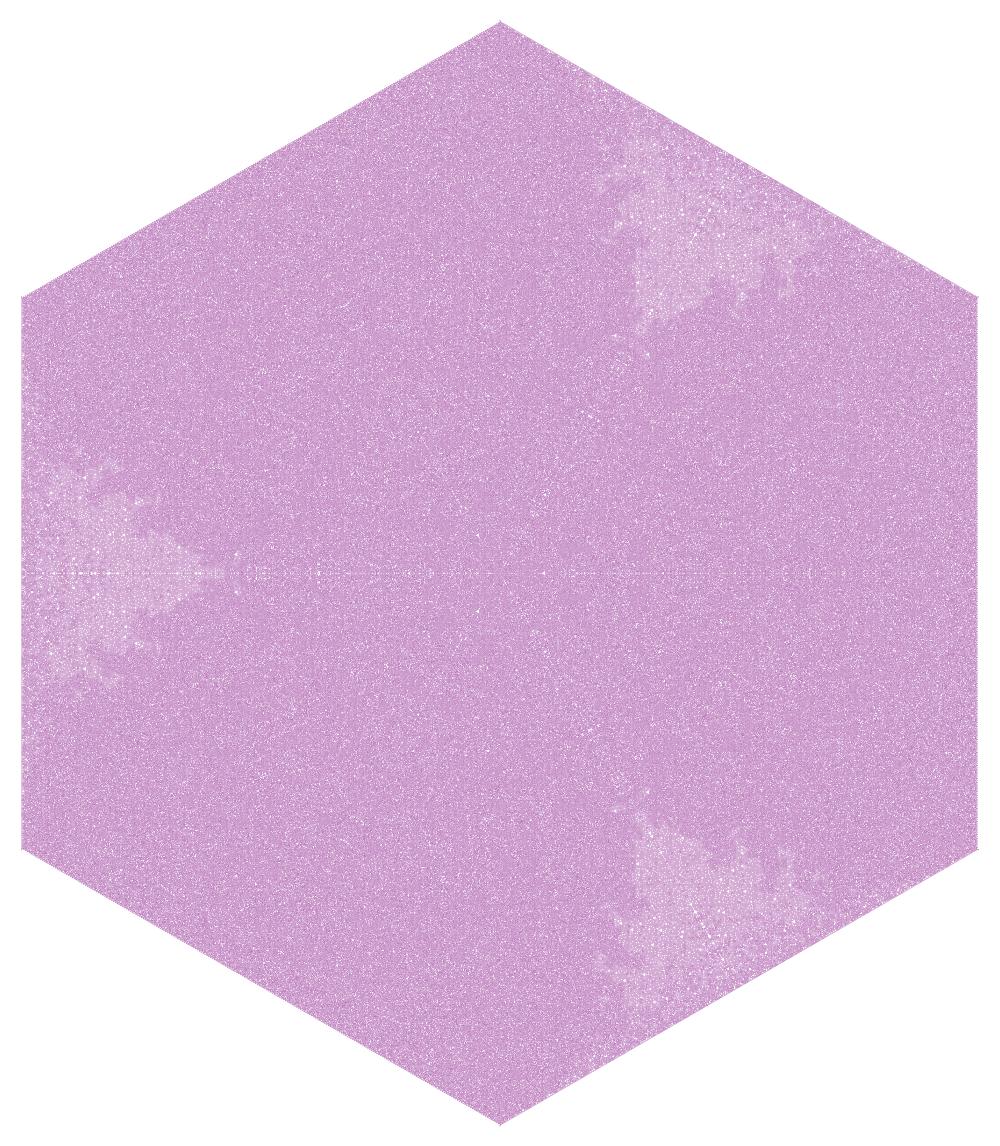}} &
            \raisebox{-.05\height}{\includegraphics[width=0.165\textwidth]{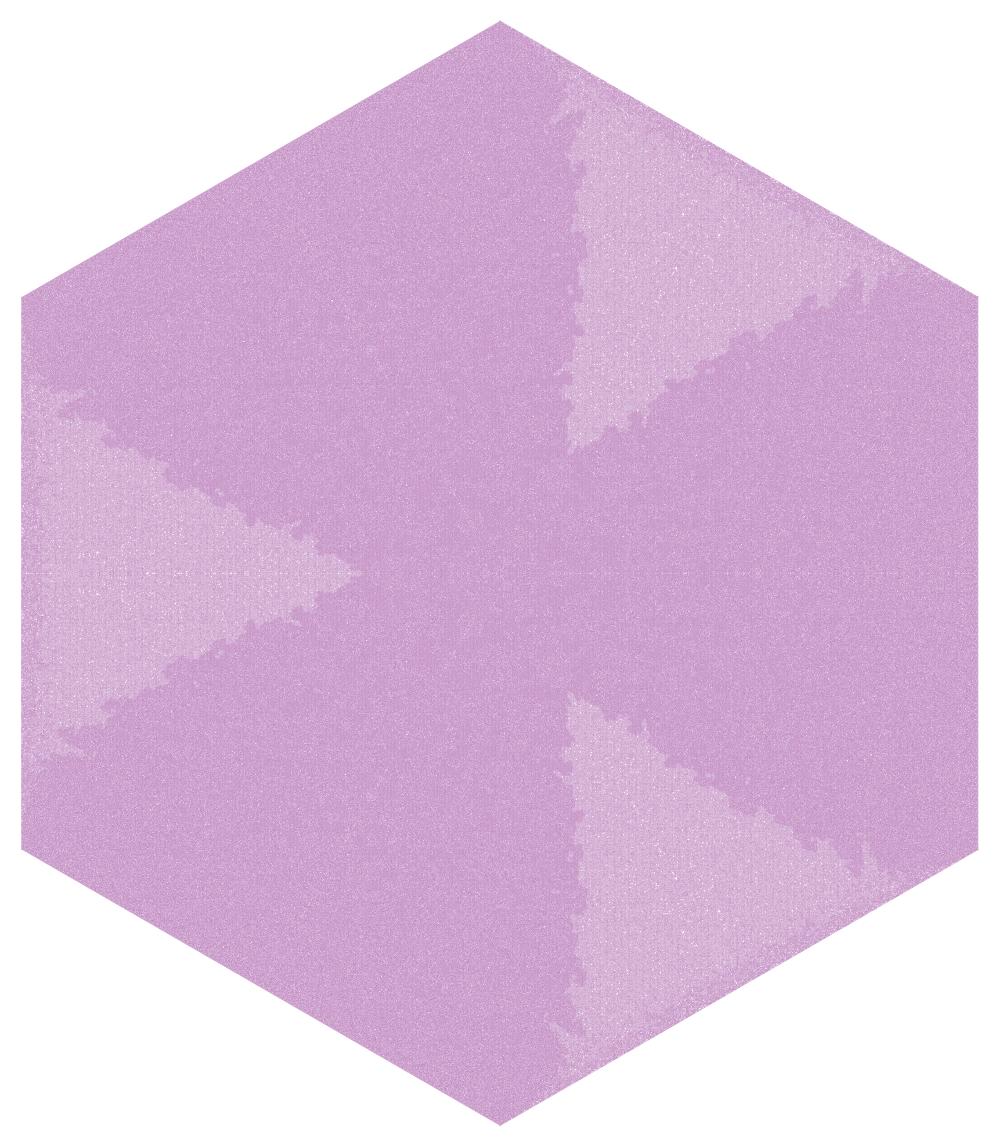}} \\
            257 & 513 & 1025 & 2049 & 4097
    \end{tblr}
    \caption{Rule 210 at $t=2^i + 1$ (the structure is best seen on odd time steps from afar)}
    \label{tab:rule-210-evolution}
\end{figure}

\begin{figure}[H]
    \centering
    \includegraphics[width=.85\textwidth]{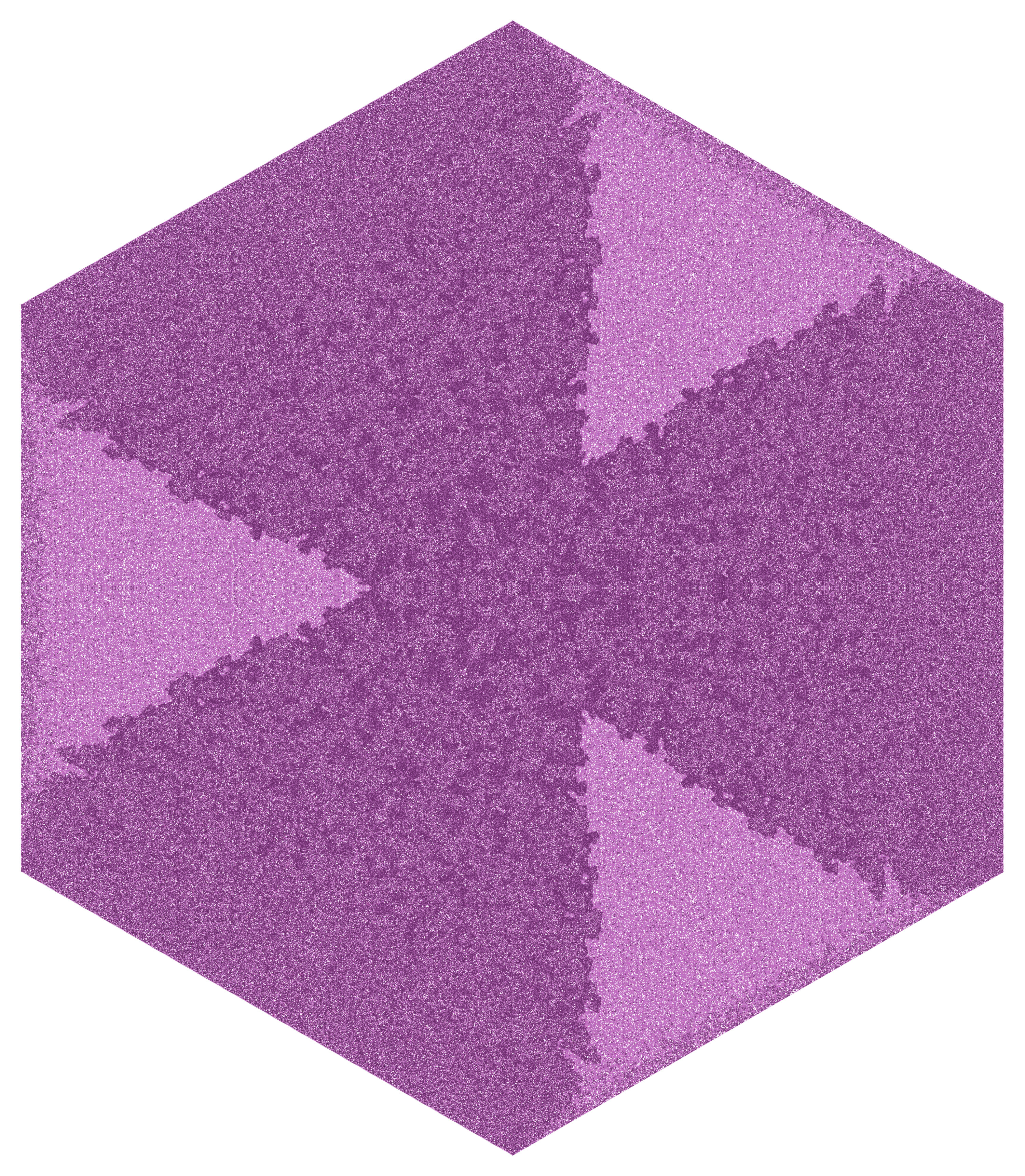}
    \caption{Rule 210 at $t=4097$ with enhanced contrast}
    \label{fig:rule-210-time-4097}
\end{figure}

\begin{figure}[H]
    \centering
    \includegraphics[width=\textwidth]{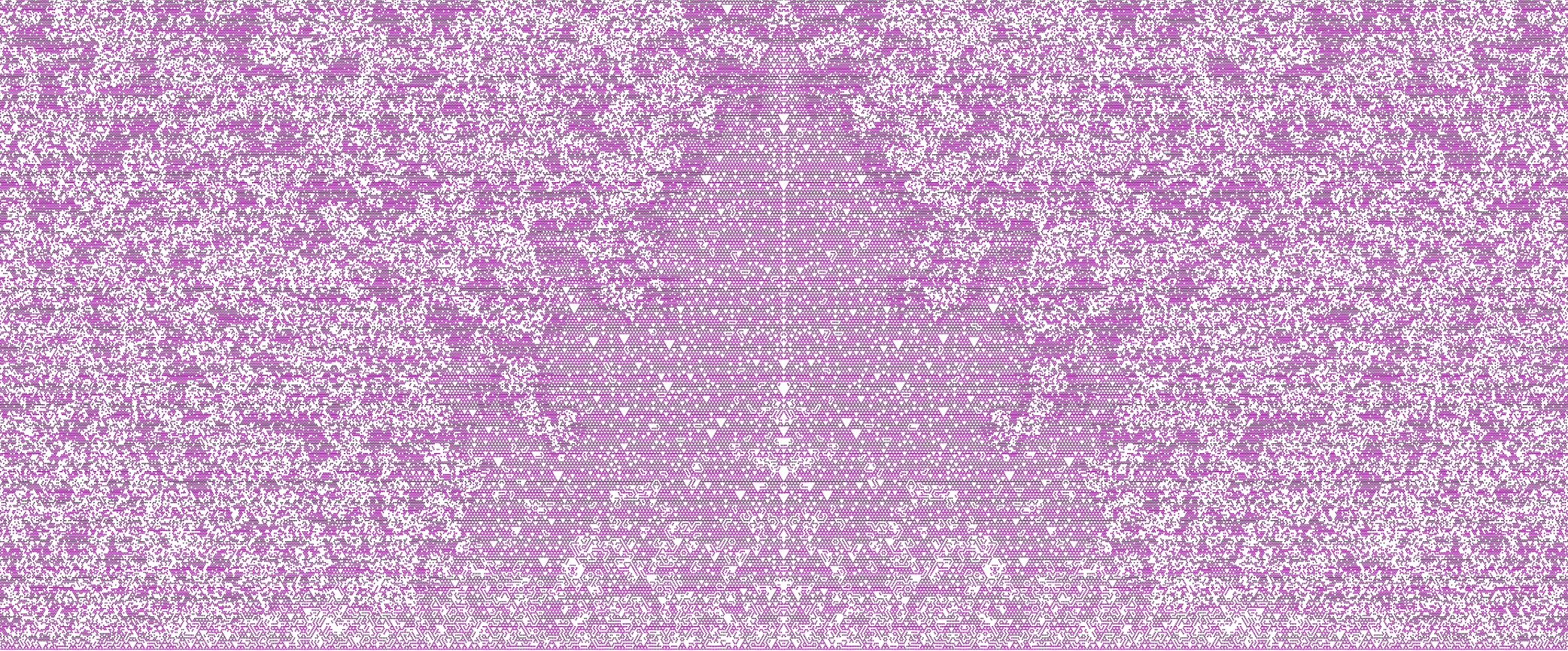}
    \vspace{-18pt}
    \caption{Close-up of rule 210 at $t=2048$}
    \label{fig:rule-210-close-up-2048}
\end{figure}
\vspace{-10pt}
\begin{figure}[H]
    \centering
    \includegraphics[width=.9\textwidth]{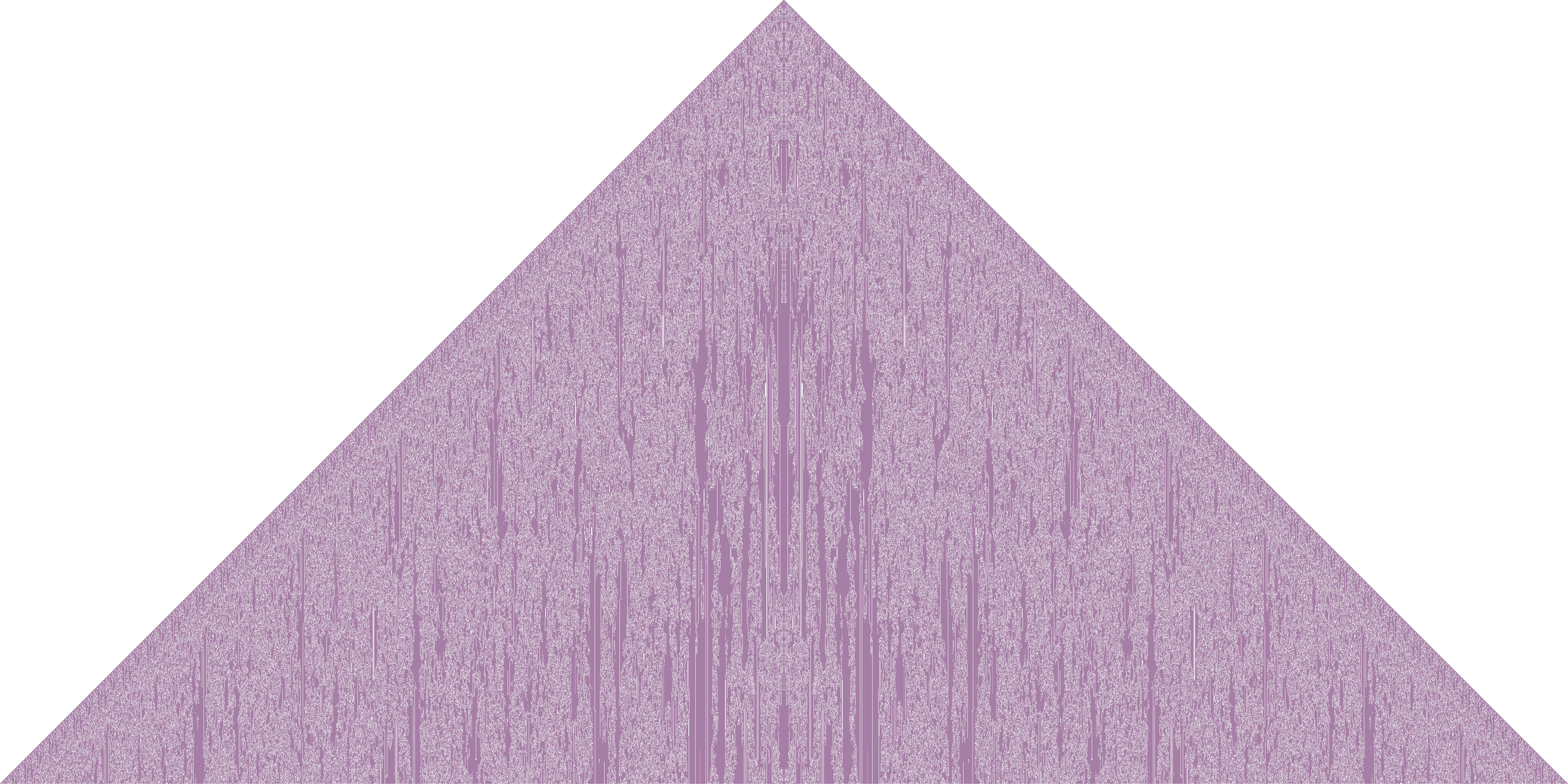}
    \vspace{-6pt}
    \caption{Slice plot of rule 210 up to $t=4096$}
    \label{fig:rule-210-slice-2816}
\end{figure}

\noindent Rule 210 initially alternates between what seems like 2 population densities (i.e., number of living cells / size of the first cell’s region of influence). When the new structure appears, around $t=1024$, the population stabilizes in a density cycle of period 4 between slowly evolving values.

\begin{figure}[H]
    \centering
    \begin{subfigure}[b]{0.51\textwidth}
        \centering
        \includegraphics[width=\textwidth]{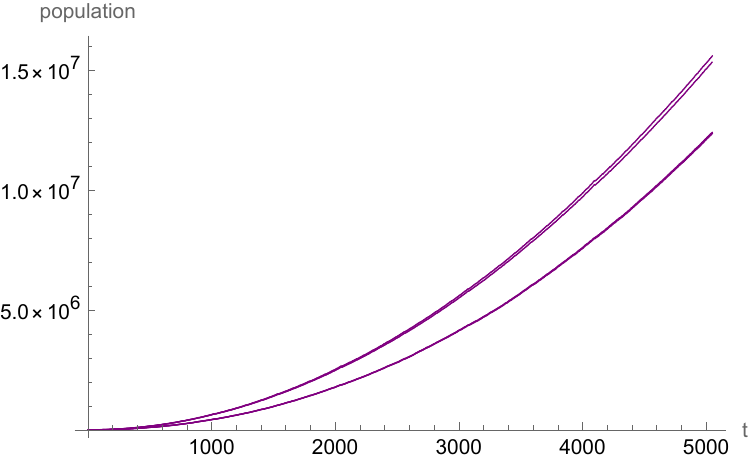}
        \caption{Population (OEIS \href{https://oeis.org/A372581}{A372581})}
    \end{subfigure}
    \begin{subfigure}[b]{0.48\textwidth}
        \centering
        \includegraphics[width=\textwidth]{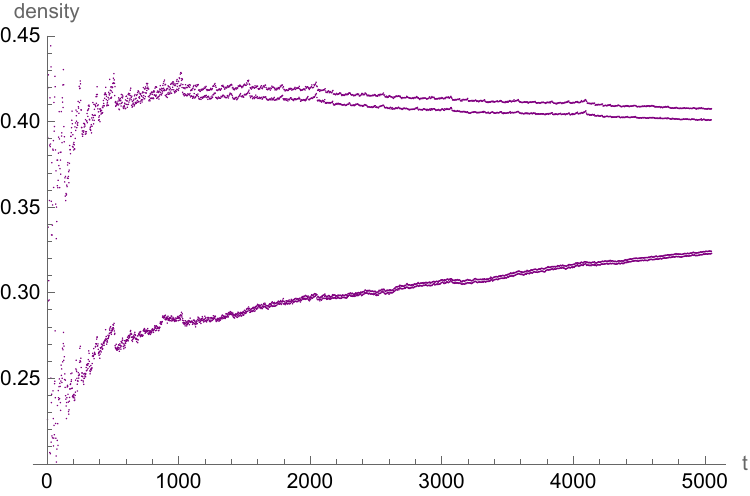}
        \caption{Population density}
    \end{subfigure}
       \caption{Evolution of two coarse-grained values under rule 210}
       \label{fig:rule-210-coarse-grained}
\end{figure}

\pagebreak
\subsection{Noise} \label{noise}
Some rules seem to generate a pretty good noise. For example, if we pick a simple starting point without symmetries, rule 37 will usually turn it into an expanding disk with a random-looking interior.

\begin{figure}[H]
    \centering
    \begin{subfigure}[t]{0.39\textwidth}
        \centering
        \raisebox{0.45\textwidth}{
            \includegraphics[width=.2\textwidth]{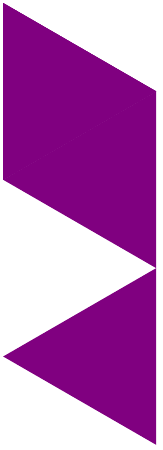}
        }
        \caption{Simple asymmetric starting point}
    \end{subfigure}
    \begin{subfigure}[t]{0.6\textwidth}
        \centering
        \includegraphics[width=\textwidth]{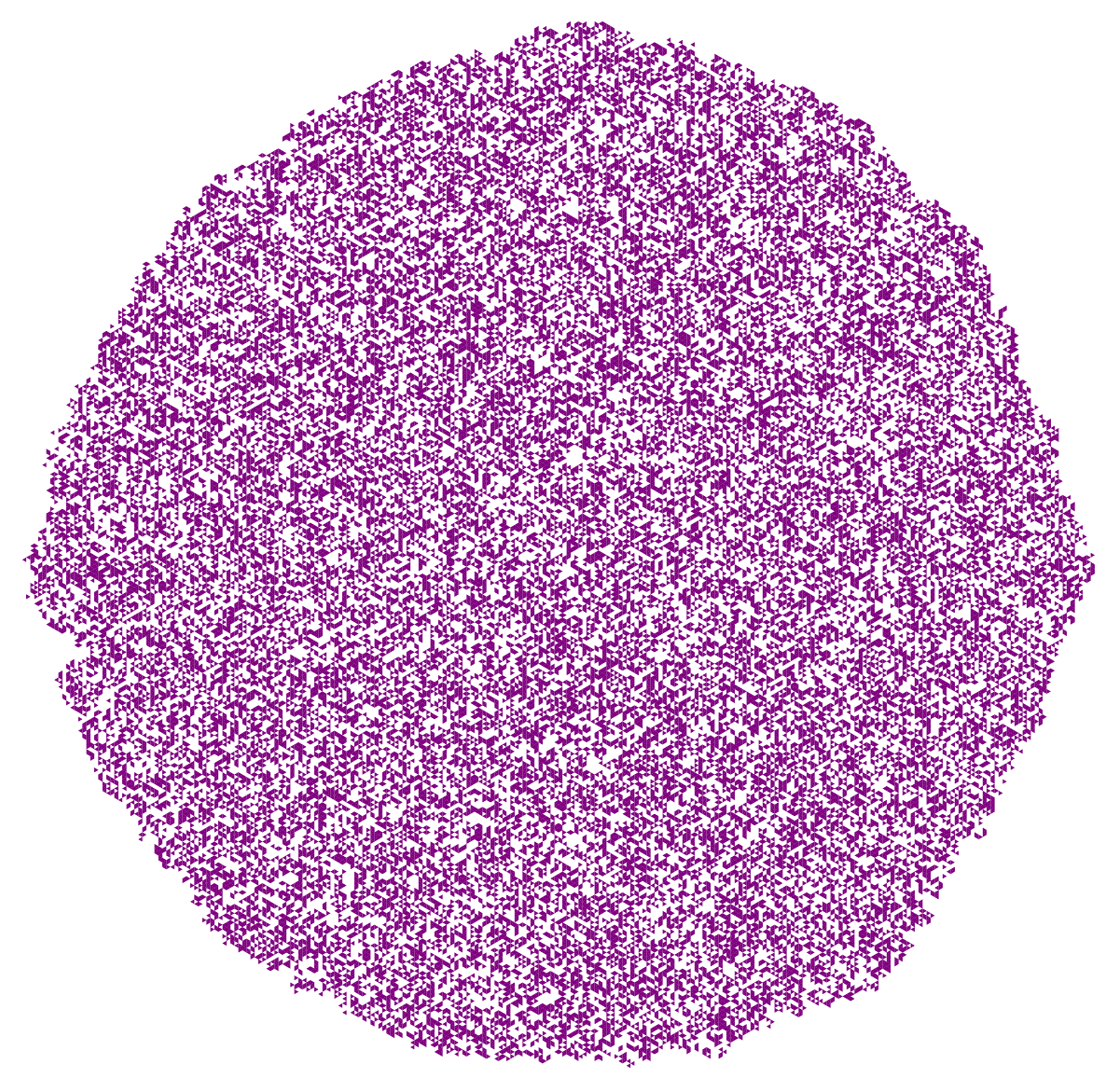}
        \caption{Result at $t=512$}
        \label{fig:rule-37-time-512-asymmetric}
    \end{subfigure}
    \caption{Generating noise with rule 37}
    \label{fig:noise}
\end{figure}

\vspace{5pt}

\subsection{Textures} \label{textures}
Organic textures can be obtained by applying other rules to this pseudorandom grid.

\begin{figure}[H]
    \centering
    \begin{subfigure}[b]{0.32\textwidth}
        \centering
        \includegraphics[width=\textwidth]{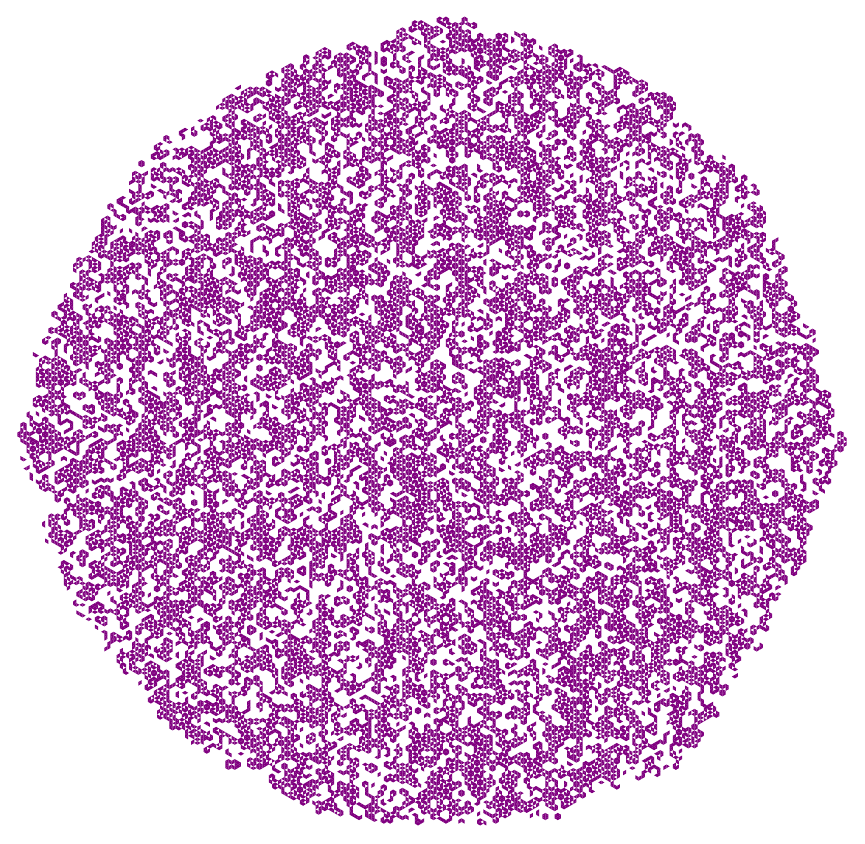}
        \caption{Rule 100 at $t=64$}
        \label{fig:rule-100-time-64-noise}
    \end{subfigure}
    \begin{subfigure}[b]{0.32\textwidth}
        \centering
        \includegraphics[width=\textwidth]{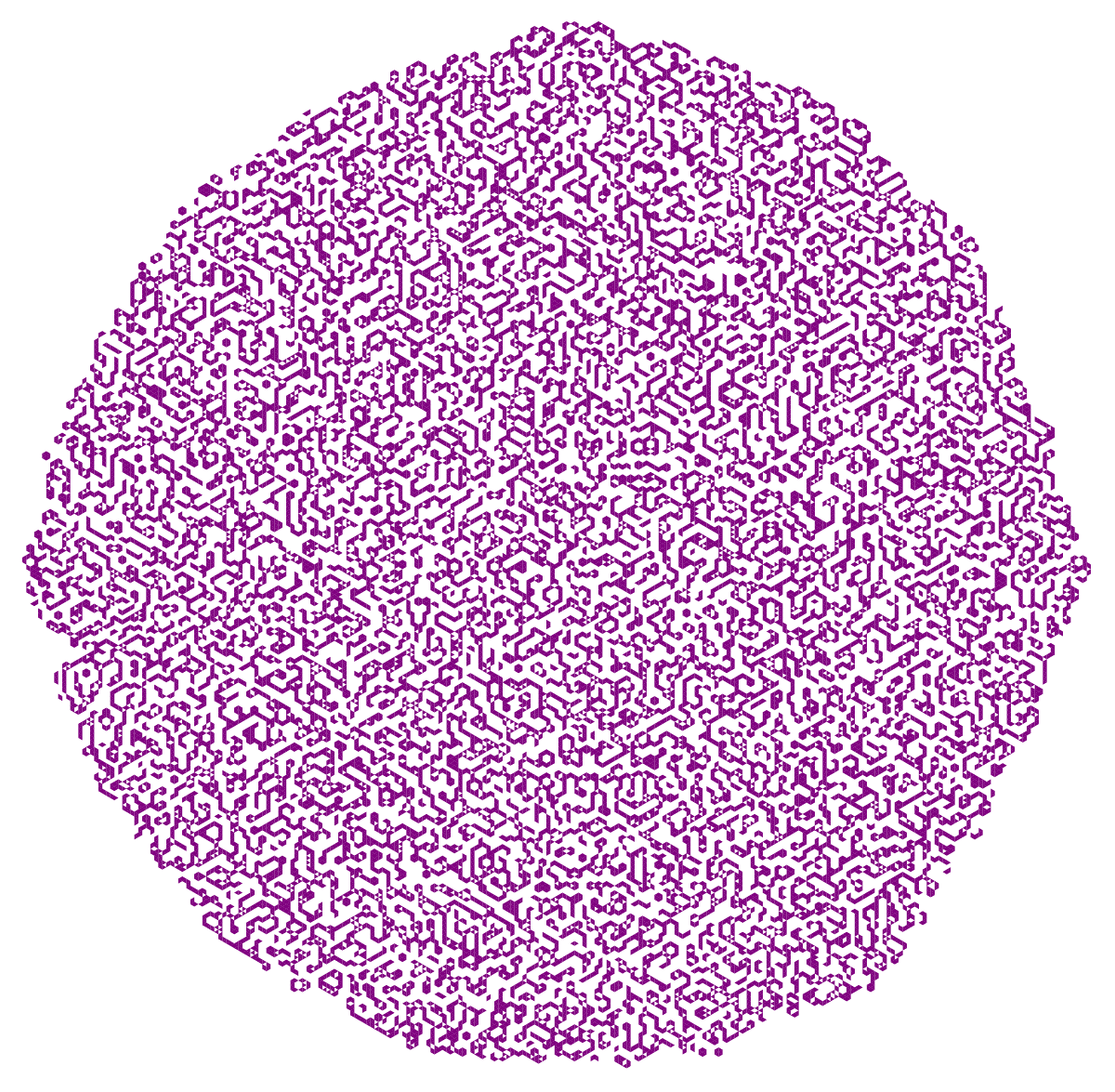}
        \caption{Rule 108 at $t=512$}
        \label{fig:rule-108-time-512-noise}
    \end{subfigure}
    \begin{subfigure}[b]{0.32\textwidth}
        \centering
        \includegraphics[width=\textwidth]{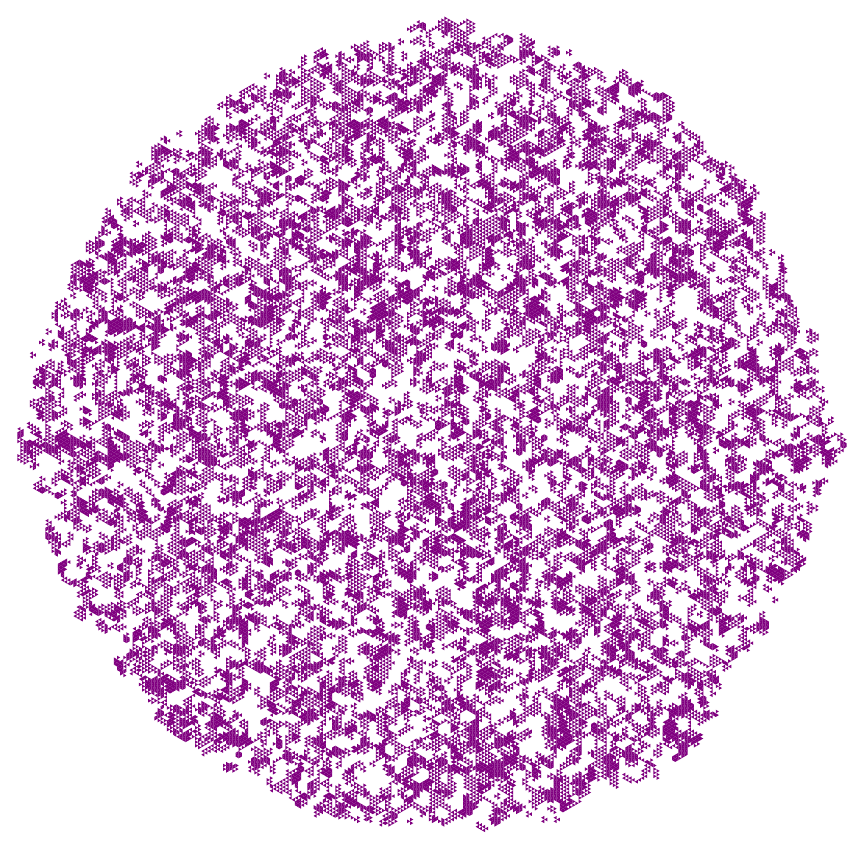}
        \caption{Rule 204 at $t=32$}
        \label{fig:rule-204-time-32-noise}
    \end{subfigure}
       \caption{Starting from \textit{Figure \ref{fig:rule-37-time-512-asymmetric}}}
       \label{fig:textures}
\end{figure}

\pagebreak
\subsection{Boring Rules} \label{boring-rules}
\noindent Rule 240 is the \textbf{identity} and leaves any grid unchanged. Rule 15 is the \textbf{negative} rule that swaps alive and dead states. Rule 0 kills everything, and rule 255 makes all cells live.

\begin{figure}[H]
    \centering
        \includegraphics[width=.55\textwidth]{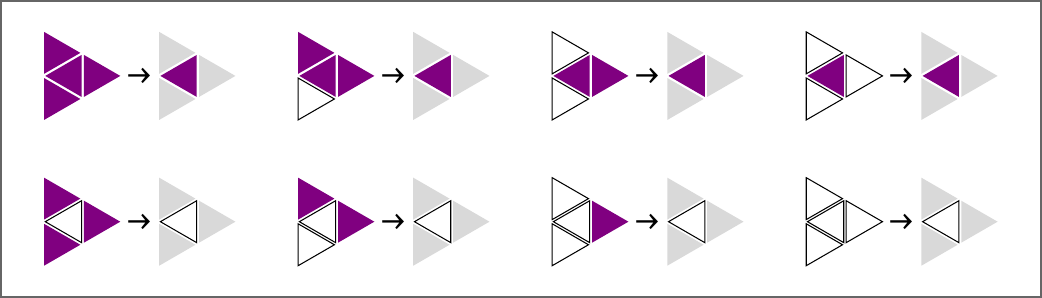}
    \caption{Rule 240}
    \label{fig:rule-plot-240}
\end{figure}

\vspace{-1em}

\begin{figure}[H]
    \centering
        \includegraphics[width=.55\textwidth]{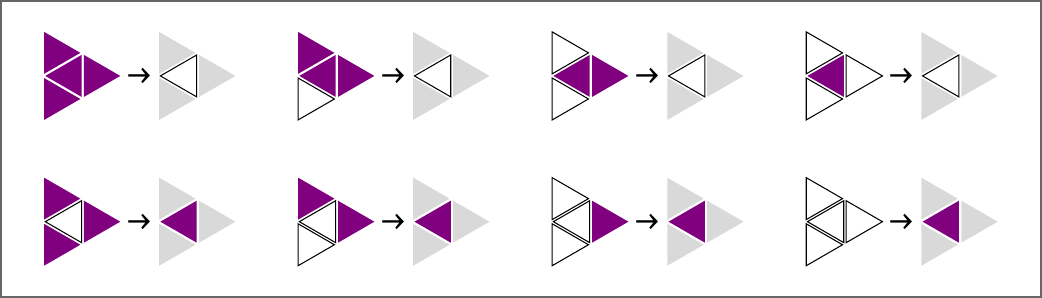}
    \caption{Rule 15}
    \label{fig:rule-plot-15}
\end{figure}

\subsection{Twins} \label{twins}
A simple procedure can be followed to find the evil twin of a rule that has the same effect but in the negative world. To find it, take the number in its binary form (with the leading zeros needed to make it 8 digits long), swap ones and zeros and read it backward. Let us take rule 214 as an example. \\

\begin{tabular}{l c c c c}
     First, find the binary form of the rule number, 
     & & $214=$ \hspace{-1.3em} & $11010110_2$ & \\
     then swap ones and zeros
     & & & $00101001_2$ & \\
     and finally reverse it.
     & & & $10010100_2$ & \hspace{-1.3em} $=148$
\end{tabular}

\begin{figure}[H]
     \centering
     \begin{subfigure}[b]{0.48\textwidth}
         \centering
         \includegraphics[width=\textwidth]{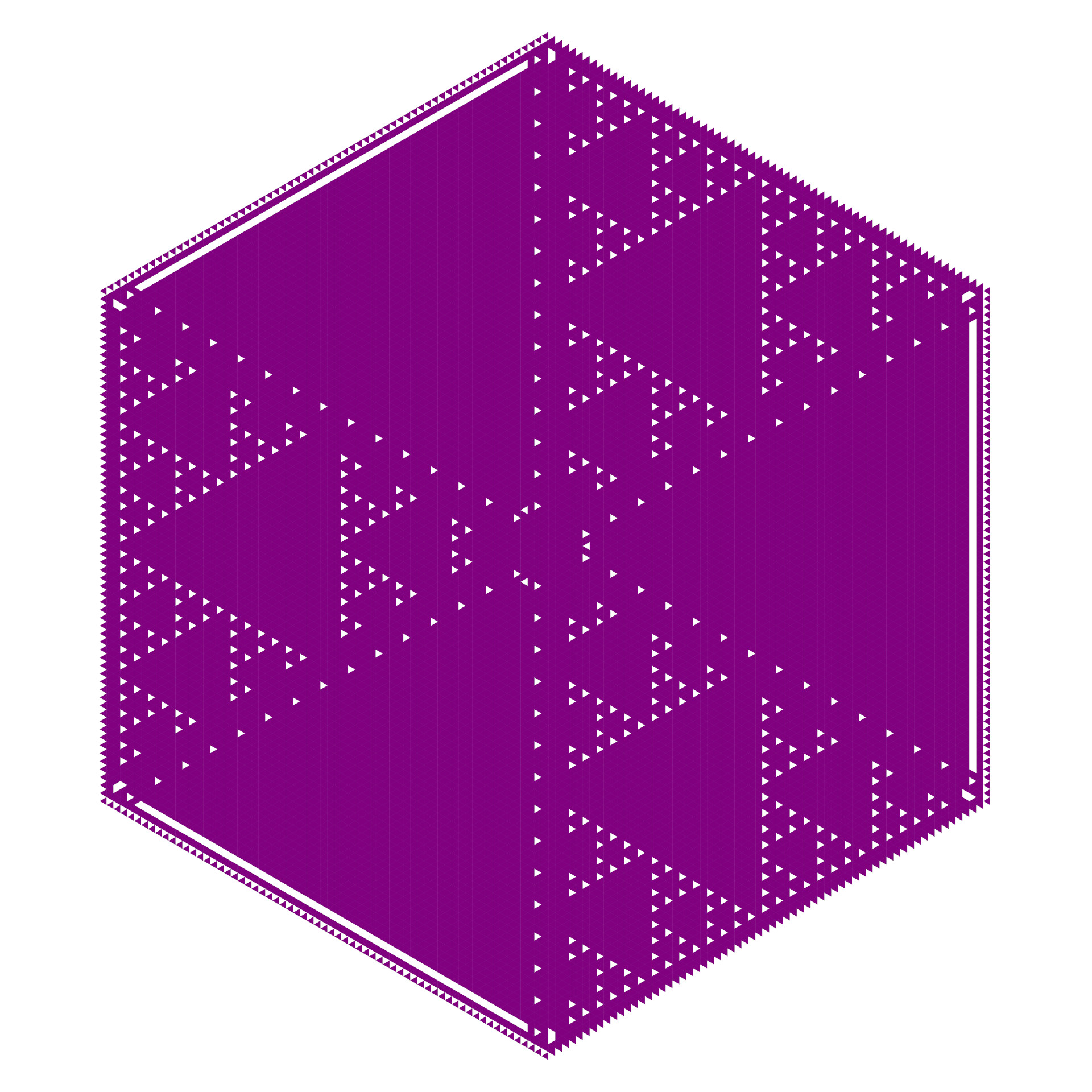}
         \caption{Rule 214 from one living cell at $t=128$}
         \label{fig:twin-1}
     \end{subfigure}
     \hspace{0.02\textwidth}
     \begin{subfigure}[b]{0.48\textwidth}
         \centering
         \includegraphics[width=\textwidth]{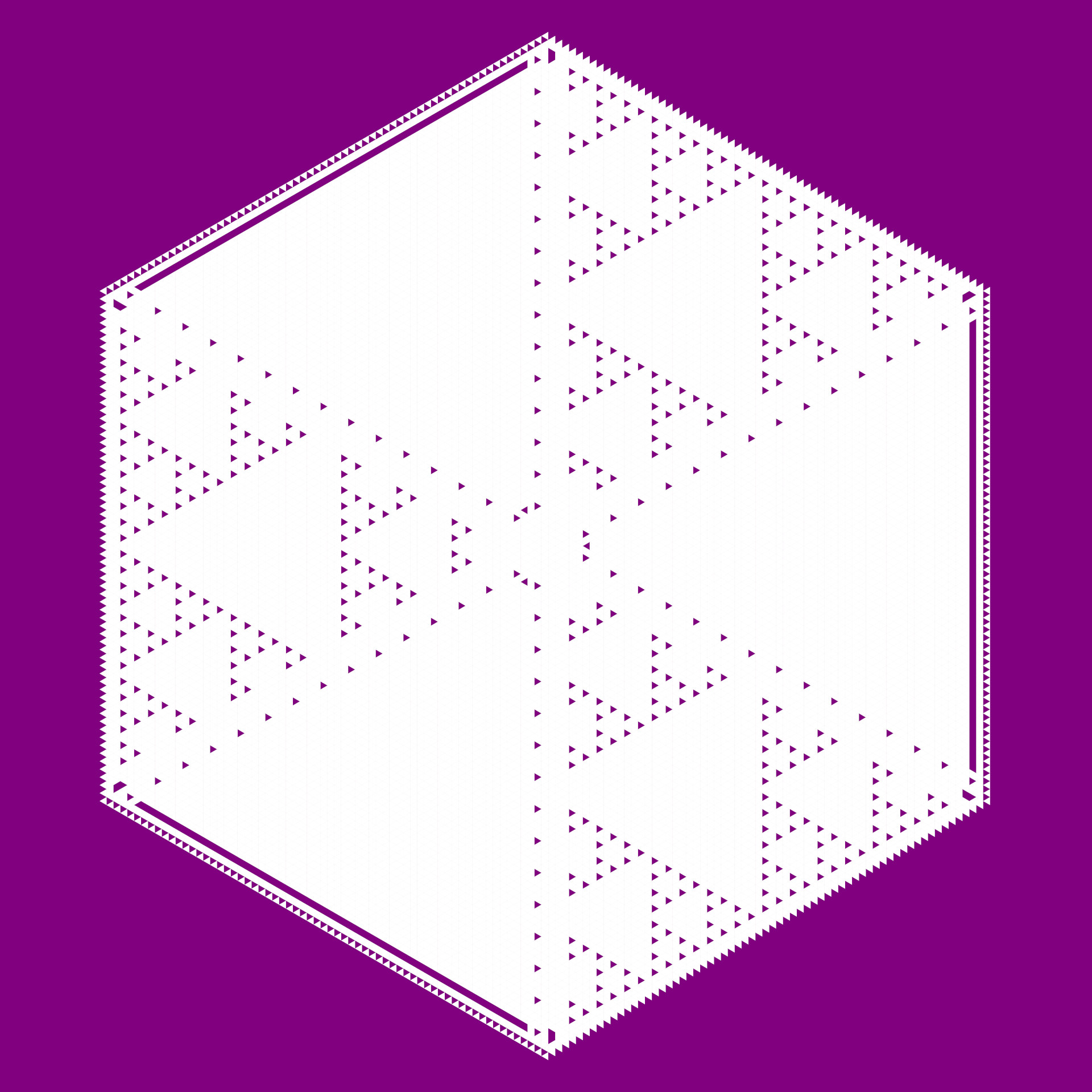}
         \caption{Rule 148 from one dead cell at $t=128$}
         \label{fig:twin-2}
     \end{subfigure}
        \caption{Twin rules}
        \label{fig:twins}
\end{figure}

\pagebreak
\section{Implementation} \label{implementation}
\subsection{Tools}
% \noindent Let us start by defining a few useful operators and matrices.

\begin{enumerate}
\item 
$\diamond$ here represents an operator joining matrices corner to corner.

\begin{equation}
\begin{pmatrix}
{\color{purple}a} 
\end{pmatrix}
\diamond
\begin{pmatrix}
{\color{red}b} & {\color{red}c} \\
{\color{red}d} & {\color{red}e}
\end{pmatrix}
\diamond
\begin{pmatrix}
{\color{orange}f} & {\color{orange}g} & {\color{orange}h} 
\end{pmatrix}
=
\begin{pmatrix}
 {\color{purple}a} & {\color{lightgray}0} & {\color{lightgray}0} & {\color{lightgray}0} & {\color{lightgray}0} & {\color{lightgray}0} \\
{\color{lightgray}0} & {\color{red}b} & {\color{red}c} & {\color{lightgray}0} & {\color{lightgray}0} & {\color{lightgray}0}\\
{\color{lightgray}0} & {\color{red}d} & {\color{red}e} & {\color{lightgray}0} & {\color{lightgray}0} & {\color{lightgray}0}\\
{\color{lightgray}0} & {\color{lightgray}0} & {\color{lightgray}0} & {\color{orange}f} & {\color{orange}g} & {\color{orange}h} 
\end{pmatrix}
\end{equation}

\item 
$^\searrow\,$ shifts diagonally the elements of a matrix and places the last row and column first.

\begin{equation}
\begin{pmatrix}
a & b & c & {\color{orange}d} \\
e & f & g & {\color{orange}h} \\
{\color{red}i} & {\color{red}j} & {\color{red}k}  & {\color{purple}l} 
\end{pmatrix}
^{\hspace{-1.5mm}\searrow}
=
\begin{pmatrix}
{\color{purple}l}  & {\color{red}i} & {\color{red}j} & {\color{red}k} \\
{\color{orange}d} & a & b & c \\
{\color{orange}h} & e & f & g \\
\end{pmatrix}
\end{equation}

\item 
@ will be an operator applying a function to every element of a matrix. 
\begin{equation}
f\text{@}
\begin{pmatrix}
a & b \\
c & d
\end{pmatrix}
=
\begin{pmatrix}
f(a) & f(b) \\
f(c) & f(d)
\end{pmatrix}
\end{equation}

\item 
$\mathbb{I}_i$ is the $i\times i$ identity matrix. $\mathbb{S}_i$ will be an $i\times (i+1)$ ``stairs" matrix.
\begin{equation}
    \mathbb{I}_1=
    \begin{pmatrix}
    {\color{purple}1}
    \end{pmatrix}
\qquad\quad\;
    \mathbb{I}_2=
    \begin{pmatrix}
    {\color{purple}1} & 0 \\
    0 & {\color{purple}1}
    \end{pmatrix}
\qquad\quad\;
    \mathbb{I}_3=
    \begin{pmatrix}
    {\color{purple}1} & 0 & 0 \\
    0 & {\color{purple}1} & 0 \\
    0 & 0 & {\color{purple}1} 
    \end{pmatrix}
\qquad\quad\;
    \hdots
\quad
\end{equation}
\begin{equation}
    \mathbb{S}_1=
    \begin{pmatrix}
    {\color{purple}1} & {\color{purple}1}
    \end{pmatrix}
\qquad
    \mathbb{S}_2=
    \begin{pmatrix}
    {\color{purple}1} & {\color{purple}1} & 0 \\
    0 & {\color{purple}1} & {\color{purple}1} 
    \end{pmatrix}
\qquad
    \mathbb{S}_3=
    \begin{pmatrix}
    {\color{purple}1} & {\color{purple}1} & 0 & 0 \\
    0 & {\color{purple}1} & {\color{purple}1} & 0 \\
    0 & 0 & {\color{purple}1} & {\color{purple}1} 
    \end{pmatrix}
\qquad
    \hdots
\quad
\end{equation}

\end{enumerate}

\subsection{Growing the triangular grid} \label{growing-the-grid}

\begin{figure}[H]
    \centering
        \includegraphics[width=.39\textwidth]{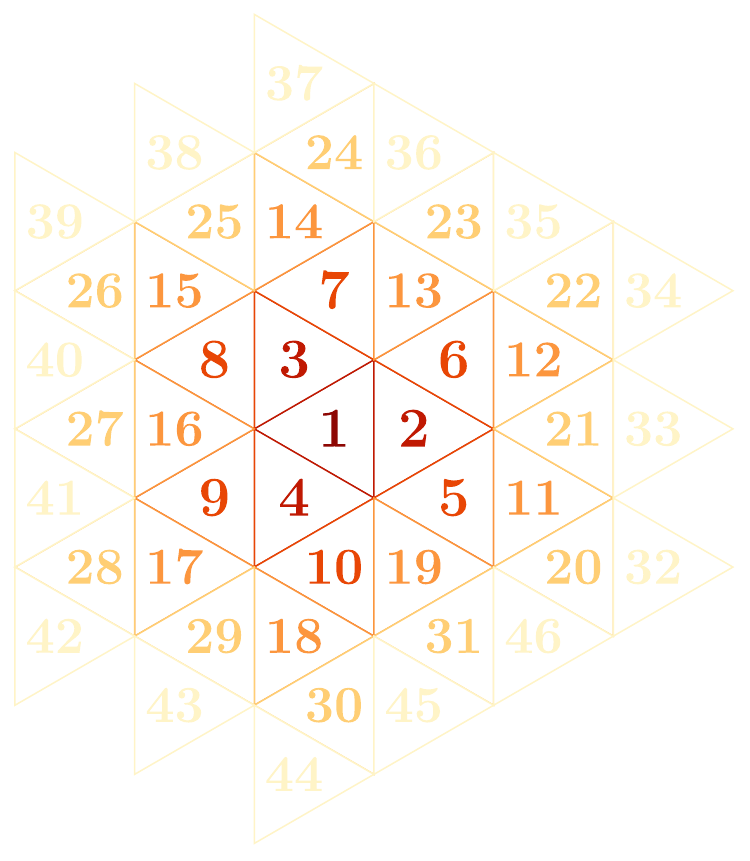}
    \vspace{-.5em}
    \caption{Ordering of cells}
    \label{fig:cell-ordering}
\end{figure}

As mentioned in \textit{Section \ref{introduction}}, the grid will be grown from a single cell by adding layers. The cells will be ordered counterclockwise, with the first vertex of each new layer placed on the southeast diagonal (see \textit{Figure \ref{fig:cell-ordering}}).

\vspace{-.5em}

This is done by using a precursor of the adjacency matrix called the \textbf{grid matrix} $\mathcal{G}$.\linebreak For the first three layers, the grid matrix is best hand-coded. The third matrix $\mathcal{G}_3$ can be seen in \textit{Figure \ref{fig:grid-matrix}}. The subsequent layers will follow a repeating pattern.

\begin{figure}[H]
\centering
\resizebox{0.7\textwidth}{!}{$
\begin{pmatrix}
    \begin{matrix}
        0 & {\color{purple}1} & {\color{purple}1} & {\color{purple}1}
    \end{matrix} 
        & 
    {\color{lightgray} \begin{matrix}
        0 & 0 & 0 & 0 & 0 & 0
    \end{matrix}}
    & 
    {\color{lightgray} \begin{matrix}  
        0 & 0 & 0 & 0 & 0 & 0 & 0 & 0 & 0
    \end{matrix}}
    \\ 
    {\color{lightgray} \begin{matrix}
        0 & 0 & 0 & 0 \\
        0 & 0 & 0 & 0 \\
        0 & 0 & 0 & 0 
    \end{matrix}}
    &
    \begin{matrix}
        {\color{red}1} & {\color{red}1} & 0 & 0 & 0 & 0 \\
        0 & 0 & {\color{red}1} & {\color{red}1} & 0 & 0 \\
        0 & 0 & 0 & 0 & {\color{red}1} & {\color{red}1}
    \end{matrix} 
    & 
    {\color{lightgray} \begin{matrix}
        0 & 0 & 0 & 0 & 0 & 0 & 0 & 0 & 0 \\
        0 & 0 & 0 & 0 & 0 & 0 & 0 & 0 & 0 \\
        0 & 0 & 0 & 0 & 0 & 0 & 0 & 0 & 0 
    \end{matrix}}
    \\
    {\color{lightgray} \begin{matrix}
        0 & 0 & 0 & 0 \\
        0 & 0 & 0 & 0 \\
        0 & 0 & 0 & 0 \\
        0 & 0 & 0 & 0 \\
        0 & 0 & 0 & 0 \\
        0 & 0 & 0 & 0 
    \end{matrix}}
    & 
    {\color{lightgray} \begin{matrix}
        0 & 0 & 0 & 0 & 0 & 0 \\
        0 & 0 & 0 & 0 & 0 & 0 \\
        0 & 0 & 0 & 0 & 0 & 0 \\
        0 & 0 & 0 & 0 & 0 & 0 \\
        0 & 0 & 0 & 0 & 0 & 0  \\
        0 & 0 & 0 & 0 & 0 & 0 
    \end{matrix}}
    & 
    \begin{matrix}
        {\color{orange}1} & 0 & 0 & 0 & 0 & 0 & 0 & 0 & {\color{orange}1} \\
        0 & {\color{orange}1} & {\color{orange}1} & 0 & 0 & 0 & 0 & 0 & 0 \\
        0 & 0 & {\color{orange}1} & {\color{orange}1} & 0 & 0 & 0 & 0 & 0 \\
        0 & 0 & 0 & 0 & {\color{orange}1} & {\color{orange}1} & 0 & 0 & 0 \\
        0 & 0 & 0 & 0 & 0 & {\color{orange}1} & {\color{orange}1} & 0 & 0 \\
        0 & 0 & 0 & 0 & 0 & 0 & 0 & {\color{orange}1} & {\color{orange}1} 
    \end{matrix}
\end{pmatrix}$}
\caption{Grid matrix $\mathcal{G}_3$ (with layers {\color{purple}1}, {\color{red}2} and {\color{orange}3})}
\label{fig:grid-matrix}
\end{figure}

\smallskip

\noindent Each layer consists of a submatrix and the grid matrix will be :
\begin{equation}
\mathcal{G}_l=m_1\diamond m_2\diamond \hdots \diamond m_{l-1} \diamond m_l
\end{equation}

\smallskip

\noindent From $m_4$, these submatrices become:
\begin{equation}
m_i=
\begin{cases}
\mathbb{S}_{\frac{i}{2}} \diamond \mathbb{I}_{(\frac{i}{2}-1)} \diamond
\mathbb{S}_{\frac{i}{2}} \diamond \mathbb{I}_{(\frac{i}{2}-1)} \diamond
\mathbb{S}_{\frac{i}{2}} \diamond \mathbb{I}_{(\frac{i}{2}-1)}
 & \text{ if $i$ is even} \\ \vspace{-2mm} \\
\left(
\mathbb{I}_{\lceil \frac{i}{2}-2 \rceil} \diamond 
\mathbb{S}_{\lceil \frac{i}{2}\rceil} \diamond
\mathbb{I}_{\lceil \frac{i}{2}-2 \rceil} \diamond 
\mathbb{S}_{\lceil \frac{i}{2}\rceil} \diamond
\mathbb{I}_{\lceil \frac{i}{2}-2 \rceil} \diamond 
\mathbb{S}_{\lceil \frac{i}{2}\rceil}
\right)^{\hspace{-1.5mm}\searrow}
 & \text{ if $i$ is odd} 
\end{cases}
\end{equation}

\medskip

Assuming that $\mathbb{I}_0$ is a $0\times 0$ matrix, this pattern actually holds for $m_2$ and $m_3$. However, depending on how this is implemented, inputting the first 3 layers by hand might be the best option. \\

\begin{figure}[H]
    \centering
    $
        \begin{pmatrix}
        \vspace{-.15cm}
        \includegraphics[width=3cm]{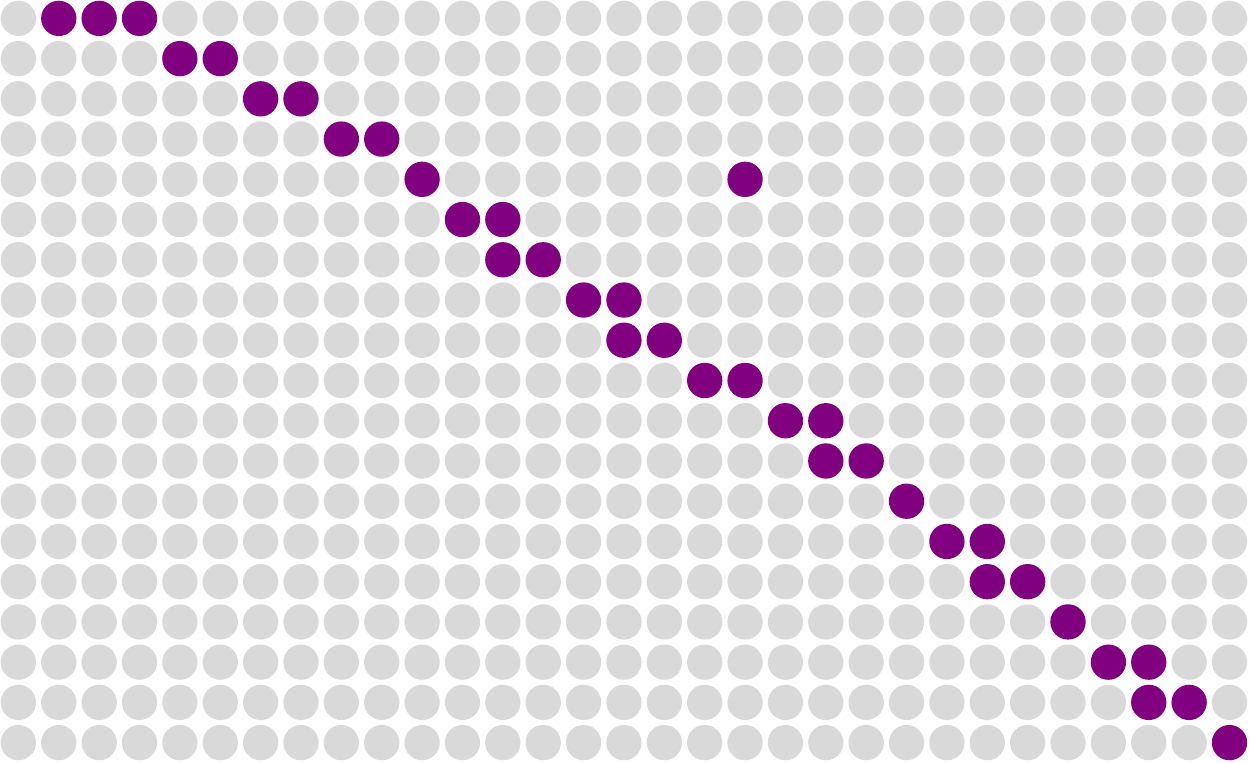}
        \end{pmatrix}
        \rightarrow
        \begin{pmatrix}
        \vspace{-.15cm}
        \includegraphics[width=3cm]{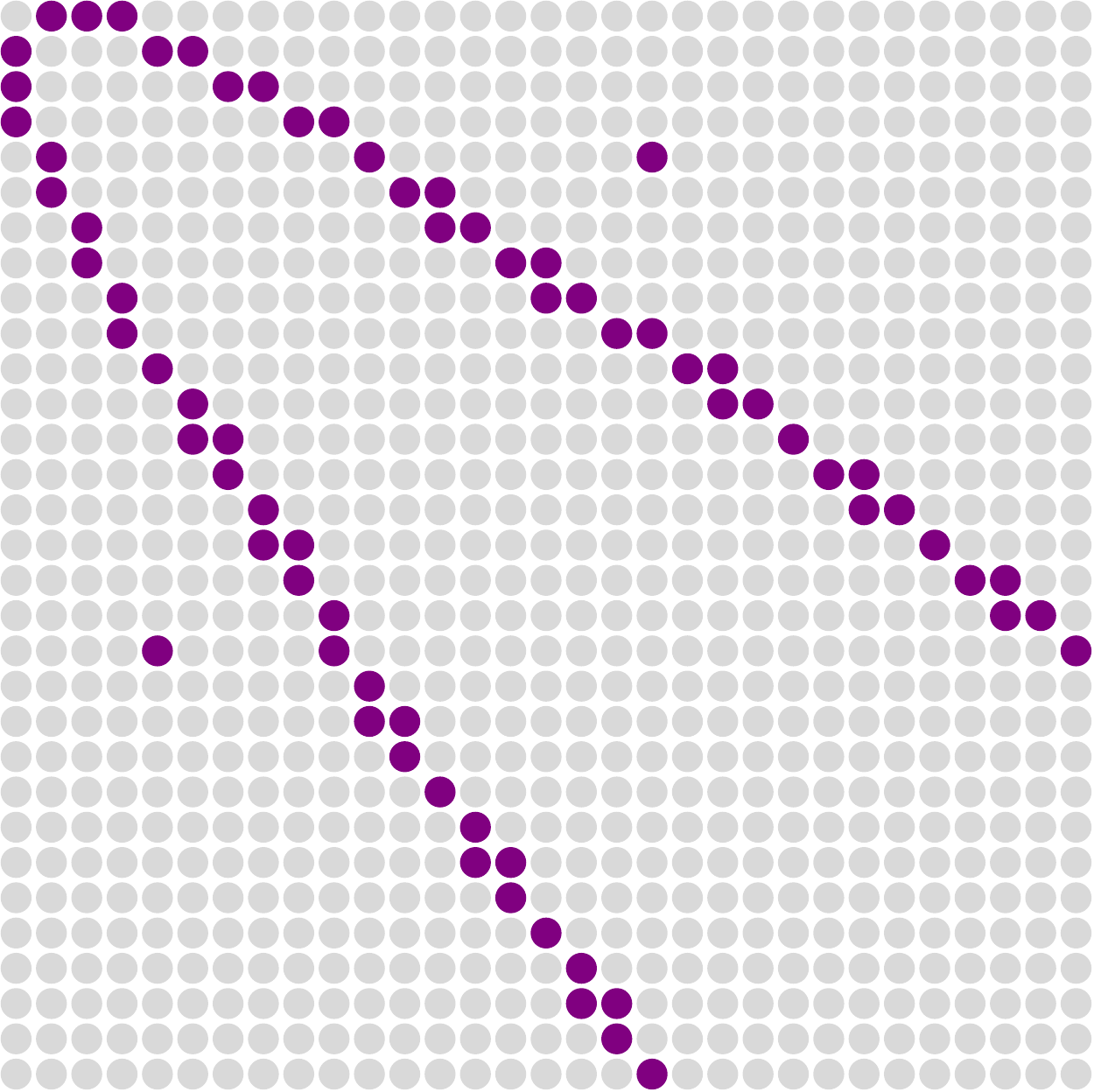}
        \end{pmatrix}
    $
    \caption{From the grid matrix $\mathcal{G}$ to the adjacency matrix $\mathcal{A}$}
    \label{fig:symmetrize}
\end{figure}

Once the grid matrix is built, it is easy to obtain the adjacency matrix by turning it into a symmetric matrix as illustrated in \textit{Figure \ref{fig:symmetrize}}.

The limit of this series of matrices $\mathcal{A}_\infty$ is the adjacency matrix of the graph corresponding to the infinite triangular grid.
 
\subsection{Evolving the state}

 The universe will be simulated using two layers around the region of influence of our initial structure. If the initial structure is a single triangle, then the computed grid will contain $t+2$ layers.

\bigskip

\noindent Updating the state of the grid will come in four steps.
\begin{enumerate}
\item 
A layer is added with the same state as the last vertex (both the grid/adjacency matrix and the state vector must be updated).

\item 
A \textbf{configuration vector} $\mathcal{C}$ is computed ($o$ is the order of the graph).
\begin{equation}
\mathcal{C}=
\begin{pmatrix} c(v_1) \\ \vdots \\ c(v_o) \end{pmatrix}
=4\times\mathcal{S}+\mathcal{A}\cdot\mathcal{S}
\end{equation}

\item 
The state vector $\mathcal{S}$ is then updated as follows.
\begin{equation}
    \mathcal{S}=R\,\text{@}\,\mathcal{C}
\end{equation}

\item 
The state of all vertices of the last layer (created in step 1) is set to the value of the last vertex of the now penultimate layer. This removes the artifacts coming from the edges of the computed grid. 
\end{enumerate}

\medskip

\underline{Remarks}
\begin{enumerate}
    \item[•] Evolving the state of the grid is where this framework pays off the most. Steps 2 and 3 are mathematically sufficient if we consider working in an infinite graph. They would also be the only steps required in a closed grid, like a triangulated surface \cite{zawidzkiApplicationSemitotalistic2D2011}, for example.
    \item[•] For this process to be efficient, it is necessary to encode the grid/adjacency matrix in a sparse array format.
    \item[•] Step 1 can be avoided if the three outermost layers have a uniform state, which is easy to check.
    \item[•] It is useful here to be able to retrieve the number of layers $l$ in the graph from its order $o$ and vice versa. 
    \begin{equation}
    l=\frac{1}{6}\Big(\sqrt{3(8\,o-5)}-3\Big)
    \qquad\qquad
    o=1+\frac{3}{2}\,l(l+1)
    \end{equation}
    \item[•] It is also useful to note that each layer $l$ contains $3l$ vertices or cells (except when $l=0$).
\end{enumerate}

\pagebreak
\subsection{Plotting the result}
2D coordinates are required to plot the resulting grid. These can be computed in a \textbf{coordinates matrix} $\mathcal{K}$. \textit{Algorithm \ref{expand-coords}} can be used to expand $\mathcal{K}$ from the coordinates of the first vertex, placed at the origin $\begin{pmatrix}0&0\end{pmatrix}$.

\begin{equation}
\mathcal{K}=
\begin{pmatrix}
x_1 & y_1 \\
x_2 & y_2 \\
x_3 & y_3 \\
\vdots & \vdots \\
x_{o-2} & y_{o-2} \\
x_{o-1} & y_{o-1} \\
x_o & y_o 
\end{pmatrix}
\end{equation}

\medskip

These coordinates will serve to translate a base triangle whose orientation depends on the layer it is in. 

\begin{table}[H]
\centering
\begin{tblr}{|Q[c,m]|Q[c,m]|Q[c,m]|Q[c,m]|Q[c,h]|}
    \hline
    Layer & \SetCell[c=3]{c} Coordinates of the base triangle& & & Illustration \\
    \hline
        even &
        $\begin{pmatrix}-\frac{1}{\sqrt{3}}& 0\end{pmatrix}$ &
        $\begin{pmatrix}\frac{1}{2\sqrt{3}}& \frac{1}{2}\end{pmatrix}$ &
        $\begin{pmatrix}\frac{1}{2\sqrt{3}}& -\frac{1}{2}\end{pmatrix}$ & 
        \raisebox{-.2\height}{\includegraphics[width=2em]{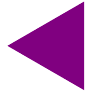}}
        \\
    \hline
        odd &
        $\begin{pmatrix}\frac{1}{\sqrt{3}}& 0\end{pmatrix}$ &
        $\begin{pmatrix}-\frac{1}{2\sqrt{3}}& \frac{1}{2}\end{pmatrix}$ &
        $\begin{pmatrix}-\frac{1}{2\sqrt{3}}& -\frac{1}{2}\end{pmatrix}$ &
         \raisebox{-.2\height}{\includegraphics[width=2em]{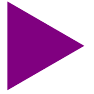}}
        \\
    \hline
\end{tblr}
\caption{Coordinates of the base triangle's vertices}
\label{table:triangles}
\end{table}

\begin{algorithm}[H]
\caption{Adding a layer to the coordinates matrix $\mathcal{K}$}\label{expand-coords}
\begin{algorithmic}[1]
    \If{$l$ is odd}
    \Comment{$l$ is the number of the new layer}
    \State $step \gets \begin{pmatrix}-\frac{1}{\sqrt{3}}&0\end{pmatrix} $
    \Else
    \State $step \gets \begin{pmatrix}-\frac{1}{2\sqrt{3}}&-\frac{1}{2}\end{pmatrix}$
    \EndIf
    \State $\mathcal{K}.append\Big(\mathcal{K}\big[-3(l-1)\big]+step\Big)$
    \Comment{$\mathcal{K}[-n] \equiv n^{th}$ coords from the end}
        
    \For{$i = 0, 3l-2$}
        \If{$i<\lfloor\frac{l}{2}\rfloor$}
            $step \gets \begin{pmatrix}0&1\end{pmatrix}$
        \ElsIf{$i<\lfloor\frac{l}{2}\rfloor+\lceil\frac{l}{2}\rceil$}
            $step \gets \begin{pmatrix}-\frac{\sqrt{3}}{2}&\frac{1}{2}\end{pmatrix}$
        \ElsIf{$i<2\lfloor\frac{l}{2}\rfloor+\lceil\frac{l}{2}\rceil$}
            $step \gets \begin{pmatrix}-\frac{\sqrt{3}}{2}&-\frac{1}{2}\end{pmatrix}$
        \ElsIf{$i<2\lfloor\frac{l}{2}\rfloor+2\lceil\frac{l}{2}\rceil$}
            $step \gets \begin{pmatrix}0&-1\end{pmatrix}$
        \ElsIf{$i<3\lfloor\frac{l}{2}\rfloor+2\lceil\frac{l}{2}\rceil$}
            $step \gets \begin{pmatrix}\frac{\sqrt{3}}{2}&-\frac{1}{2}\end{pmatrix}$
        \ElsIf{$i<3\lfloor\frac{l}{2}\rfloor+3\lceil\frac{l}{2}\rceil$}
            $step \gets \begin{pmatrix}\frac{\sqrt{3}}{2}&\frac{1}{2}\end{pmatrix}$
        \EndIf
        \State $\mathcal{K}.append\Big(\mathcal{K}\big[-1\big]+step\Big)$
    \EndFor
\end{algorithmic}
\end{algorithm}

\pagebreak
\section{Conclusion} \label{conclusion}
The triangular tessellation plays an important role in many disciplines, from computer graphics to architecture. Triangular Automata (TA) are a way of populating the triangular grid with aesthetic patterns. Beyond the possible applications, Elementary Triangular Automata (ETA) are somewhat fundamental cellular automata, making them an elegant model of complexity. With the framework presented here, the 256 ETA rules can now be thoroughly explored, even with limited computational resources. \\

\noindent Here are some possible directions for future work:
\begin{enumerate}
\item ETA rules could be classified according to some common criteria.
\item The approach taken in this paper could easily be applied to a wider class of TA.
\item A GPU-accelerated implementation could be used to explore longer timescales, as has already been done for Graph-Rewriting Automata \cite{GRAweb}.
\item Turing completeness and other interesting properties could be searched for in ETA rules.
\end{enumerate}

% References
\bibliographystyle{IEEEtran}
\bibliography{main.bib}

@inproceedings{cousin2022organic,
  title={Organic Structures Emerging From Bio-Inspired Graph-Rewriting Automata},
  author={Cousin, Paul and Maignan, Aude},
  booktitle={2022 24th International Symposium on Symbolic and Numeric Algorithms for Scientific Computing (SYNASC)},
  pages={293--296},
  year={2022},
  organization={IEEE},
  doi={10.1109/SYNASC57785.2022.00053}
}

@book{wolfram2002new,
  title={A new kind of science},
  author={Wolfram, Stephen and others},
  volume={5},
  year={2002},
  publisher={Wolfram media Champaign}
}

@article{weisstein2002elementary,
  title={Elementary cellular automaton},
  author={Weisstein, Eric W},
  journal={MathWorld},
  year={2002},
  publisher={Wolfram Research, Inc.},
  url= {https://mathworld.wolfram.com/ElementaryCellularAutomaton.html}
}

@misc{GRAweb,
title = {Graph-Rewriting Automata},
author = {Paul Cousin},
url = {https://paulcousin.github.io/graph-rewriting-automata}
}

@article{gerlingClassificationTriangularHoneycomb1990,
  title = {Classification of Triangular and Honeycomb Cellular Automata},
  author = {Gerling, R.W.},
  year = {1990},
  month = jan,
  journal = {Physica A: Statistical Mechanics and its Applications},
  volume = {162},
  number = {2},
  pages = {196--209},
  issn = {03784371},
  doi = {10.1016/0378-4371(90)90438-X},
  urldate = {2023-10-22},
  langid = {english}
}

@article{baysCellularAutomataTriangular1994,
  title = {Cellular {{Automata}} in the {{Triangular Tessellation}}},
  author = {Bays, Carter},
  journal={Complex Systems},
  volume={8},
  number={2},
  pages={127},
  year={1994},
  publisher={[Champaign, IL, USA: Complex Systems Publications, Inc., c1987-}
}

@article{imaiComputationuniversalTwodimensional8state2000,
  title = {A Computation-Universal Two-Dimensional 8-State Triangular Reversible Cellular Automaton},
  author = {Imai, Katsunobu and Morita, Kenichi},
  year = {2000},
  month = jan,
  journal = {Theoretical Computer Science},
  volume = {231},
  number = {2},
  pages = {181--191},
  issn = {03043975},
  doi = {10.1016/S0304-3975(99)00099-7},
  urldate = {2023-10-22},
  langid = {english}
}

@incollection{baysCellularAutomataTriangular2009,
  title = {Cellular {{Automata}} in {{Triangular}}, {{Pentagonal}} and {{Hexagonal Tessellations}}},
  booktitle = {Encyclopedia of {{Complexity}} and {{Systems Science}}},
  author = {Bays, Carter},
  year = {2009},
  pages = {892--900}
}

@incollection{baysGameLifeNonsquare2010,
  title = {The Game of Life in Non-Square Environments},
  booktitle = {Game of {{Life Cellular Automata}}},
  author = {Bays, Carter},
  year = {2010},
  pages = {319--329},
  publisher = {{Springer}}
}

@incollection{brecklingCellularAutomataEcological2011,
  title = {Cellular Automata in Ecological Modelling},
  booktitle = {Modelling {{Complex Ecological Dynamics}}: {{An Introduction}} into {{Ecological Modelling}} for {{Students}}, {{Teachers}} \& {{Scientists}}},
  author = {Breckling, Broder and Pe'er, Guy and Matsinos, Yiannis G},
  year = {2011},
  pages = {105--117},
  publisher = {{Springer}}
}

@inbook{linApplicationUnstructuredCellular2009,
  title = {Application of {{Unstructured Cellular Automata}} on {{Ecological Modelling}}},
  booktitle = {Advances in {{Water Resources}} and {{Hydraulic Engineering}}},
  author = {Lin, Yuqing and Mynett, Arthur and Chen, Qiuwen},
  year = {2009},
  pages = {624--629},
  publisher = {{Springer Berlin Heidelberg}},
  address = {{Berlin, Heidelberg}},
  doi = {10.1007/978-3-540-89465-0_108},
  urldate = {2023-09-11},
  collaborator = {Zhang, Changkuan and Tang, Hongwu},
  isbn = {978-3-540-89464-3 978-3-540-89465-0},
  langid = {english}
}

@incollection{naumovGeneralizedCoordinatesCellular2003,
  title = {Generalized Coordinates for Cellular Automata Grids},
  booktitle = {International {{Conference}} on {{Computational Science}}},
  author = {Naumov, Lev},
  year = {2003},
  pages = {869--878},
  publisher = {{Springer}}
}

@incollection{ortigozaACFUEGOSUnstructuredTriangular2016,
  title = {{{ACFUEGOS}}: {{An Unstructured Triangular Cellular Automata}} for {{Modelling Forest Fire Propagation}}},
  shorttitle = {{{ACFUEGOS}}},
  booktitle = {High {{Performance Computer Applications}}},
  author = {Ortigoza, Gerardo M. and Lorandi, Alberto and Neri, Iris},
  editor = {Gitler, Isidoro and Klapp, Jaime},
  year = {2016},
  volume = {595},
  pages = {132--143},
  publisher = {{Springer International Publishing}},
  address = {{Cham}},
  doi = {10.1007/978-3-319-32243-8_9},
  urldate = {2023-09-11},
  isbn = {978-3-319-32242-1 978-3-319-32243-8}
}

@article{pavlovaUsingCellularAutomata2020,
  title = {Using Cellular Automata in Modelling of the Fire Front Propagation through Rough Terrain},
  author = {Pavlova, A V and Rubtsov, S E and Telyatnikov, I S},
  year = {2020},
  month = oct,
  journal = {IOP Conference Series: Earth and Environmental Science},
  volume = {579},
  number = {1},
  pages = {012104},
  issn = {1755-1307, 1755-1315},
  doi = {10.1088/1755-1315/579/1/012104},
  urldate = {2023-09-11}
}

@article{saadatCellularAutomataApproach2018,
  title = {Cellular {{Automata Approach}} to {{Mathematical Morphology}} in the {{Triangular Grid}}},
  author = {Saadat, MohammadReza and Nagy, Benedek},
  year = {2018},
  journal = {Acta Polytechnica Hungarica},
  volume = {15},
  number = {6},
  pages = {45--62},
  issn = {17858860, 20642687},
  doi = {10.12700/APH.15.6.2018.6.3},
  urldate = {2023-07-05}
}

@mastersthesis{saadatCellularAutomataTriangular2016,
  title={Cellular {{Automata}} in the {{Triangular Grid}}},
  author={Saadat, MohammadReza},
  year={2016},
  school={Eastern Mediterranean University (EMU)-Do{\u{g}}u Akdeniz {\"U}niversitesi (DA{\"U})}
}

@article{saadatCopyMachinesSelfreproduction2023,
  title = {Copy {{Machines}} - {{Self-reproduction}} with 2 {{States}} on {{Archimedean Tilings}}},
  author = {Saadat, Mohammad Reza and Benedek, Nagy},
  year = {2023},
  journal={Journal of Cellular Automata},
  volume={17},
  pages={221--249}
}

@inproceedings{saadatGeneratingPatternsTriangular2021,
  title = {Generating {{Patterns}} on the {{Triangular Grid}} by {{Cellular Automata}} Including {{Alternating Use}} of {{Two Rules}}},
  booktitle = {2021 12th {{International Symposium}} on {{Image}} and {{Signal Processing}} and {{Analysis}} ({{ISPA}})},
  author = {Saadat, Mohammad Reza and Nagy, Benedek},
  year = {2021},
  month = sep,
  pages = {253--258},
  publisher = {{IEEE}},
  address = {{Zagreb, Croatia}},
  doi = {10.1109/ISPA52656.2021.9552107},
  urldate = {2023-07-05},
  isbn = {978-1-66542-639-8}
}

@article{uguzStructureReversibility2D2017,
  title = {Structure and Reversibility of {{2D}} von {{Neumann}} Cellular Automata over Triangular Lattice},
  author = {Uguz, Selman and Redjepov, Shovkat and Acar, Ecem and Akin, Hasan},
  year = {2017},
  journal = {International Journal of Bifurcation and Chaos},
  volume = {27},
  pages = {1750083}
}

@inproceedings{wainerIntroductionCellularAutomata2019,
  title = {An Introduction to Cellular Automata Models with Cell-{{DEVS}}},
  booktitle = {2019 {{Winter Simulation Conference}} ({{WSC}})},
  author = {Wainer, Gabriel A.},
  year = {2019},
  pages = {1534--1548},
  publisher = {{IEEE}}
}

@article{zawidzkiApplicationSemitotalistic2D2011,
  title = {Application of {{Semitotalistic 2D Cellular Automata}} on a {{Triangulated 3D Surface}}},
  author = {Zawidzki, M.},
  year = {2011},
  month = jan,
  journal = {International Journal of Design \& Nature and Ecodynamics},
  volume = {6},
  number = {1},
  pages = {34--51},
  issn = {1755-7437, 1755-7445},
  doi = {10.2495/DNE-V6-N1-34-51},
  urldate = {2023-09-12},
  langid = {english}
}

\end{document}